\documentclass[12pt,preprint]{aastex}

\def\arcs{$''$}

\def\be{\begin{equation}}
\def\ee{\end{equation}}

\begin{document}

\title{Strong Lensing Analysis of A1689 from Deep Advanced Camera Images}

\author{Tom Broadhurst\altaffilmark{1}, Narciso Benitez\altaffilmark{2,3}, Dan Coe\altaffilmark{2}, Keren Sharon\altaffilmark{1}, Kerry Zekser\altaffilmark{2}, Rick White\altaffilmark{7}, Holland Ford\altaffilmark{2}, Rychard Bouwens\altaffilmark{4},John Blakeslee\altaffilmark{2}, Marc Clampin\altaffilmark{5}, Nick Cross\altaffilmark{2}, Marijn Franx\altaffilmark{11}, Brenda Frye\altaffilmark{6}, George Hartig\altaffilmark{7}, Garth Illingworth \altaffilmark{4}, Leopoldo Infante\altaffilmark{8}, Felipe Menanteau\altaffilmark{2}, Gerhardt Meurer\altaffilmark{2}, Marc Postman\altaffilmark{7}, D.R. Ardila\altaffilmark{2}, F. Bartko\altaffilmark{9}, R.A. Brown\altaffilmark{7}, C.J. Burrows\altaffilmark{7}, E.S. Cheng\altaffilmark{10}, P.D. Feldman\altaffilmark{2}, D.A. Golimowski\altaffilmark{2}, T. Goto\altaffilmark{2}, C. Gronwall\altaffilmark{12}, D. Herranz\altaffilmark{13}, B. Holden\altaffilmark{4}, N. Homeier\altaffilmark{2}, J.E. Krist\altaffilmark{7}, M.P. Lesser\altaffilmark{14}, A.R. Martel\altaffilmark{2}, G.K. Miley\altaffilmark{10}, P. Rosati\altaffilmark{15}, M. Sirianni\altaffilmark{5}, W.B. Sparks\altaffilmark{5}, S. Steindling\altaffilmark{1}, H.D. Tran\altaffilmark{16}, Z.I. Tsvetanov\altaffilmark{2},W. Zheng\altaffilmark{2}}

\affil{
\altaffiltext{1}{School of Physics and Astronomy, Tel Aviv University, Tel Aviv 69978, Israel; tjb@wise3.tau.ac.il }
\altaffiltext{2}{Physics and Astronomy Dept, Johns Hopkins University, Baltimore Maryland, USA}
\altaffiltext{3}{Instituto de Astrof\'\i sica de Andaluc\'\i a (CSIC), C/Camino Bajo de Hu\'etor, 24, Granada, 18008, Spain}
\altaffiltext{4}{UCO/Lick Observatory, University of California, Santa Cruz, CA 95064.}
\altaffiltext{5}{NASA Goddard Space Flight Center, Code 681, Greenbelt, MD 20771.}
\altaffiltext{6}{Princeton University, Peyton Hall - Ivy Lane, Princeton, NJ 08544}
\altaffiltext{7}{STScI, 3700 San Martin Drive, Baltimore, MD 21218.}
\altaffiltext{8}{Departmento de Astronom\'{\i}a y Astrof\'{\i}sica, Pontificia Universidad Cat\'{\o}lica de Chile, Casilla 306, Santiago 22, Chile.}
\altaffiltext{9}{Bartko Science \& Technology, 14520 Akron Street, Brighton, CO 80602.}
\altaffiltext{10}{Conceptual Analytics, LLC, 8209 Woburn Abbey Road, Glenn Dale, MD 20769}
\altaffiltext{11}{Leiden Observatory, Postbus 9513, 2300 RA Leiden,Netherlands.}
\altaffiltext{12}{Department of Astronomy and Astrophysics, The Pennsylvania State University, 525 Davey Lab, University Park, PA 16802.}
\altaffiltext{13}{Istituto di Scienza e Tecnologie dell'Informazione "Alessandro Faedo", Via G. Moruzzi 1, 56124 Pisa, Italy}
\altaffiltext{14}{Steward Observatory, University of Arizona, Tucson, AZ 85721.}
\altaffiltext{15}{European Southern Observatory,Karl-Schwarzschild-Strasse 2, D-85748 Garching, Germany.}
\altaffiltext{16}{W. M. Keck Observatory, 65-1120 Mamalahoa Hwy., Kamuela, HI 96743}
}

\begin{abstract}

 We analyse deep multi-colour Advanced Camera images of the largest
known gravitational lens, A1689. Radial and tangential arcs delineate
the critical curves in unprecedented detail and many small
counter-images are found near the center of mass. We construct a
flexible light deflection field to predict the appearance and
positions of counter-images. The model is refined as new
counter-images are identified and incorporated to improve the model,
yielding a total of 106 images of 30 multiply lensed background
galaxies, spanning a wide redshift range, 1.0$<$z$<$5.5. The resulting
mass map is more circular in projection than the clumpy distribution
of cluster galaxies and the light is more concentrated than the mass
within $r<50kpc/h$. The projected mass profile flattens steadily
towards the center with a shallow mean slope of $d\log\Sigma/d\log r
\simeq -0.55\pm0.1$, over the observed range, r$<250kpc/h$, matching
well an NFW profile, but with a relatively high concentration,
$C_{vir}=8.2^{+2.1}_{-1.8}$. A softened isothermal profile
($r_{core}=20\pm2$\arcs) is not conclusively excluded, illustrating
that lensing constrains only projected quantities. Regarding
cosmology, we clearly detect the purely geometric increase of
bend-angles with redshift. The dependence on the cosmological
parameters is weak due to the proximity of A1689, $z=0.18$,
constraining the locus, $\Omega_M+\Omega_{\Lambda} \leq 1.2$. This
consistency with standard cosmology provides independent support for
our model, because the redshift information is not required to derive
an accurate mass map. Similarly, the relative fluxes of the multiple
images are reproduced well by our best fitting lens model.

\end{abstract}

\keywords{clusters lensing, Cosmological parameters}

\section{Introduction}\label{Introduction}

The puzzling ``dark matter'' phenomenon is strikingly evident in
the centers of massive galaxy clusters, where large velocity
dispersions are measured and gravitationally lensed arcs are often
formed.  Central cluster masses may be estimated by several means,
leading to exceptionally high mass-to-light ratios, $M/L \sim
100-300h(M/L_B)_{\odot}$, far exceeding both the stars responsible for
the light of the cluster galaxies, and the mass of plasma implied by
X-ray data. Reasonable consistency is claimed between dynamical,
hydrodynamical and lensing-based estimates of cluster masses,
supporting the conventional understanding of gravity. However, the
high ratio of mass-to-light implies an unconventional non-baryonic
dark material dominates the mass of clusters.

 In detail, a discrepancy is often reported between the strong lensing
 and X-ray mass measurements, in the sense that X-ray masses are lower
 in the cluster center. This may be attributed to gas dynamics in
 unrelaxed cluster (Allen 1998) or perhaps to the current restriction
 on X-ray spectroscopy to energies below $\sim 8$KeV, which often falls
 short of the Bremsstrahlung cutoff for the massive lensing clusters,
 making temperature measurements uncertain and less direct. In many
 cases luminous X-ray clusters display merger induced effects
 (Markevitch etal 2002, Reiprich etal 2004) and surprisingly detailed
 structure in SZ maps has also been reported for RX1347-1145 (Kitayama
 etal 2004), although sometimes lensing and X-ray derived mass
 profiles are claimed to agree eg. MS1358+6245 (Arabadjis, Bautz \&
 Garmire 2002). Lensing masses are often made uncertain by obvious
 substructures in the cores of massive clusters like A2218 (Kneib et
 al 1996), and weak lensing measurements are subject to observational
 problems (Kaiser, Squires \& Broadhurst 1995) and an inherent mass
 profile degeneracy (Kaiser 1995, Schneider \& Seitz 1995).

 Simulations of massive clusters based on interaction-less cold dark
matter are reliable enough to make statistical predictions for the
mass profiles of galaxy clusters. A relatively shallow central mass
profile is expected for cluster-sized haloes at the typical Einstein
radius of $20-100h^{-1}kpc$. The gradient of an ``NFW'' profile
(Navarro, Frenk \& White 1996) continuously flattens towards the
center, and for the most massive haloes is considerably flatter than a
pure isothermal profile interior to the characteristic radius, but
does not possess a constant density core in the center. The limiting
inner slope of this profile seems to depend on the resolution of
simulations, with somewhat steeper inner profiles claimed for more
detailed simulations (Ghigna et al. 1998, 2000, Fukushinge\& Makino
1997, Okamoto \& Habe 1999, Power etal. 2003, Navarro
etal. 2004), with an intrinsic variation in slopes predicted, related
to variations in the assembly histories (Jing \& Suto 2000, Tasitsiomi
etal 2004). Other more radical suggestions include self-interacting
dark matter (Spergel \& Steindhardt 2000), for which the largest
deviations should occur at high density, and hence this idea is
amenable to investigation via strong lensing (Miralda-Escud\'e 2002).

 Weak lensing has not yet provided any useful constraint on the 
mass profiles of galaxy clusters, with the best current data unable to
distinguish a singular isothermal profile from the NFW model (eg. Clowe etal
2000). This unfortunately follows from the near degeneracy
of weak lensing to the gradient of the mass profile. The shallower the
profile the more magnified images become, but their shapes are hardly
influenced, because the major and minor axes are stretched by nearly
identical factors, virtually independent of the gradient of the mass
profile. A firmer constraint may be made with magnification
information, breaking the mass-sheet degeneracy, but this requires
very deep imaging to overcome the intrinsic clustering of background
galaxies or redshift information to filter the clustering along the line
of sight (Broadhurst, Taylor \& Peacock 1995).  Measurements of
magnification have so far proved noisy with ground-based data, so that
the detection of this effect is currently restricted to only the most
massive clusters (Fort et al. 1998, Taylor et al. 1998, Croom \&
Shanks 1999, Mayen \& Soucail 2000, R{\" o}gnvaldsson et al.(2001),
Athreya et al.  2002, Dye et al. 2002).

 Strong lensing, leading to multiple images, occurs when the projected
mass density of a body exceeds approximately $1.0g/cm^2$ (Turner,
Ostriker \& Gott 1981), producing elongated images of extended
background galaxies (Paczynski 1986). This limit is surpassed it
seems for nearly all distant clusters identified in deep survey work
(Gioia et al. 1997, Gladders et al. 2003, Zaritski \& Gonzalez 2003,
Rosati et al. 2003) where giant arcs are commonly seen. However, at
low redshift, z$<$0.1, many fewer clusters are known to display giant
arcs. Here, lensing is harder to recognise because the isophotes of a
low redshift central cD galaxy extend over a larger angular scale,
exceeding the Einstein radius, which has only a weak dependence on the
cluster redshift, thereby burying the main arcs.  Only a few examples
of giant arcs have been identified in nearby clusters, requiring
ingenuity and careful extraction (Allen, Fabian \& Kneib 1996,
Blakeslee \& Metzger 1999, Blakeslee et al 2001, Cypriano et
al. 2001). Further work from space should perhaps be attempted given
the detailed complementary information available on the internal
dynamics and X-ray properties of the well studied nearby clusters.

 It is important to appreciate that the vast majority of strongly
lensed background galaxies do not resemble giant arcs and although
images near the critical radius must be highly elongated, in practice
they are often only marginally resolved, even in high resolution HST
data. This is because the faint galaxy population is intrinsically
small, FWHM $\sim 0.2$\arcs at faint magnitudes, suffering pronounced,
$\sim (1+z)^{-1.5}$, angular size evolution (Bouwens,Broadhurst \&
Silk 1998a,b, Bouwens et al. 2003). In the limit of point sources,
like quasars, resolving the elongation is infeasible.  Additionally,
substructure in the cores of clusters complicates the appearance of
arcs, so that images formed in regions between sub-clumps may be
stretched roughly equally in all directions, leaving the shape of an
image little affected, even if the source is highly magnified.  This
is also the case for images formed in an annulus lying between the
radial and tangential critical curves, on the contour where the
surface mass density is equal to the critical density (see for
example, Figures 14,15\&16), separating an inner region where images
are radially directed from the outer area where they are predominantly
tangential in shape.

 The mass contained within the Einstein radius is given by fundamental
constants and with knowledge of the distances involved and is
therefore independent of the mass profile, provided the critical curve
is approximately circular. Typically, significant asymmetry is evident
and corrections must be made, though these are small in some
favourable cases, for example the well studied clusters Cl0024+17
(Colley et al. 1995, Broadhurst et al. 2000), A370 (B{\' e}zecourt etal
1999) and MS2137 (Hammer et al. 1996, Kneib et al. 2002, Sand et al,
2002, Gavazzi et al 2003, Treu et al. 2003).

 The clearest example of multiple lensing around a galaxy cluster is
arguably the symmetric system identified around Cl0024+17, consisting
of four tangential images lying on a nearly circular ring of radius
$\approx 31$\arcs ~(Smail et al. 1994) and a small additional central
image (Colley et al. 1995). These, together with a measurement of the
redshift of the lensed system have been used to produce an accurate
central mass for the region enclosing the Einstein radius, yielding a
precise ratio for the center, $M/L(r<100kpc/h)=320\pm20 h(M/L_B)_{\odot}$
(Broadhurst et al. 2000). In modeling this cluster, Broadhurst et al.
(2000) find that the mass distribution closely follows that of the
central luminous galaxies and the lensed images are readily reproduced
in detail with only a modest number of parameters. A second pair of
multiple images is identified for this cluster and predicted to lie at
$z=1.3$, given their smaller relative deflection compared with the
main system of arcs at $z=1.67$. Careful work has also been performed
on a number of other clusters, most notably A2218, MS0404 and MS2137
(Kneib et al. 1996, Hammer et al., Treu et al. 2003, Gavazzi et al. 2003) for
which two or three sets of multiple images are identified in each
cluster, with varying conclusions regarding the mass profile depending
on assumptions about the symmetry of the dark matter and the relative
contribution of the central cD galaxy, which is particularly prominent
in the case of MS2137.

 Here we concentrate on A1689, which has the largest known Einstein
radius of all the massive lensing clusters, of approximately 50\arcs~
in radius, based on the radius of curvature of a giant low
surface-brightness arc. Relatively little work on this cluster has
been carried out with HST, and the field of WFPC2 is too small to
cover the full area interior to the Einstein radius. Although no
actual multiple images had been identified prior to this
investigation, we were confident that the exceptional depth and high
resolution of the Advanced Camera would lead to the detection of many
sets of multiple images, more than possible around other clusters, by
virtue of the large Einstein radius.

 We begin by describing the target selection ($\S$\ref{Target}),
observations ($\S$\ref{Observations}), and photometric analysis
($\S$\ref{Photometry}).  We then visually identify the most obvious
multiply lensed systems ($\S$\ref{IMII}), allowing us to develop an
initial mass model and refine the model in an iterative process.
Section $\S$\ref{MIS} describes the 30 multiply-imaged sources we have
identified.  We integrate the cluster light ($\S$\ref{LightMap}) for
comparison with our mass model.  Section $\S$\ref{Modeling} describes
the actual procedure to construct the model.  The resulting mass map,
comparisons to other work, and cosmological implications are described
in sections $\S$\ref{Modeling}-\ref{Cosmo}.  Finally, we summarise our
conclusions ($\S$\ref{Conc}).  Note, throughout we adopt $H_0 = 100\,
\textrm{km/s/Mpc}$ to allow comparison with earlier work.

\section{Target Selection}\label{Target}

 The ACS Guaranteed Time Observations (GTO) program includes deep
observations of several massive, intermediate redshift galaxy
clusters. Our aims are to determine the distribution of the matter in
clusters, to place new constraints on the cosmological parameters and
to study the distant lensed galaxies, taking advantage of the large
magnifications.

 In selecting a target for a deep lensing study, we are not tempted to
pursue the well known systems with notorious giant arcs. Instead we
select our target principally by the size of the Einstein radius, in
order to uncover as many multiply lensed background galaxies as
possible. All background galaxies whose images fall within an area of
approximately twice the Einstein radius will belong to a set of
multiply lensed images, thus a larger Einstein radius will provide
greater numbers of multiple images. With sufficiently deep
multi-colour images of high spatial resolution, we may confidently
integrate for long periods, secure that for systems of large Einstein
radius many examples of multiply lensed background galaxies will be
registered. For this reason A1689 is the preferred choice being the
largest known lens, with an Einstein radius of approximately 50\arcs,
much larger than other better studied lensing clusters with typically
only $\sim 15$\arcs ~radius, corresponding to a of factor of $\sim 10$
times more sky to a fixed magnification and hence a similar gain in
the numbers of expected lensed galaxies around the Einstein radius,
depending on the slope of the faint galaxy counts. In addition, the
relatively low redshift of A1689 means we are not concerned so much
with foreground contamination, thereby minimizing potential confusion
when identifying counter-images.

We prefer to image deeply only a few clusters rather than make a
larger survey, to secure significantly new information regarding the
nature of dark matter. An approximate rule for cluster mass modeling
is that the `resolution' with which we can map the mass distribution
depends on the surface density of multiply lensed images. To find many
of these systems, the observations have to be deep. However, it is
very difficult to get redshifts for galaxies fainter than $I\gtrsim
24.5-25$ even with the best instrument/telescope combinations
available from the ground. In order to overcome this limitation, we
have split our observations into 4 filters, $F475W$, $F625W$, $F775W$
and $F850LP$, in order to obtain reliable photometric redshift
information, which may be compared with the known redshifts of some of
these arcs that we have obtained already (Frye, Broadhurst \& Benitez
2000).

\section{Observations}\label{Observations}

A1689 was observed in June 2002 with the newly installed ACS on HST.
The ACS images are aligned, cosmic-ray rejected, and drizzled together
using the ACS GTO pipeline (Blakeslee et al.\ 2003). We utilise the
full spectral range of the Advanced Camera, matching the relative
depths of exposures in the g,r,i,z passbands to the relative
instrumental sensitivity. In total we imaged 4 orbits in G and R and 3
in I and 7 in Z (see Table 1).  We use the (09/25/02) CALACS
zeropoints, offset by small amounts necessary for the errors present
in this calibration. We reach $10\sigma$ magnitudes for point sources
(inside a $4\times$FWHM aperture) of $27.5$ in the $g$ band, $27.2$ in
the $r$ and $i$ bands, and $26.7$ in the $z-$band.  Table
1 summarises the observations.  The depth of the data and colour
coverage are unprecedented for deep lensing work, and as we see below,
sufficient for reliably identifying many sets of multiply lensed
images.  .

  We complemented our ACS-based observations with $U-$band
observations obtained with the DuPont telescope at Las Campanas
Observatory, with a final PSF FWHM$=$0.59\arcs ~and also $J,H,K$ data
at La Silla with the NTT telescope, with PSFs width of $\approx 0.8$
\arcs and AB limiting magnitudes of 24.21, 22.83 and 22.2
respectively.  These datasets will be discussed in more detail in an
upcoming paper (Coe et al. in preparation). Although they are far from
matching the depth of our ACS data, nevertheless they are useful in
improving our photometric redshifts for the brighter or redder
galaxies.

\section{Photometry}\label{Photometry}

Photometry of faint objects in a cluster crowded field presents
several challenges.  The standard software for this task, SExtractor
(Bertin \& Arnouts 1996), cannot be applied directly to the
images. The presence of bright, extended galaxies complicates the
estimation of the real background, and even if this problem is solved
by including an external background estimation, the software is not
able to de-blend even moderately faint objects from the central cluster
galaxies. To overcome this problem, we have carefully fitted and
subtracted the central cluster galaxies. This process will be
described in detail in an upcoming paper by Zekser et al. (in
preparation).  We combine the $g$, $r$, $i$ and $z$ images, weighting
by the inverse of the variance of each image to create a detection
image. After the bright galaxy subtraction, SExtractor produces a
detection of most of the faint objects in this image, although in some
difficult cases the apertures have to be defined manually. We measure
isophotal magnitudes within these apertures, which have been shown to
produce more accurate colours and photometric redshifts (Ben\'\i tez et
al. 2004).  We use the same ACS-defined apertures for measuring
magnitudes in the ground based images, correcting for the differences
in the PSF using a new software we have developed (Coe
et al., in preparation).

\subsection{Photometric redshifts}\label{Photo-z}

  We estimate photometric redshifts using the Bayesian based analysis
code, BPZ (Ben\'\i tez 2000) and the new set of templates introduced
in Ben\'\i tez et al. (2003).  BPZ produces a full redshift
probability distribution of the form: 

\be\label{bpz} p(z|C)\propto \sum_T p(z,T|m_0)p(C|z,T) \ee 

where $p(D|z,T)$ is the redshift likelihood obtained by comparing the
observed colours $C$ with the redshifted library of templates $T$. The
factor $p(z,T|m_0)$ is a prior which represents the redshift/spectral
mix distribution as a function of the observed $I-$band magnitude. We
use a prior which describes the redshift/spectral type mix in the
HDFN, which has been shown to significantly reduce the number of
``catastrophic'' errors ($\Delta z >1$) in the photometric redshift
catalog (see Ben\'\i tez et al 2004 and references therein).  We have
modified this prior to adapt it to this particular catalog of objects,
which are known to be background to the cluster and strongly
magnified. The prior excludes the redshift range $z<0.7$, and assumes
that the lensing corrected fluxes of the galaxies can be up to 20
times fainter than observed.  The resulting redshifts for each
individual arc are listed in Table 2.  To obtain the redshift of the
system, we take advantage of the Bayesian framework, and combine all
the individual redshift probability distributions into a single
probability for the system, $p(z|C)\propto \prod p_i(z|C)$, where
$i=1,...,n$ corresponds to each of the multiple images. This may seem
to have the same of effect of simply adding together all the observed
fluxes and then estimating the redshift of the combined system. This
would be the case if the only source of error in our photometric
measurements were random noise, but unfortunately this is not so, in
some cases, due to the presence of nearby residuals from the bright
galaxy subtraction, our apertures are contaminated by spurious light,
a problem especially common for the ground based
observations. Combining the redshift probabilities together
automatically ``prunes'' some of the spurious peaks, and quickly shows
when one of the individual arc photometry is seriously contaminated
and should be excluded from the redshift estimation for the whole
system.  The final redshifts for the whole systems are presented in
Table 2.

\section{Initial Multiple Image Identification}\label{IMII}

 Inspection of the colour image of the cluster provides, after some
hours of scrutiny, several convincing cases of multiple imaging, which
are later verified using the model. For example, the very red galaxy
identified by Frye, Broadhurst \& Benitez (2000) at z=4.9 (object
$7.1$, Figure 5) has an obvious symmetrically placed counterpart of
the same unusual colour in the form of a radial arc close to the
center of the cluster (object $7.2$, Figure 5). The relative rarity of
such bright red images adds confidence in this case. For bluer
galaxies it is harder to choose between the many faint similarly blue
background images and experience warns us that counter images can form
in the most unlikely places. Small changes in the scaling of the
bend-angle from the unknown source distance alters the location and
even the formation of counter images, hence it becomes clear that the
guidance of a useful model is needed to make reliable progress in
reducing confusion. Redshift information is very useful as it helps
narrow the range of allowable values of $d_{ls}/d_s$ when searching
for counter-images. The high spatial resolution of ACS data enables
morphological detail and internal colour variations to be used in
identifying counter images. The generally complex morphology of faint
galaxies means we can often identify unique internal features that
must reproduce between images of the same source and must obey parity
inversions.  Usually some model guidance is needed here, as different
arcs can be stretched along different directions, emphasizing
different features.

 Some sets of images visible around the cluster are so unique that
they are undoubtedly related. For example, five sets of images
are visible as shown in Figure 1\&2, labeled as objects 1\&2. This
system comprises a close pair of galaxies which is repeated four times
around the cluster. A fifth pair of images is subsequently identified
with the help of the model (see below, Section \ref{MIS}). One member has a
secure redshift of z=3.05 (Frye et al. 2000). We will show below that
our modeling places the other neighbouring image at a slightly lower
redshift of z$\sim$2.5. Other images near to this pair can then be
identified as multiple images, and then we may start on this limited
basis to build a model.

\section{Multiple Image Systems}\label{MIS}

 Here we make notes regarding the individual sets of multiple images
identified in the process of modeling (the modeling procedure is
described in detail below). We include a table of their locations,
photometric properties, and spectroscopic redshifts where
measured. About one fifth of the lensed galaxy images described here
were initially found by eye after careful scrutiny of the full colour
image (see Figure 5), allowing an initial mass model to be constructed
which in turn allows us to predict and verify the existence of more
lensed images.  When identified, these new images are added to refine
the model in an iterative process, finally resulting in a robust mass
model that reproduces nearly all of the lensed galaxy images
accurately, in terms of their positions, morphology and relative
magnifications. All the images identified by eye are confirmed by the
model and the plausibility of model predicted images are examined
carefully by eye.

Each set of images is shown in colour as a set of ``postage stamps''
in Figures 1--4, demonstrating rather obviously their relation to the
parent source galaxy in most cases. Also shown alongside these
observed images are the model generated images used to help identify
counter images and scaled to the best fit deflection angle,
$d_{ls}/d_s$. Also included is the de-lensed appearance of the source
in the source plane, based on the most magnified image of a given
source, which is placed in column 1, of Figures 1--4.  The model
images are generated by simply de-lensing one of the members of a set
of images with the best-fit model for the deflection field, treating
each pixel of the chosen input image as a set of pixels and not by
mapping it first onto fixed grid in the source plane. This has the
advantage that when re-lensing the source into the image plane to
generate the model counter-images that the resolution is maximally
preserved, as each pixel in the observed image can be re-mapped. This
is obviously preferable to re-lensing a binned source plane, which
would lead to a loss of resolution in the image place.

The stamps are shown with the same physical scale so they may be
compared readily, except in the case of small images where we magnify
the scale of the stamp, by an amount which we label on the stamp, in
order to see the detail better.  The source images are also shown
magnified, usually by a factor of 5-10, so that their detailed
internal structure revealed by lensing may be appreciated
better. Corresponding observed and model-generated images typically
agree to within 1-3\arcs ~for the best-fit model, so the position of
each model stamp has been centered on the model image in figures 1--4,
for a better comparison.

 The resolution of our re-lensing procedure is matched to that of the
data (i..e performed on a 4K grid, which is an effective limitation
imposed by the FFT of the mass distribution when constructing the
deflection field) and consequently the reproduction can look a little
blurry in some cases. The image chosen for de-lensing is the first one
in each row of multiple images shown in figures 1--4, and usually has
a slightly sharper appearance than the rest, as it is simply a
re-lensed version of itself. Usually we choose the largest image for
this purpose since then information is generally only lost in creating
the other smaller counter images. This does not work quite so well
when the predominant lens stretching of a counter-image is not matched
well in direction to that of the input image, especially if the source
is intrinsically elongated orthogonally to the main direction in which
an image is stretched as then the magnification of such images will
emphasise different internal features, that may not be well resolved
in the input image. Similarly, we have avoided where possible using
input images which lie very close to caustics, since then the location
of the caustic relative to the image must be known exquisitely well in
order that the counter images and the source have meaningfully
predicted shapes.

Photometric redshifts are estimated for all images and usually these
agree well, with the exception of very faint counter-images, or in
cases where there is light contamination from a neighbouring cluster
galaxy.  In the seven cases where we have spectroscopic redshifts, the
agreement with the photometric estimate is good. These are of course
generally bright images allowing successful spectroscopic redshift
measurements, so the precision of the photometry is relatively good.

\subsubsection*{Sources 1 \& 2}

 This is a close pair of galaxies separated by $\simeq 1$\arcs ~and
identified four times around the critical curve with both radial and
tangential parity reversals (see Figures 1,2,\&5).  The model is able
to reproduce this pair of objects very well.  Source 1's images have a
slightly larger deflection angle, requiring that it be slightly more
distant than source 2.  In addition, a de-magnified pair is predicted
by the model to fall close to the cluster center of mass, lying within
the radial critical curve. These two images are in fact identified
close to the predicted locations with the correct sizes, relative
orientations, and colours (stamp $1.5$ of Figure 1).  The brightest
image, $1.1$, is actually two closely split images due to the
proximity of a luminous cluster member which locally perturbs the
critical curve of the cluster. In total there are 7 images of source~1
and 5 of source~2, all fully accounted for by the model and with no
additional images predicted by the model.

The redshift of image $1.1$ is known from many low ionization
absorption lines in a very high quality Keck spectrum to be $z=3.05$
(Frye, Broadhurst \& Benitez 2000) and is in good agreement with the
photometric redshift estimate z=3.2$\pm$0.3. The spectrum also shows,
rather unusually, 2 damped systems at lower redshift at $z=2.5$ and
$z=2.3$. The photometric redshift for object~2 lying close to object
1, is estimated to be z=$2.5\pm0.2$, and so $z=2.5$ is adopted when
comparing the data with predictions of the model. One of the two
damped systems in the spectrum of $1.1$ is very likely caused by
extended gas associated with object 2, given their proximity. It is
interesting to appreciate that this extended gas must cover a good
fraction of the critical curve of A1689, following the distribution of
the images of sources 1 \& 2 around the whole lens. Deep narrow-band
imaging tuned to Ly$_\alpha$ at the redshift of this damped system may
be rewarded with a continuous emission map of the critical curves!

\subsubsection*{Source 3}

 There are 3 images of a very red dropout galaxy with a photo-z of
z=5.3$\pm0.4$. The most magnified images are a merging pair 3.1 \& 3.2
crossing the critical curve. Two more images are predicted by the
model. After subtracting an elliptical cluster member galaxy, we locate
3c, which is much fainter, as predicted, and lies close to our
detection limit. A tiny counter image is also predicted on the
opposite side of the cluster, however, this image is not identified
and seems to lie below our threshold as we may expect based on its
predicted flux which is a factor of 3 less than the barely-detected
image 3.3. 

\subsubsection*{Source 4}

 This is a distinctive high surface brightness galaxy with a clear
internal colour variation, a point-like nucleus and an accompanying spot
lying off one end. Four images are apparent, including the obvious
mirror symmetric pair with reverse parity 4.1 \& 4.2 and two smaller
images confirmed by the model 4.3 \& 4.4. This set of images corresponds
to relatively small deflections, $\sim$15\% smaller than that of
images 1 \& 2, and indeed the photo-z estimate is lower z=$1.1\pm
0.15$. This is an important set of arcs for the lens model, expanding
the redshift range of the background sources and because its images
lie all around the lens providing a tight constraint on the model. A
central de-magnified image is predicted, 4.5, and is clearly identified
by its characteristic colours (half red and half white) with the
predicted parity. It lies close to the brightest cD galaxy.

\subsubsection*{Source 5}

 A pair of radially directed mirror images are seen close to the
center of the cluster bisected by the radial critical curve.  One
counter image, 5.3, is predicted rather far off to one side of the lens
and is readily identified near this position with the
corresponding orientation, morphology and colour. The photometric
redshift for this system is $3.2\pm0.4$, indicating some flux has
``dropped out'' in the G band.

\subsubsection*{Source 6}

 Three very similar large images of an obvious disk galaxy with large
internal colour variation and structure are seen associated with the
main subgroup of the cluster. A fourth image, 6.4, is predicted by the
model within the isophotes of a luminous cluster galaxy in this
subgroup, and is readily found after subtraction of the cluster light
(see Figure 4). This set of images has a reliable photometric
redshift of $z=1.2\pm0.15$.  This set of arcs helps to constrain well
the relative mass of the main sub-group of the cluster.

\subsubsection*{Source 7}

 A high surface brightness high redshift galaxy produces two prominent
red images, one tangential (7.1) and one radial (7.2). The measured
redshift of the brighter tangential image, 7.1, is z=4.9 (Frye,
Broadhurst\& Benitez 2000), in close agreement with the photometric
redshift $z=4.8\pm0.4$. The redshift of the radial counter image has
also been measured and confirmed to lie at the same redshift (Bernard Fort 
priv. comm). A tiny central image is predicted and tentatively detected
as the only red speck, 7.3, very close to the predicted position, near
the most luminous cD galaxy. Deeper images would help clarify this
along with many other small counter images in the center.

\subsubsection*{Source 8}

 This is a 4 image system with a similar arrangement to Source 4, but
with a slightly larger set of bend-angles corresponding to a higher
redshift source. A measured redshift of z=1.8 has been measured
(Bernard fort, Priv. Comm).  Images 8.1 \& 8.2 are a continuous mirror
symmetric pair of images forming the only giant arc around this
cluster and bisected by the tangential critical curve at an oblique
shallow angle (see Figure 1). A fifth image is predicted in the very
center of the cluster and a likely candidate is identified by colour
and shape near the predicted position (stamp 8.5 of Figure 1)

\subsubsection*{Source 9}

This is another high-z dropout galaxy with a photometric redshift of
$z=4.7\pm0.4$. Image 9.1 is highly elongated and given its location
requires that other multiple images should be formed. The model is
essential here in finding the counter images. Three other images are
predicted in total and all are securely identified close to
the predicted locations. The shapes of the predicted images are not
accurately matched which we blame on the proximity of the brightest 
image to the tangential critical curve in one of the most magnified
regions of the lens (see Figure 1) and therefore the degree of 
stretching is enormous and not well aligned with the other images.
This problem is very similar to that of object 12 which lies close by.

\subsubsection*{Sources 10,15,18}

Two very similar looking blobby white images are seen on diametrically
opposed locations and confirmed by the model, flanked by a pair of
resolved blue objects.  The photo-z of the brighter white object is
$z=2.02\pm0.3$. Two much fainter bluer neighbouring objects (objects
15 \& 18) are also visible with photo-z consistent with z=2.02 which
we may take to be independent galaxies with very similar redshifts to
No. 10.  The model shows that under this assumption the relative
orientation and locations of this close triplet is consistent with all
three objects being at the same distance, and thus nearly identically
deflected. A third appearance of this set of three galaxies is
predicted near the very center of the lens, as small images near
the central cD galaxy and these are quite readily identified by their
relative orientations and colours, and labeled 10.3,15.3 \& 18.3.

\subsubsection*{Source 11}

This is a pair of images of a nicely resolved spiral galaxy lying on
opposite sides of the lens, with a photometric redshift of
$z=2.9\pm0.2$. One image is tangential and the other lies in the
region between the radial and tangential critical curves where images
are stretched roughly equally in all directions, producing a
relatively undistorted image of the source which lies in the rather
highly magnified region between the critical curves (see Figure 2 for
relative magnification map) depending on the local slope of the mass
profile. The model reproduces these very well, 11.1,11.2. A central
image 11.3 is predicted and identified, close to the predicted
position, with matching size, shape, colour and surface
brightness. This may be the most distant known example of spiral
galaxy, whose morphological identification is helped by the factor of
$\approx10$ in magnification.

\subsubsection*{Source 12}

A giant blue arc with a mirror symmetric pattern forming two images
12.2 \& 12.3 has a reliable redshift of z=1.83 from many emission lines
with an AGN-type spectrum (Bernard Fort, priv. Com.). We predict this object to have 3 other
images spread around the lens. One of these images, 12.1, has a redshift
measurement in agreement with 12.2,12.3, with the same unusual
emission line spectrum.  The two other images predicted do
not have redshift estimates, being much fainter but are readily
identified. Note that this source has two components in the source
plane which are resolved in images 12.1,12.4, and which are evidently
elongated normal to a line connecting the two components and therefore
form a continuous-looking image. Again the photo-z is in good
agreement with the spectroscopic redshift (Table 2).

\subsubsection*{Source 13}

 A giant arc is located at the apex of the main subgroup of galaxies
and on close inspection is resolved into three images straddling the
tangential critical curve with a photo-z of $z=1.6\pm0.2$. No other images
are predicted by the model, as the source is evidently too far from
the center of mass of the cluster. This source, together with source 30, 
helps fix the location of the critical curve around the outskirts of the
the secondary group.

\subsubsection*{Source 14}

 A pair of closely separated highly elongated images 14.1 \& 14.2 shows the
location of the critical curve which they straddle. The photo-z for
this object is $3.46\pm0.3$, dropping out in the B-band. The close
separation of these two images means that it does not significantly constrain
the lens model, but it is very sensitive to the location of the critical 
curve which must pass between them.

\subsubsection*{Source 16}

 Two images of a very lumpy blue source are readily identified by their 
common morphology. Image 16.3 is highly radially stretched, but the model
verifies that an elongated image can form at this location. The
photo-z for this is $z=1.8\pm0.37$.

\subsubsection*{Source 17}

 A close pair of radially elongated images of similar morphology are
found straddling the radial critical curve, with a photo-z of
$z=2.27\pm0.25$. The predicted third image, 17.3, is readily
identified close to the expected position with a morphology very close
to the model prediction in shape and position angle on the opposite
side of the lens.

\subsubsection*{Source 19}

 This source has two clearly related images 19.1,19.2 with a photo-z
estimate of $z = 2.7\pm 0.2$. The source is elongated in the direction
of the shear for both of these two images. A close pair of images is
predicted and stretched perpendicularly to the intrinsic elongation of
the source. Close to the location of this predicted image is an
obvious pair of closely separated mirror symmetric images, 19.3 \&
19.4, but with colours which are slightly greener than 19.1 \& 19.2
certainly due to contamination with images 8.1 \& 8.2.  A fifth image,
19.5, is also predicted to lie inside the radial critical curve close
to the center of mass, and is readily identified by its orientation
and elongated morphology. The photometric redshift is $z=2.7\pm0.2$
(estimated for the uncontaminated images 19.1,19.2).

\subsubsection*{Source 20}

Two relatively large images with similar internal colours and
structure are found close to each other. The model convincingly
reproduces the internal structure and relative locations. The 
photometric redshift is relatively low, $z\sim1.57\pm 0.2$.

\subsubsection*{Source 21}

A spotty blue radially stretched image is linked to two
other images after the application of the model. The lumpy internal
structure is replicated by the model and helps to correctly identify
the counter images. The photometric redshift is rather uncertain for
this faint blue object, with a most likely value of $z=1.86\pm0.35$

\subsubsection*{Sources 22 \& 23}

Two high surface brightness poorly resolved sources lie close to each
other and modeling shows that this pair is linked to two other sets of
very similar images. The faintest pair lies interior to the radial
critical curve. The outer pair is the brightest, and although these
are considerably tangentially magnified they remain only marginally
resolved, indicating the sources of these two objects are
intrinsically very small (Figure 3). The photometric
redshifts are consistent $z=2.24\pm0.2$, in agreement with the model
which finds their deflection angles have very similar scales.

\subsubsection*{Sources 24 \& 29}

A somewhat unusual pair of extended objects are confirmed by the model
to form 5 pairs of images around the lens.  The higher surface
brightness object appears to be a barred galaxy and the low surface
brightness accompanying smudge is not particularly galaxy-like in
appearance. One image of this latter object falls amongst the massive
galaxies of the main subgroup of cluster galaxies and forms a very
highly magnified large trail spread between two of the massive
galaxies. A small pair of counter images is predicted to lie within
the radial critical curve and is readily identified close to the
predicted location. A photometric redshift of $z=2.4\pm0.3$ is
measured. The appearance of this pair of images in the source plane
suggests the source is one relatively large spiral galaxy with a large
central bar.

\subsubsection*{Source 25}

A large green radial arc bisecting the radial critical curve is
identified using the model with a small barely resolved image of the
same colour lying well outside the tangential critical curve.  No more
images of this source are predicted by the model. This case
illustrates well the difficulty that one faces in identifying
prominent radial arcs with shallow imaging data - the source is
fortuitously located so that one image falls precisely on the radial
critical curve forming a very magnified image whose counter image is
much more modest, with a much smaller magnification. This object drops
out in G and has a photometric redshift $z=3.79\pm0.15$.

\subsubsection*{Sources 26 \& 27}

A close pair of pink and blue images are found at three locations
forming 6 images. In one case the two images are coincident and seem
to form one object. The distances of these two sources are not quite
identical and indicate that these sources are unrelated, so that
their relative positions can even overlap as seems to occur in the
case, 26.3 \& 27.3. For the combined object a photo-z of $z=1.7\pm0.2$
is estimated, though from the modeling a very small difference in redshift
may be expected between the blue and pink parts, corresponding to a
difference of $\Delta{d_{ls}/d_s}\simeq 0.003$

\subsubsection*{Source 28}

A very faint red galaxy with a counter image. Red lensed objects are
scarce enough that identification is not very tricky - by running
through a range of $d_{ls}/d_s$, the counter image is securely
identified and has an estimated photometric redshift of
$z=5.07\pm0.3$.

\subsubsection*{Source 30}

Three small green images lie close to the apex of the main subgroup
following the tangential critical curve, the locations of which are
accurately confirmed by the model. No other counter image of this
source is predicted. A photometric redshift is estimated, $z=3.35\pm0.2$.

\section{Light Map}\label{LightMap}
 
 We have built a light map of the cluster in 2D so that we may make
direct comparisons with the mass map. We start with the main cluster
galaxies modeled as described in Zekser et al. (2004, in
preparation). By making 2D fits to the cluster sequence galaxies we
ensure that their haloes are modeled to large radii. In addition, we
manually include those objects which, because of their colours or
magnitudes, either belong to the cluster or may be in the foreground.
We run SExtractor (in association mode) to generate an image which
contains the flux belonging to these galaxies and is set equal to 0
elsewhere.  Figure 8 shows the light map and the residual, background
galaxy image, both obtained from the F475W image. To convert the
observed AB F475W fluxes to B-band in the Johnson-Cousins systems we
use the E/S0 and Sbc templates from Benitez et al. 2004, which yields
filter corrections of 0.59 and 0.46 magnitudes respectively. Since we
estimate that $80\%$ of the cluster galaxies have colours similar to
E/SO galaxies, we use a weighted average correction of 0.565 for all
the galaxies in the cluster, so $m_B=m_{F475W}+0.565$. Of this
quantity, 0.12 mags correspond to the AB to Vega correction for the
Johnson B filter. In a $\Omega_m=0.3$, $\Omega_{\Lambda}=0.7$
cosmology, the distance modulus at z=0.18 is 39.71, and the weighted
k-correction for the $B_{Johnson}$ filter is 0.76 magnitudes, so
$M_B=m_{f475w}-39.90$. We use $M_\sun=5.48$ to convert to solar
luminosities: $2.179\times 10^{12}h^{-2}L_{\odot}$ within the ACS
field of view.  As a check, this may be compared with the catalog
created by the ACS pipeline: after excluding all the stars and objects
fainter than g=25, we obtain a total magnitude of $m_{F475W}=14.58$.
This converts to $2.1\times10^{12}h^{-2}L_{\odot}$, differing by only
3\%.

\section{Mass Modeling}\label{Modeling}

The usual approach to lens modeling of galaxy clusters relies on many
assumptions to describe the unknown cluster mass distribution and
sub-cluster components. For any supposed cluster component, a center
of mass must be designated, along with some ellipticity, positional
angle, mass profile and normalization. This generates many largely
unconstrained parameters and a degree of subjectivity in deciding what
constitutes the main cluster and sub-components. The galaxy
contributions may be better estimated from their location and
luminosities, but the complications of dynamical interaction between
galaxies and the cluster on the form of their mass profiles are of
course unaccounted for in such idealised approaches.

 The contribution of luminous cluster galaxies is not insignificant
and must be included in any accurate modeling.  The mass associated
with a typical massive cD galaxy with a velocity dispersion of
typically $\sim300km/s$, may alone be expected to account for $\approx
(300/1200)^2$ of the total mass, or $\sim 5\%$. The central location
of such objects ensures they will influence the appearance of
lensing. Cluster galaxies which happen to lie close to the Einstein
ring can often be seen to perturb the location of lensed images,
indicating masses amounting to a few percent contribution to the total
mass interior to the Einstein ring (Franx et al. 1987, Frye \&
Broadhurst 1998). Of course, all galaxies in the cluster must be
included at some level.

It is customary in lens modeling to specify the mass profiles of
cluster galaxies in detail - including a profile, a core, a truncation
radius, ellipticity and a scaling of these parameters with
luminosity. Our preference, detailed below, is not to get bogged down
in detail when specifying the galaxy contribution. As will become
clear, it is in fact difficult in practice to motivate much more than
simply the mass of a galaxy and its position in this context.  This is
principally because the deflection of light $\alpha(\theta)$, depends
on the projected gradient of the gravitational potential, $\phi(r)$,
so that any mass truncation radius is not distinctive, as the
projected potential will drop off relatively slowly with angular
separation from the center of mass, tending to the point mass limit,
$1/\theta$, beyond any truncation radius. Interior to the lensing
galaxy, the bend-angle will be approximately independent of radius if
the mass distribution is approximately isothermal:
$\alpha=4\pi{({\sigma\over{c}})^2}{d_{ls}\over{d_s}}$. Any core is
very hard to constrain as images are nearly always deflected well
beyond any reasonably sized small core. The ellipticity of the galaxy
might be worth including, however this is usually small for the
typical round shaped luminous cluster members and in any case the
shape of the deflection field is a convolution of the projected mass
distribution with angular separation, $\Sigma(\theta) \times
1/\theta$, hence the deflection field is intrinsically smoother and
hence rounder than the mass distribution. Additionally, one has the
uncertainty of converting the ellipticity of the galaxy isophotes into
contours of surface mass density, along with any dependence of
ellipticity on radius.

The hard part of the modeling is deciding how to deal with the
general ``dark matter'' distribution of the cluster, which is
understood to be the dominant component and is of course undetectably
faint or invisible (by definition), so we have no direct visible guide to its
distribution. We may begin with the simple expectation that mass
should roughly follow light and develop an approach which avoids
parameterization with idealised forms, but is based on the empirical
distribution of the light in the cluster.

 From previous experience with Cl0024+17 we have learned that a
surprisingly good starting point for the shape of the general mass
distribution may be obtained by simply using the luminous cluster
galaxies and summing up extended profiles assigned to each one,
producing a continuous 2D surface-density distribution (Broadhurst et
al. 2000). We generalise this approach by dividing up the mass in to
high and low frequency components, allowing a structured ``galaxy''
contribution and a smooth ``cluster'' component to be modeled
independently.  We do not iterate individual galaxy masses (following
Broadhurst et al. 2000) as there are far too many galaxies in the
strongly lensed region of A1689 to permit this. Instead we calculate
the deflection field of the low frequency mass component and add a low
order perturbation, since it has an intrinsically smoothly varying
field light deflection field, unlike the more structured galaxy
distribution. The contribution from the galaxy component is taken to
be the difference between the smooth cluster component and the initial
sum of galaxy mass profiles and is allowed freedom only in its
normalization to approximately mimic the effect of changing the M/L
ratio for the galaxy component. 

Together, the perturbed low frequency ``dark matter'' component and
the higher frequency ``cluster galaxy'' contributions constitute our
lens model, which has the advantages of being very simple and flexible
and is able to make accurate predictions for the location and
appearance of counter images, so that the majority of the multiple
images are identified by applying the model and all are confirmed using
the model.

\subsection{Starting Point for Lens Model}

 We begin by simply assigning an extended power-law profile to all
cluster galaxies lying close to the E/SO colour-magnitude
sequence. These are virtually all elliptical galaxies with only a
handful of obvious disk galaxies. In total we select the brightest 246
objects to a magnitude limit of $I=23$, faintward of which, the
background galaxies add confusion and are in any case at least 8
magnitudes fainter than the brightest cluster galaxy and do not add
significantly to the overall mass. We have also verified that the
exact choice of limit has negligible effect on the model. The profiles
of the selected galaxies are not truncated but are run out to the edge
of the field and beyond, to form a smooth general mass distribution
tending to a 2D power-law profile at large radius. It is important to
extend the mass distribution beyond the boundary of the data since the
lensing deflection field that we wish to construct is not local but an
integral over the surface of the cluster, falling off only slowly in
projection by the angular separation, $1/\theta$. We start with this
surface based on the light distribution of the cluster members and
proceed as follows to break it into a smooth ``dark matter'' component
which we iterate in shape, and a lumpy galaxy residual which we
associate with the cluster galaxies and which is allowed to vary only
in amplitude. We avoid additional unconstrainable parameters that only
add to uncertainty in the model. The steps taken are as follows:

 For each cluster galaxy a symmetric power-law surface density
profile, $\Sigma(r)=Kr^{-q}$, is integrated to give the interior mass,
$ M(<{\theta})={{2\pi K}\over{2-q}}{d_l\theta}^{2-q} $, leading to a
bend-angle of light:

\be\label{defl1}
\alpha(\theta)={4GM(<\theta)\over{c^2}\theta}{d_{ls}\over{{d_s}{d_l}}}
\ee

or 

\be\label{defl2}
\alpha(\theta)={CD\over{2-q}}\theta^{-q+1}
\ee

where we separate the normalization, $C=8\pi K G/c^2$, from the
distances, $D=d_{ls}d_l^{1-q}/d_s$.

 We simply scale the normalization, $C$, to the luminosity of each
galaxy.  (One could experiment with more complex scaling of M/L with M,
though a linear relation seems to be indicated by recent statistical
work (Sheth et al. 2002, Padmanabhan et al. 2002, although see Guzik \&
Seljak 2003). Once the surface mass distribution is constructed from
the sum of the above profiles then the deflection field can be derived
and is initially used to crudely represent the combination of galaxies
and the cluster together, implicitly assuming that the mass of the
cluster traces the light:

\be\label{defllight}
 \vec\alpha_T(\vec\theta)=\Sigma_i \vec\alpha_i(\vec\theta)
\ee

 This deflection field is our starting point and is now broken down
into a smooth component, $\vec\alpha_s(\vec\theta)$ to represent the
overall cluster mass distribution, and a lumpy residual component, to
represent the cluster sequence galaxies
$\vec\alpha_g(\vec\theta)=\vec\alpha_T(\vec\theta)-\vec\alpha_s(\vec\theta)$.
These contributions are varied separately in conjunction with the
positions of the multiple images to generate the model fit, as
described below.

\subsection{Cluster and Galaxy Components}\label{ModelClusGal}

The locations of the cluster galaxies should serve as a rough tracer
of the overall mass distribution and so we take the low order 2D
function to generate a smooth surface mass distribution as a starting
point for modeling the dominant smooth cluster component. We have
tested a variety of smooth surface fits of varying resolution and
settled on cubic splines of order 5-8. Higher order fits are
unreasonably structured and produce many more multiply lensed images
than are observed, and with lower order a virtually featureless
circular mass profile is produced requiring large perturbations to
match the data that are hard to control.

 The deflection of light received at any angular position can then be
calculated by a convolution of the surface mass profile with a
$1/\theta$ kernel. The contribution to the angle,
$\vec\alpha(\vec\theta)$, by which light is deflected at angular
position, $\vec\theta$, by a small mass element,
$dm=\Sigma(\vec\theta{'}){d^2\theta{'}}$, separated by ,
$\Delta\theta$, can be approximated as a point mass deflection,
$d\alpha={dm\over{\Delta x}}{d_{ls}\over{d_s}}$, where, $\Delta
x={d_l}\Delta\theta$, is the projected separation in the lens plane,
i.e.  $d\alpha={dm\over{\Delta\theta d_l}}{d_{ls}\over{d_s}}$, and
integrating this over the mass distribution in the full lens plane,
allows the full deflection field to be calculated:

\be\label{deflsmooth}
\vec\alpha_s(\vec\theta)={4G\over{c^2}}{d_{ls}d_l\over{{d_s}}}\int_R
\Sigma(\vec\theta^{'})
{\vec\theta-\vec\theta{'}\over{|\theta-\theta{'}|^2}}d^2\theta{'} 
\ee

In practice this deflection field is calculated by FFT by first
binning the mass distribution onto a 1024$\times$1024 grid and
repeating it 16 times in a 4$\times$4 array so that the deflection
field calculated by FFT is completely free of any evidence of spurious
boundary effects. The resulting deflection field is then numerically
interpolated by a factor of 16, using the gradients of the local
deflection field, to match the spatial resolution of the data. This
high resolution is essential for making a proper comparison of the
observed images with the model generated lensed images, as described
below. Note, we calculate the deflection field over
a wider area than covered by the data, extending the summation of
the mass profiles to cover an area twice the size of the data field
in order to deal with the edge effects since the deflection field is not
local and so this way we minimise the edge effects in the calculation
described by Eq. \ref{deflsmooth}.

 The full deflection field is required for iterating the relationship
between the unknown source position, $\vec\beta$, and the image via
the lensing equation, i.e.

\be\label{theta}
\vec\theta=\vec\beta+\vec\alpha{(\vec\theta)}{d_{ls}\over{d_s}}
\ee

 The most practical part of our procedure is the use of the deflection
field in the model iteration. The reason we choose this rather than
the input mass distribution is twofold. Firstly, the deflection field
is always smoother than the mass, as can be readily seen from the
convolution above, $\Sigma(\theta) \times {1\over{\theta}}$
(Eq. \ref{deflsmooth}) and hence more smoothly perturbed than the mass
distribution. In addition, the deflection field must be calculated in
the process of minimization as it is required so that the bend-angles
of the model generated multiple image locations can be predicted.

The angular diameter distance ratio of each source,
$D_{k}(z)=d_{ls}/d_s$, scales with redshift and acts simply to scale
the amplitude of the deflection fields.  However since the
normalization of the surface density distribution also acts to
linearly scale the bend-angles then distances can not be known
separately from the density normalization i.e. $\alpha \propto \Sigma_o
d_{ls}/d_s$, we must therefore work only with relative bend
angles. Hence we define a convenient fiducial value of
$D_o \equiv d_{ls}/d_s(z=3)$, corresponding to the mean redshift of the
background galaxies $z=3$, and define the relative distance ratios to
be:

\be\label{fk1}
f_k \equiv D_{k}(z)/D_o
\ee

This relative ratio is referred to in the minimization below and also
used for comparison with cosmological models discussed in section 13.

\subsection{Iterating the Deflection Fields}\label{ModelIter}

We treat the perturbation as a modification of the gravitational
potential and use the orthogonal derivatives of this perturbation
to modify the deflection fields to avoid introducing any curl, so that
the smooth components of the deflection fields become:

\be\label{asx}
\alpha_{s_x}{'}(\vec\theta)=\alpha_{s_x}+\frac{\partial P(\vec\theta)}{\partial\theta_x} 
\ee

\be\label{asy}
\alpha_{s_y}{'}(\vec\theta)=\alpha_{s_y}+\frac{\partial P(\vec\theta)}{\partial\theta_y}
\ee

The coefficients of $P(\theta)$ are free parameters in modeling the
shape of the smooth cluster component. We use plain polynomials as
only a low order perturbation is required (3rd or 4th order) and we
prefer not work in polar coordinates as this imposes a spurious
multi-pole pattern on the deflection fields at low order.

Two additional free parameters are required, to allow the galaxy
contribution to vary in amplitude, $R$,

\be\label{ag}
\vec\alpha_g{'}(\vec\theta)={R}{\vec\alpha_g(\vec\theta)}
\ee

with an overall normalization, $A_o$, of both components, i.e.

\be\label{at}
\vec\alpha_T{'}(\vec\theta)=A_o(\vec\alpha_s{'}(\vec\theta)
+\vec\alpha_g{'}(\vec\theta))
\ee

Finally, the resulting surface mass density distribution,
$\Sigma{'}(\vec\theta)$, responsible for the iterated deflection
field, $\vec\alpha{'}$, is obtained very simply via Poisson's equation,
since the bend-angle is the gradient of the projected potential and
therefore the surface density is simply:

\be\label{sigma1}
\Sigma{'}(\vec\theta)={1\over2}\Sigma_{crit}\vec\nabla\cdot\vec\alpha_T{'}(\vec\theta)
\ee

This rather general approach to parameterising the model leads to
results which are relatively free of assumptions. We emphasise again
that division of the surface mass distribution into low and high order
components is not meant to correspond literally to a division between
cluster dark-matter and the galaxy masses, but rather that their
combination is a description of the entire mass distribution.  This
resulting surface mass distribution may if desired be subsequently
divided into galaxy and cluster contributions following a plausible model,
which we do not attempt here. A follow-up paper treats this subject,
beginning with an NFW profile for the smooth cluster component (Zekser
et al. 2004).

\subsection{Minimization}\label{ModelMin}

 The simplest measure of the accuracy of the model is to compare model
predicted image positions, $\vec\theta^{m}_{i,k}$ with the observed
image locations, $\vec\theta_{i,k}$ and thereby avoid the pitfalls of
working in the source plane (Kochanek 1991). For each of the N sources
we have M multiple images, forming in total N$\times$M images to
constrain the model. Ranking the M images of each source, we subtract
in turn the model deflection angle from the each image of each source,
k, and assign these a value of k, on a grid in the source plane, all
other grid points are zero by default. Then the source plane is
re-lensed but only for a relatively small box around each of other,
M-1, observed image positions, saving time, and the model positions
are recorded if all the re-lensed images of all the sources appear in
these boxes. This process is repeated for each of the other M-1 images
of each source and finally we sum over the differences between all the
pairs of measured image positions and their corresponding model
generated positions to form a measure of the accuracy of the model,
which is simply expressed as:

\be\label{chisq}
\chi^2_{im}=\sum_{k}\sum_{i,j,i>j}\left({\vec\theta_{j,k}
-\vec\theta^{m}_{j,k}}\right)^2
\ee


The above flagging of the source belonging to each set of images means we can 
identify the model image location of the correct source, avoiding confusion
between counter images of different sources. The whole image
plane does not need to be scanned for images as we are only interested
in cases where the predicted images form close to the actual
observed positions.  In practice a box of side 5\arcs ~is sufficient to
include nearly all the counter image positions while converging to a
solution in a reasonable time. The number of iterations is reduced
considerably by terminating an iteration when the variation between
the source positions of a given set of multiple images is unreasonably
large and moving to the next iteration step.

In this process we set all the relative distances to be equal,
$f_k=1$, since, as we shall show below the best-fit model is
negligibly affected by the choice of $f_k$. The search box must also
be large enough to allow for some variation in the unknown distances,
$d_{ls}/d_s$, which as we show below are reasonably expected differ by
up to $\sim$ 15\% in angular scale over the redshift range of the
background sources. This range is small because of the relatively low
redshift of A1689, $z=0.185$ so that only slow redshift dependence of
$d_{ls}/d_s$ is expected for $z>1$ (see section 8.5), where the bulk of
the faint lensed galaxies lie, i.e $d_{ls}\simeq d_s$ for z$>$1.

We have applied the ``downhill simplex'' algorithm (Press et al.) to
find a minimum and this solution has been compared with a crude grid
search as a sanity check.  We have found that 9-14 coefficients are
preferred for the perturbing potential to achieve considerable model
flexibility (equation N), corresponding to third or fourth order with
cross terms. Adding more terms seems not to be justified, with little
noticeable improvement of any relevant quantities.

 We have plotted $\chi^2_{im}$, calculated above for different choices
of the input slope, $q$, and plotted this against the mean slope of
the resulting mass profile, $\bar{p}=d\log\Sigma(\theta)/d\log\theta$,
as this is the major quantity of interest. We find a clear minimum in
the difference between the model and observed image locations for a
slope around $\bar{p}\sim-0.55$, corresponding to $q=-1.2$. For
steeper slopes the error increases and the disagreement with the data
is pronounced in the center where the radial critical curve breaks up
into small islands. For flatter slopes, $\chi^2$ is larger and the
elongation of the lensed images becomes large and additional
unobserved tangential images form readily as the area of the image
plane close to the critical density is enlarged and upward
fluctuations from the galaxies lead to exaggerated arcs. In the
center, a flatter slope also produces much longer and fatter radial
arcs than are observed, with a deficit of central de-magnified images,
these being pushed outward too far in radius compared with the observed
images. We include 3 model figures to demonstrate this behaviour,
showing the locations, sizes and shapes of the lensed images compared
to the critical curves for three different choices of input profile slope
(Figures 14,15\&16).

Note that the input slope of the galaxy profiles is generally steeper
than the resulting mass profile of the cluster, simply because the
galaxies are spread over the surface of the cluster and are not
concentrated at one point in the middle.

We emphasise here that the slope of the mass profile is the main
parameter of interest for comparison with predictions of N-body codes
etc. The overall gradient of the profile has a noticeable effect on
the appearance and location of images, as described above and, as we
shall see later, on the relative distances and relative magnifications
predicted for the background lensed galaxies, which we will compare
with the observed redshift information (in sections 9 and 13). The
parameters of the model are coupled somewhat in that the initial
choice of the slope, q, of the galaxy profiles used to build the
starting surface will determine what fraction of the mass is assigned
to the ``galaxy'' and ``cluster'' mass components, and hence the
relative fractions should not of course be taken as literal division
into cluster and galaxy mass contributions.

We express our results in terms of the resulting profile of the mass
distribution because the profile is the quantity of interest for
testing physical models for the dark matter. Note that N-body
simulations include the full spectrum of density perturbations, so
that galaxies are included by default in the derived profile of the
overall mass profile of clusters and therefore when making comparisons
of mass profiles with the results of N-body simulations, it is
preferable not to remove the galaxy contributions - with the exception
only of the small baryonic component. If desired the resulting model
surface mass distribution can of course be decomposed into ``galaxy''
and ``cluster'' contributions, following reasonable assumptions and
will be explored in a subsequent paper (Zekser et al.  2004, in
preparation).

\subsection{Role of lensing distance ratio}\label{ModelD}

We can make use of the best fitting model deflection fields to
calculate the corresponding values of $f_k$ predicted by the
model. The scale of the deflection field grows with increasing source
distance and hence we can form a statistical measure of $f_k$ by
calculating the angular separations between images:

\be\label{fk2}
f_k={1\over{(n_k-1)^2}}\sum_{i,j,i>j}
{{(\vec\theta_{i,k}}-{\vec\theta_{j,k})}\cdot{
(\vec\alpha{(\vec\theta_{j,k})}-\vec\alpha{(\vec\theta_{i,k})})}\over{
(\vec\alpha{(\vec\theta_{j,k})}-\vec\alpha{(\vec\theta_{i,k})})^2}}
\ee

This useful expression shows that $f_k$ scales in amplitude as the
ratio of the observed image separations divided by the predicted model
image separations, as one might anticipate, since the larger the value
of $f_k$, the larger the deflection angle $\propto
{\vec\alpha(\theta)}{d_{ls}/d_s}$. The output values of $f_k$ span a
plausible range of relative distances $0.8<f_k<1.1$, with the fainter
red dropout galaxies lying at larger predicted distances than the
brighter, lower redshift objects (Figure 12).

The output values of $f_k$ shown in figure 12 are produced by setting
all of the distances equal. Here we investigate the dependence of the
output $f_k$ on the choice of input values for $f_k$. We generate
random values for the input $f_k$ and look at the corresponding output
values. The size of the scatter on the input values $f_k$ was varied
by 30\%, more than covering the range anticipated from the
cosmological variation (discussed in section 8.5). Figure 12 shows
multiple trials of random sets of initial values assigned to the k
sets of multiple images, showing clearly that the scatter on the model
derived $f_k$ is small $\simeq 0.01$ per image, much smaller than the
range of output, $f_k$, calculated as above (eqn 14), and with an
average value very close to the input value of unity. Hence, we can be
confident that we do not need to leave the values of $f_k$ free in our
modeling, and incur many more free parameters in the minimization,
but we can fix them at a desired mean level initially and use the
resulting model to calculate accurate $f_k$ using eqn 14. This
shortcut is possible because of the large number of background sources
and because the mass distribution obtained is more based on the
average redshift when the lens plane is covered by many multiple
images. The mass distribution of the lens is of course independent of
$d_{ls}/d_s$, and so must be the same no matter the distances of the
background sources.

\section{Relative Magnifications}

We can compare the relative magnifications of the multiply lensed
images with the model predicted values using the results of the above
section. The relative magnifications are quite sensitive to
the profile, as shown in figures 14,15,\&16, and
since this is the main quantity of interest for comparison with
physical descriptions for the dark-matter distribution, we make a
detailed check of our model for model generated profiles covering a
wide range of slope.

 The relative magnifications are calculated from the data by simply
taking the ratio of fluxes determined from our photometry and
comparing this with the magnification map generated by the model at
the centroid positions of the observed images. For this purpose we use
only the I-band magnitudes, which allows the reddest galaxies (distant
dropout galaxies) to be included. Not all the multiple images are
suitable for our purpose because many images lie very close or bisect
the critical curves where the magnification diverges. It is
unreasonable to demand that modeling define the location of the
critical curves accurately enough so that the magnification of such
images can be meaningfully predicted. Also the magnification varies
strongly along the most magnified arcs and not characterised by a
single value.

The most reliable values of the relative magnification are for images
that lie in the regions well away from the critical curves and well
away from perturbing galaxies. We compare sets of images of given
sources for which images lie in regions well outside the tangential
critical curve, or safely between the radial and tangential critical
curve or well inside the radial critical curve. In the case of the
5-image systems such images may fall in all three locations, and for
the remaining cases only two of these regions may be covered. Also, we
exclude duplications where neighbouring sets of images lie very close
to each other and so add little new information for the model. For example
sources 2,15,23 \& 27 lie very close to sources 1,10,22,\& 24
respectively and are excluded leaving only the relatively brighter
neighbours (1,10,22,\& 24) for which the photometric error is more
reliable. In addition, images which are obviously locally influenced
by a massive cluster galaxy, like 1.1 and 1.2, are also excluded as
their magnifications are very sensitive to the modeling.

In total, approximately one third of the images are useful for our
purpose, and their relative magnifications are compared with the model
for three different profiles (figure 29). We take the outermost images
as the reference for calculating the relative magnifications in each
case, as this helps better to understand and radial behaviour. Figure
30 shows the best fitting model derived above using the minimization
made on the basis of the image positions and here the relative
magnifications cluster about the value of unity over the full range of
radius as one might have hoped supporting strongly our best model.
For shallower profiles (figure 29, left panel), the model predicts
rather larger relative magnifications in the center (as can also be
seen clearly in Figure 15), where the model profile is relatively flat,
dragging down the mean value of the ratio between the data and the
model. For the case of steeper profiles, a wide scatter is found,
particularly in the center, where in many cases the model predicts
much less magnified images than are observed (Figure 29, right-hand
panel).

The mean ratio of the data to the model values of the relative
magnifications varies smoothly with the mean profile slope (Figure 29)
increasing with increasing slope as anticipated from the comparison of
figures 14,15\&16. The smallest scatter is for a mean slope of, $d\log
\Sigma/d\log r \simeq -0.55$, where the ratio of the data to the model
as close to unity and this slope is in good agreement with the best
fit model derived above on the basis of the independent image location
information, adding considerable confidence to our modeling. Note
that the best fitting relative values of $f_k$ are used in making this
comparison as the magnification is a function of distance but
the level of this small inter-dependence can evaluated by simply
referring everything to a mean distance, i.e.  setting $f_k=1$, and is
found to have a negligible effect on the relative magnifications as
may be expected given the similarity of $d_{ls}/d_s$ for the multiply
lensed sources.

Note that the errors shown in Figure 29 are just the photometric
errors (added in quadrature between the reference image and the
comparison counter image) and although these errors are significant,
they do not account for the full dispersion of the relative
magnifications about unity, as can be seen in figure 29. Clearly there
remains some additional scatter between the model and the data of
around 15\%, but with no systematic radial trend.

We can also incorporate the relative magnifications, where reliable,
into the lensing code. This is essential for models in which the
minimization is made in the source plane, due to tendency otherwise to
generate spuriously large magnifications by minimizing the separations
between de-lensed images projected back to the source plane and hence a
smaller source plane area relative to the image plane (see Kochanek
1991). However since we have gone to the considerable trouble of
minimizing in the image plane, we suffer no such bias and we can only
hope to gain by at most a factor of $\sqrt2$ in precision by including
the extra flux information, if every image has an accurate flux. In
fact since only one third of the images have useful flux information,
as described above, and because the minimization is proportional to
the number of independent pairs of lensed images, then we gain only a
modest improvement to the lens solution this way. It is also not clear
how best to weight the relative fluxes relative to the positional
information when minimizing. Inverse variance weighting cannot be
applied here as the positional accuracy is so much greater than the
precision of the flux measurements, leading to no improvement in the
model fit. If we go to an extreme case and simply weight all points
equally, a modest improvement in the uncertainty on the model profile
is produced of around 10\% in precision (Figure 27), though the
best-fit profile derived is almost identical, indicating that in the
context of the parameters used to describe our fit, the same solution
is found with or without the flux information. Clearly the relative
flux information, best serves as a consistency check on the lens model
derived from the image locations - which as we noted above clearly
supports the validity of our best-fitting solution.

 This tendency to find the same solution despite the addition of extra
information is easily understood as a consequence of the large amount
of information we have at our disposal compared to the number of free
parameters we use to describe the model. In fact we may experiment in
excluding some positional information, and the effect on the solution
is not noticeable until more than half of the points are left out. This
may argue for adding some more parameters to the model, although we
have seen earlier that the main conclusions regarding the are not
changed by the addition of more free parameters, meaning that the model
is flexible enough to describe well the mass distribution.

\section{Critical Curves}\label{Crit}

 The location of the radial and tangential critical curves are clearly
defined in our data and therefore we examine the constraint they
provide on the mass distribution of the lens.  The critical curves are
obtained from the magnification field of the lens, and readily
identified as loci of divergent values, corresponding to positions
where images are maximally stretched either tangentially or radially.
The radial critical curve is traced by long spoke-like images and by
close pairs of images aligned radially and lying on either side of the
radial critical curve. Images formed close to this radial ring point
towards local concentrations of mass and usually form interior to
tangential critical curves. This behaviour is shown in Figures
14,15,\&16 , where a grid of evenly spaced circular sources is lensed
by our best fitting lens model and projected onto the magnification
field to show the relationship between the caustics and the distortion
of the lensed images. A colour-coded map shows roughly where images of
the same source will be located around the lens, to help develop an
intuition for where counter-images (with the same colour) will appear
over the surface of the lens and with what orientation and shape to
expect (Figure 17).
 
The magnification at a given position, $\mu(\vec\theta)$, is given by
the Jacobian of the lens mapping (e.g. Young '81), and may be
conveniently expressed in terms of derivatives of the deflection
field:

\be\label{u2D}
\mu(\vec\theta)^{-1}=1-\nabla\cdot\vec\alpha
+{d\alpha_x\over{dx}}{d\alpha_y\over{dy}}-\left[{d\alpha_x\over{dy}}\right]^2
\ee

Hence, the deflection field iterated above in our modeling procedure
can be used to generate the magnification field directly.  The
resulting magnification field is plotted in a series of Figures
14,15\&16. The bright lines in this plot correspond to divergences in
the mapping following critical curves.  The main tangential critical
curve forms where the mean interior surface mass density exceeds the
critical mass density ($0.95g/cm^2$ for A1689 at $z=0.185$, with a
source at $z=2$), and is not particularly circular in shape,
stretching around a sub-group, and is also perturbed by cluster
members. A typical cluster galaxy may have a critical radius of
1-3\arcs ~and this is added to the large scale deflection of the lens,
generating obvious excursions so that the formation of images lying
close to the critical curve is strongly influenced by these
perturbations.

 An inner critical curve is also apparent; this is where images are
stretched maximally in the radial direction leading to radial ``arcs''
pointing towards the cluster center. This is the main radial critical
curve and forms if the central mass profile is shallower than
isothermal, shrinking to zero radius for a pure isothermal profile
(see below).  Single radial arcs have been observed in other massive
clusters, but here, for the first time, we observe many radial arcs
that trace out the entire radial critical curve.  We attribute this
accomplishment to the unprecedented quality of our data and the
powerful magnification of A1689.  And we expect that similar high
quality images will trace out radial critical curves in other massive
clusters as well.

The radial magnification is also of interest below when we 
examine the azimuthal mass profile and is very simply related to
the radial deflection angle:

\be\label{urad}
\mu(\theta)^{-1}=\left(1-{\alpha(\theta)\over{\theta}}\right)\left(1-{d \alpha\over{d \theta}}\right)
\ee 

The first term is the tangential stretch factor and the second is the
radial stretch factor.  The tangential and radial critical curves are
defined by where these terms diverge: at two discrete radii in this
azimuthal average.  This form for the magnification proves more useful
for model comparisons than the 2-D form of Eq. 15, which diverges at
many radii due to the asymmetry of the lens.

\subsection{Radial vs. Tangential Critical Radii}\label{CritRadTan}

 A striking and simple result follows from comparing the ratio of the
Einstein radius, $\simeq 50$ \arcs, to the radius of the radial
critical curve, $\simeq 17$\arcs, Figure \ref{one_over_mu}). This
ratio is $\simeq 3$, close to the minimum value for a power law
profile, which we now show tends to the base of natural logs,
$e=2.718$ for a flat profile, $p=0$. To see what constraint this ratio
places on the inner projected slope of the mass profile, consider a
power-law mass distribution.  Using equation~3 above for a power-law
profile, the tangential $a(\theta)$ and radial $b(\theta)$ stretch
factors become:

\be\label{a}
a(\theta)^{-1}={1-{CD\over{2-q}}}\theta^{-q},
\ee

and

\be\label{b}
b(\theta)^{-1}=1-CD{{1-q}\over{2-q}}\theta^{-q},
\ee

These stretch factors diverge at the tangential critical (Einstein)
radius $\theta_e$ and at the radial critical radius $\theta_r$, so
that the ratio of critical radii is simply:

\be\label{e/r}
\theta_e/\theta_r=(1-q)^{-1/q},
\ee

Hence, the radial critical radius saturates at a maximum value of
$\theta_e/e$ as $q \rightarrow 0$ and disappears altogether for the
isothermal case, as $q \rightarrow 1$. The observed ratio of the
critical radii noted above, $\theta_e/\theta_r\simeq 3$, corresponds
to $dlog\Sigma(\theta)/dlog(\theta)\simeq -0.3$, using eqn 19. Clearly
then the existence of a well-defined radial critical curve and its
large radius relative to the tangential critical curve indicates that
the inner mass profile of A1689 is much flatter than the purely
isothermal case and will be shown later to compare very well with the
more physical models discussed in section 11.5, which are predicted to
have shallow central profiles.

\section{Mass, Light and Model Profiles}\label{Prof}

 In this section we examine the radial profiles of the mass and light
and make comparisons with models. It is necessary to bin the data
radially for such comparisons and so we must discuss the role of
substructure and ellipticity. The effect of the main subgroup on the
mass profile can be seen in the above mass profile (Figures 24,25) as
a small flat excess of mass in the mass profile just outside the
critical radius. This subgroup is most likely a chance projection
lying $\sim 70Mpc/h$ in the background, as argued below. Our model
solution down-weights this group as can be seen by comparing the
initial mass distribution (Fig. 18) with the model output solution
(Fig. 19), where the secondary peak is much less prominent, implying a
lower M/L for the subgroup compared with the main cluster. The mass
associated with this group can be ignored when determining the radial
profile, by excluding a generous area centered on the subgroup. If we
exclude a circular area with a radius of 20\arcs centered on the four
luminous galaxies comprising the subgroup, at a distance of 80\arcs
~from the center of the cluster, the excess bump in the radial mass
profile disappears - Figure 19.  Hence the main subgroup has only a
minor effect on the mass or light profiles but we can easily exclude
it in the subsequent analysis.

\subsection{Mass vs. Light }\label{ProfML}
 
 The radial distribution of the mass and light can be compared
directly by binning both the model generated mass map and the observed
2D light map of the cluster sequence galaxies.  The exact choice of
center for constructing radial profiles is not very important. The
center of mass must lie close to the center of the radial critical
curve, which is approximately circular. This point is nearly
coincident with the most luminous central cD galaxy, separated by
$\simeq 6$\arcs, so that for most purposes we may conveniently use the
centroid of the cD galaxy as a representative center for the cluster
as a whole. Figure 21 shows a comparison of the mass and light
profiles, in circular bins, suitably scaled. The profiles of mass and
light are quite similar for $r>50kpc/h$, but with an obvious
difference on smaller scales, $r<50kpc/h$. The central luminous
galaxies are tightly bunched within this radius, leading to an 
excess of light over mass, compared with larger radius. One might
wonder to what extent the central mass profile is composed of mass
associated with these galaxies. We may simply remove the mass
associated with the stars in these galaxies by adopting a standard M/L
for the luminous regions where velocity dispersion measurements lead
to $M/L_B \sim 5hM_\odot/L_B\odot$, (Sheth et al. 2001, Padmanabhan et
al.  2002) for luminous elliptical galaxies. If we subtract this mass
from the total, the resulting mass profile is offset by a negligible
amount over the full range of radius because the M/L of the cluster as
a whole is so much larger than that of the starlight, $\sim 400h(M/L_B)_{\odot}$.

Only if the M/L associated with the galaxies is increased to greater
than $30h(M/L_B)_{\odot}$, is there any significant effect on the
profile (Figure 21), producing a slight flattening in the center,
$r<20kpc/h$. A large $M/L$ for the galaxies may be reasonable but must
include mass well in excess of the baryonic component from the stars
and is also in excess of the dynamical masses estimated for luminous
elliptical galaxies, $\sim 10(M/L_B)_{\odot}$. In N-body work, the
resulting dark matter profile is not separated into cluster and galaxy
components, so for comparison with the N-body work, we include all the
mass and do not make any correction for the galaxies. It is not
correct to subtract dark matter associated with galaxies when
comparing with the N-body work, since this and other substructure is
included by default in the N-body based radial mass profiles.

 Careful work on other clusters in which the central cD galaxy is much
more prominent and the overall M/L of the cluster is lower, so
more care is required to evaluate the role of the central galaxy on
the strong lensing effects (Sand et al. 2002, Gavazzi et al. 2003,
Treu et al. 2003). 

It is also worth quoting here an M/L for the region interior to the
Einstein radius as this number is usually the only reliable
lensing-related quantity for more typical quality data, especially for
ground-based data, where one giant arc is often the only strongly
lensed image identified. The mass-to-light ratio measured in a
circular aperture of 50\arcs ~centered close to the cD galaxy, is
$(M/L)_{Ein}=320\pm20h(M/L_B)_{\odot}$, with a small uncertainty as
the mean redshift of the critical curve is defined accurately at
$\bar{z}=3$ from the average photometric redshift of the multiply
imaged background galaxies. This value is identical to that derived
for the mass interior to the Einstein ring of CL0024+17, where
$M/L(r<100kpc/h)=320\pm20 h(M/L_B)_{\odot}$ (Broadhurst etal 2000),
and corresponds to almost the same physical radius $\sim 100kpc/h$ in
projection.

\subsection{Radial Magnification Profile}\label{ProfR}

 The location of the radial critical curve relative to the tangential
critical curve is sensitive to the inner profile of the mass
distribution, the ratio of the radii being larger for a shallower
profile, as described above.  We readily identify the asymmetry of the
tangential critical curve which enhances the azimuthally-averaged
magnification in circular bins. A corresponding slight excess is seen
in the mass profile too but is much less pronounced as the mass
distribution is more circularly symmetric than the tangential critical
radius (Figure 23). To minimise the effect of the divergent critical
lines on the magnification, we can azimuthally average
$1/\mu(\vec\theta)$ over all pixels, and invert this sum to produce a
radially binned magnification profile $\mu(r)$. This profile is a much
better fit to an NFW profile for the region outside the tangential
critical curve, but at the expense of well-defined critical radii.

\subsection{Critical Curves and the Light Profile}\label{ProfCrit}

 We can ask where the critical curves should lie if the mass strictly
follows the light. By taking the map of the light distribution of
galaxies associated with the cluster, described in section 7, and
scaling by an average M/L and binned radially we can readily calculate
the corresponding magnification profile and identify critical
curves. With $M/L\sim 400h(M/L_B)_{\odot}$, the Einstein radius,
although a little noisy, matches the 50\arcs ~radius observed.
However, no obvious radial critical curve is formed.  This is because
the central light profile is too steep, $\simeq \theta^{-1}$ (close to
an isothermal slope) and as pointed out above, equation 19, the radial
critical curve shrinks to zero as the index of the profile tends to
$p=1$, so that no radial critical curve is expected. Therefore the
presence of a large radial critical curve in the data immediately
makes clear that the central mass profile is not as steep as that of
the light, or equivalently M/L must be lower than average in the 
center.

Notice that although we began our modeling by taking the light profile
as a starting point, the resulting best fit has a shallower
mass profile than the light (Figure 21). This, together with the
more model-independent conclusion based on the relative ratio of the
critical radii (Eqn 19), implies that the mass does not trace the light
in detail, and instead the light is more concentrated than the
mass. Plausibly, dynamical friction may operate on the cluster
galaxies, concentrating them relative to the mass in general, as can
be inferred from El-Zant et al. (2003).

\subsection{Softened Isothermal Profile}\label{ProfIso}

A softened isothermal profile is often discussed as a physical
description of a gravitationally relaxed system. Since lensing relates
only to the projected mass profile, a distinctive core in 3D
will be less obvious in projection. Integrating a softened isothermal
potential with a projected core radius of $\theta_c$, and Einstein
radius, $\theta_E$, yields a projected mass profile:

\be\label{sigma2}
{\Sigma(\theta)/{\Sigma_{crit}(z)}}={1\over{2}}{{\theta_c
+\sqrt{\theta_c^2+\theta_E^2}}\over{\sqrt{\theta^2+\theta_c^2}}}
\ee

Using Eq. 1, the corresponding bend-angle for a softened potential becomes,

\be\label{bendangle}
\alpha(\theta)={1\over{\theta}}(-\theta_c+\sqrt{\theta^2+\theta_c^2})
(\theta_c+\sqrt{\theta_c^2+\theta_E^2})
\ee

This expression tends to a constant bend-angle of $\alpha=\theta_E$
when the core radius is zero, equivalent to the singular isothermal
case. For small angles, $\theta<<\theta_c$, the bend-angle tends to
zero, so that light passing through the core is relatively undeflected
by the presence of a core.
 
If we fix $\theta_E$(z=3)=49.5\arcs ~to match the observations, and
vary the core radius, $\theta_c$, then a radial critical curve can be
generated and matched to the data where $\theta_r(z=3)=17$\arcs, as
shown in Figure 28 (lower right box). The corresponding surface mass profile for this
choice of core radius is also shown in Figure 28 (upper right box). This
profile is steeper than derived from our observations for angles
larger than the core radius, and flattens to a constant density
interior to this radius. This profile is not favoured by the
observations, although it is surprisingly similar in form, with a mean
slope of $d\log\Sigma/d\log(\theta)=-0.8$. The corresponding form of
the radial magnification profile is also unlike that derived from the
data, being more sharply peaked around $\theta_E$ and falls well below
predictions for $\theta \leq \theta_r$.

We may instead adjust the core radius to match our best fitting mass
profile, requiring $\theta_c=20$\arcs, which produces a significantly
larger radius for the radial critical curve, 22\arcs, than observed ,
as shown in Figure 28. This sensitivity to the location of the
critical curves is produced by the flattening of the density profile
in the core enlarging the radius of the radial critical curve relative
to the Einstein radius, so that even a relativity small core has a
marked effect on the radius of the radial critical curve.

It is perhaps premature to conclude that the softened isothermal is
entirely excluded. More work on the effect of ellipticity and the
relative brightness of central images compared with the corresponding
counter images at larger radius will allow a fuller exploration of
this profile.

The softened isothermal profile is favoured over a generalised NFW
profile in the analysis of Gavazzi et al. (2003) for the cD dominated
cluster MS1417+22. The authors point out this conclusion hinges on a
plausible identification of a faint central image as part of a
multiple lensed system and on an accurate description of the central
cD galaxy and assumes that the centers of mass of the cD and the
cluster mass profile are spatially coincident.

\subsection{Comparison with NFW profile}\label{ProfNFW}

 A universal parameterization of CDM-based mass profiles has been
advocated by Navarro, Frenk \& White (96; NFW hereafter). This is an
azimuthally averaged profile summed over sets of haloes identified in
N-body simulations and variations between haloes of similar mass are
found in detail. This N-body based profile is summed over all the mass
contained within the main halo, including the galaxy sized haloes.
Hence, we compare our integrated observed mass profile directly with
the NFW predictions without having to invent a prescription to remove
the cluster galaxies.

 NFW have shown that massive CDM haloes are less concentrated with
increasing halo mass, a trend identified with collapse redshift, which
is generally higher for smaller haloes, following from the increased
cosmological density of matter at higher redshift. Cluster haloes, being the most
massive bound structures, form later in hierarchical models like CDM, and
are therefore have a relatively low concentration, quantified by the ratio of
$C_{vir}=r_{virial}/r_s$, and found to follow a density profile
lacking a core, but with a much shallower central profile ($r\leq
100kpc/h$) than a purely isothermal body. This predicted inner
flattening, although a distinctive prediction of CDM-based
cosmologies, has not been critically examined by observation for want
of discriminating data. Weak lensing measurements, for example, are
inherently ambiguous with respect to the mass profile and not able to
distinguish the NFW profile from other quite different profiles
(e.g. Clowe et al. 2000, King, Clowe \& Schneider 2001).

 The fit to an NFW profile is made leaving both parameters, $r_s$, and
$\rho_s$ free - the characteristic radius and the corresponding
density. These are separately constrained by the critical radii - both
the tangential and the radial critical curves and by the form of the
projected mass profile. Integrating the mass along a column, z, where
$r^2=({\xi_r}{r_s})^2+z^2$ gives:

\be\label{Mcol}
\rm{M}(\xi_r)={\rho_s}{r_s^3}(\xi_r)\int_o^{\xi_{r}}d^2\xi_r
\int_{-\infty}^{\infty} {\frac{1}{(r/r_s)(1+r/r_s)^2}}{dz\over{r_s}},
\ee

Using this mass, a
bend-angle of $\alpha=\frac{4GM(<\theta)}{c^2\theta{d_l}}
\frac{d_{ls}}{d_s}$ is produced at position ${\theta}=\xi_r{r_s}/d_l$.

The mean interior mass within some radius $r_x=x C_{x} r_s$ can be obtained 
by integration of the NFW profile giving:

\be\label{McolNFW}
M(<r)={4\pi\over{3}}r_x^3{\Omega_m(o)\Delta_c(1+z)^3\over{x^3\Omega_m(zl)}}
{\left[{\log(1+xC_x)-{xC_x\over{1+xC_x}}\over{\log(1+C_x)-{C_x\over{1+C_x}}}}\right]} 
\ee

 An accurate fit to an NFW profile is achieved with a characteristic
radius of $r_s=310^{+140}_{-120}kpc/h$ and concentration parameter
$C_{vir}=8.2^{+2.1}_{-1.8}$, (figures 25\& 27), where the above
integral is carried out to the virial radius, $r_{vir}$ (note,
sometimes the concentration parameter is quoted for a smaller radius at
which the mean interior density is 200 times the cosmological mean
density, which gives a smaller value for the concentration
parameter, $C_{200}=6.5^{1.9}_{-1.6}$). Note that
by incorporating the relative magnifications (section 9) a small improvement in
the accuracy of the concentration parameter is obtained (figure 27),
$C_{vir}=7.3^{1.6}_{-1.4}$.  This corresponds to a virial mass of of
$M_{vir}=2.6\times 10^{15}hM_{\odot}$ and a velocity dispersion of
$\sigma\sim 1700km/s$ at $r_s$ (Figure 31).  The fit is remarkable for the
way it not only follows the form of the profile all the way to the
center of mass, but also accurately reproduces the locations of both
the radial and tangential critical curves. One may make a
comparison with the entire magnification profile not just the
divergent critical radii and this is also shown in Figure 25 (lower
left panel). The magnification profile is rather well reproduced
overall, with a small excess compared to NFW outside the tangential
radius.  If we reduce $r_s$ to only $100kpc/h$ for example, the
critical curves are underestimated by a factor of $\sim 2.5$. This
less massive model reproduces the outer weak lensing measurements
discussed below but is clearly inconsistent with the strong lensing
results.

The concentration parameter derived from this fit is relatively high
for such a massive halo. A value of $C_{vir} \sim 5$ is generally
anticipated for massive clusters, although the scatter predicted in
concentration at a given mass is considerable (eg. Bullock etal 2001).
It is important to measure the outer profile of the mass distribution
to constrain more clearly the degree of concentration.  This we now
are tackling with high quality Subaru imaging (Broadhurst etal, in
preparation).

\section{Weak Lensing, Dynamical, and X-ray Results}\label{WL}

  We have only sampled the profile to a radius of $250kpc/h$, which
corresponds to $r\sim r_s$ for the best fit NFW profile obtained
above. We may compare the outer profile of A1689 with the weak lensing
distortion measurements made by various groups. Weak lensing
measurements of A1689 analysed by Clowe \& Schneider (2001) and King
etal (2002), under-predict the Einstein radius by over a factor of
$\sim2$, predicting an Einstein radius of only 20\arcs ~compared to
the 50\arcs~ observed and are unable to discriminate significantly
between an NFW profile and a purely isothermal profile. The weak
lensing signal is known to be heavily diluted in typical ground based
data (Kaiser, Squires \& Broadhurst 1995), requiring large corrections
(a correction factor upward of 3 is typical for the seeing alone) for
the seeing and the contamination by undistorted galaxies belonging to
the cluster and the foreground. Although this latter effect has been
addressed in detail by Clowe \& Schneider (2001), they have only one
optical passband and thereby suffer maximal contamination. The authors
conclude that the mass derived from their weak lensing observations
falls well short of the Einstein radius, by a factor $\sim2.5$, a
discrepancy which is reinforced with recent deeper weak lensing work
of Bardeau etal 2004 for this cluster using CFHT who find good
agreement with the weak lensing measurements of Clowe \& Schneider.
Tyson \& Fischer (1995) on the other hand favour a mass profile which
is steeper than isothermal, ($p=-1.4$) for the region $r>100kpc/h$,
using modest quality 4m CTIO imaging.

 For the small overlap region of the strong lensing and weak lensing
data, our best fit model over-predicts the weak lensing strength
by $\sim 2.0\pm0.3$ (Figure 25, bottom right plot). If we take the
most concentrated profile consistent with the strong lensing data this
discrepancy is reduced by only 30\%. Clearly it is important to
attempt weak lensing measurements with better seeing and with deep
multi-colour images, to minimise the contamination by blue foreground and
cluster galaxies.
 
 A redshift survey of the luminous cluster galaxies by Teague, Carter,
\& Grey (1990), reveals a broad distribution of cluster galaxy
velocities and a tail to higher velocities (Figure 31). The large
value of $\sigma=2355\pm200$ km/s derived by Teague, Carter \& Grey
(1990) is probably overestimated due to the presence of a background
sub-group, (Miralda Escud\'e \& Babul 95, Girardi et
al. (1997)). Lower values of 1400km/s are preferred by a wavelet
analysis of Girardi et al. (1997), though inspection of the 1D
distribution of velocities shows that a Gaussian of 1700km/s is a good
fit when excluding the only obvious subgroup at +7000km/s (see Figure
31 where we have overlayed this fit).  This background group is
estimated to have a velocity dispersion of only 700km/s (Girardi et
al. 1997) and contains the bright galaxies of the sub-group, which
from our strong lensing analysis, is obviously a separate entity
distinct from the main body of the cluster. The relatively low
velocity dispersion of this sub-group is consistent with our modeling
which shows that this group has a noticeably lower M/L compared with
the main cluster, by a factor of $\sim 3$.

 Xue \& Wu (2002) obtained a detailed Chandra X-ray map and limited
spectral information for the plasma in A1689. These authors prefer a
high gas temperature of $\sim10-13$KeV, with a trend to somewhat lower
temperature towards the center, based on a double beta model fit which
aims to account for soft X-ray absorption. The X-ray contours of A1689
are close to circular at all radii, out to $1Mpc/h$, the limit of the
data, with no obvious excess corresponding to the main sub-group
described above. Naively taking the above gas temperature and assuming
the gas and the dark matter are isothermal and that the gas is fully
ionized with primordial abundance, then the predicted Einstein radius
becomes $\theta_e=4\pi(KT/{0.6m_pc^2})d_{ls}/d_s = 45$ \arcs, close to
the observed value of $\sim 50$ \arcs. More recent X-ray observations
of A1689 with XMM (Andersson \& Madejski 2004) indicate a maximum
temperature of $\sim8$ KeV, at about the Einstein radius, $100kpc/h$,
and the best fit to an NFW profile gives $r_s=1.1\pm0.2Mpc/h$ which is
about one third of the value derived here. However, Andersson \&
Madejski (2004) point out there is evidence of asymmetry of the gas
velocity profile and also asymmetry in the temperature map, so that
the simple assumption of hydrostatic equilibrium required to derive a
cluster mass profile from X-ray temperature and density profile
information may not be justified. 

\section{Cosmological Parameters}\label{Cosmo}

 In principle, lensing may provide a measurement of the cosmological
curvature by simply comparing the scale of the deflection field for
multiply lensed sources as a function of source redshift behind a
lens. This follows from the simple linear dependence of the bend-angle
with the distance ratio, $d_{ls}/d_s$, which appears as a
scaling term in the general expression for the bend-angle (equation
5). For a given source distance, the deflection field,
$\vec\alpha(\vec\theta)$, is proportional to the the distance ratio,
$D=d_{ls}/d_s$. The bend angles also scale linearly with the amplitude
of the surface mass distribution, $\Sigma_o$, but this is not measured
independently by lensing. Therefore in the absence of any external
constraint, the relative scaling of the bend-angle field is
proportional to the product of both, $\Sigma_o D $. Hence, only the
relative scaling of the deflection field can be used when examining
cosmological curvature via lensing.

 The relative distance ratios of sources lensed by the same cluster
are only a weak function of cosmology and are best measured for a
given lens using many sources over a wide range of source redshift. In
constructing the model mass distribution (section 8) we set the
normalisation at a reference redshift of $z=3$, i.e $f_k\equiv
D(z)/D_o$, $D_o\equiv D$(z=3).  This redshift is preferred as it is
the average redshift of the background sources. The function $D(z)$
can be expressed in terms of redshift for a given set of cosmological
parameters. In the main case of interest, that of a flat model with a
nonzero cosmological constant, the relation is given by:

\be\label{D(z)}
D(z) ={c\over H_0}\;
\frac{1}{(1+z)} \int^z_0
\frac{dz}{[\Omega_\Lambda+\Omega_M(1+z)^3]^{1/2}}
\ee 

General expressions for the dependence of this distance ratio
on arbitrary combinations of $\Omega_m$ and $\Omega_\Lambda$ are 
lengthy and can be found in Fukugita et al. (2000).

 A comparison of relative deflection angles and hence relative
distance ratios has not been feasible to date using strong lensing, as
this requires many multiple sources behind a given lens and a reliable
model for a statistically meaningful measurement. Here we take our set
of models described above and simply determine the value of $f_k$
using equation N. Figure 29 (upper panels) shows the values of
$f_k\equiv D_k(z)/D_o(z=3)$ for three choices of resulting mass
profiles (also shown in Figures 14,15\&16) covering the range $0.4<d\log(\Sigma(r)/d\log(r)<1$ and we
compare the corresponding predictions for the redshift dependence of
$D_k(z)/D_o(z=3)$, for various choices of cosmological parameters.

 The difference between cosmological model for the interesting range
$0<\Omega_M+\Omega_\Lambda\leq 1$ is smaller than the scatter between
the data points (Figure 29). However, the form of this function,
$D(z)/D_o$, does have a distinctive dependence on redshift, virtually
independent of the cosmological parameters for the above range. So we
may turn the question around and adopt the reasonable assumption that
the cosmological parameters lie in the standard range,
$0<\Omega_M+\Omega_\Lambda\leq1$, and then evaluate the fits to $f_k$
to see which profile is preferred. Importantly, this evaluation of the
mass profile is independent of our model fitting to the image
locations because we do not make use of the redshift information when
fitting the model to the locations of the multiple images, since as we
showed in section 8.5 (see Figure 12), the output mass distribution is
virtually independent of the choice of input values of $f_k$ when
fitting the image positions.

The values of $f_k$ are found to increase with redshift as predicted
for the above range of cosmological parameters. In particular, the
model resulting in a mean slope of $\bar{p} \simeq 0.55$ (Figure 29, central
panel) is preferred by the data over the extreme slopes considered
(left and right panels of Figure 29) with a tight
scatter about the cosmological relation, accurately following the
distinctive trend with redshift.

That the best fitting lens model should produce this sensible
geometric trend with redshift adds considerable credibility to the
model and need not have turned out so well, as neither the redshift
information nor any cosmological model is involved in making the mass
model, as noted in section 8.5. Note the scatter on the
distance-redshift relation found this way is not small enough to
distinguish between the interesting cosmological models, however
combinations with $\Omega_M+\Omega_\Lambda>1.2$ are clearly excluded
by the data.

We have evaluated the scatter about the standard cosmological
relations as a function of the mean output mass profile finding a
smooth trend with a minimum at $\bar{p} \sim 0.55^{+0.1}_{-0.15}$,
where the $1\sigma$ scatter is estimated empirically from dispersion
on the scatter of the best model $\sigma(D_i(z)/D_o(z=3))=0.009$. This
best fit value for the slope is also the slope preferred by fitting the
multiple image positions as described above. Note that the 3 objects
with measured redshifts are no better at describing the cosmological
redshift dependence than those with photo-z's, lending, by
consistency, further credibility to our method of estimating
photometric redshifts in addition to the good agreement with the
spectroscopic redshifts.

 The lens model can of course be re-run by setting these relative
distances to the their best-fit value and rerunning the model to
improve the relative distances. However, this iteration produces a
negligible change in the mass profile, and only a tiny change
in the re-derived relative distances.

 With independent mass information for the cluster we may use the
lensing information to better discriminate between the models, since
the normalization using $D_o$(z=3) is not then required. If we take
the measured velocity dispersion of the cluster
$\sigma=1700\pm100$ km/s and adopt a circular critical radius
based on the pattern of the most tangentially stretched arcs to be
approximately $\theta_e=40-55$\arcs, and we assign the mean redshift of
$\bar{z}\sim 3$, we can adjust the amplitude of the mass distribution
accordingly, assuming that motions are approximately isothermal, which
may not be consistent with the flatter than isothermal mass profile
obtained from our modeling. By plotting the unnormalised values of
$D_i(z)$ compared with the above choices of cosmological parameters, we
then see that within the additional uncertainties introduced by the
measured velocity dispersion and the uncertain critical radius, little
is gained in terms of precision and may be significantly undermined
by the assumption of isothermality. Perhaps careful dynamical
modeling or hydrodynamic modeling can help to make use of the
dynamical and X-ray measurements will help in establishing the benefit
of this external information for this cosmological test.

Finally, we mention the benefit of extending this work to higher
redshift clusters. For the low redshift of A1689, the distance ratio
is close to saturation by $z\sim 1$, so that $d_s\simeq
d_{ls}$. The dynamic range of $d_{ls}$ can be expanded for higher
redshift clusters. For a cluster at z=0.5 for example, the range in
$f_k$ for source redshifts covering the range $1<z<5$ is 30\%, a
factor of 3 larger than for A1689 at z=0.185.  However, some of this
improvement is offset by the naturally smaller critical radius
produced for a more distant cluster of fixed mass, as the lens-source
separation is smaller and hence the numbers of multiply lensed images
will be correspondingly fewer, unless we obtain even deeper images
to compensate.

\section{Discussion and Conclusions}\label{Conc}

We have obtained the highest quality images of a lensing cluster to
date, surpassing previous work both in terms of depth and spatial
resolution and have been rewarded with an order of magnitude increase
in the numbers of multiply lensed galaxies identified around an
individual cluster. This substantial improvement has permitted
detailed modeling of the cluster, which for the first time allows us
to trace a radial critical curve and to measure the mass profile of
the cluster all the way to the center, inside the radial critical
curve, using the many small counter-images found projected on the
center of mass.

 Also for the first time, we have measured the purely geometric change
of the bend-angle with source distance, finding consistency with a
flat universe for a wide range of redshift, $1.0<z<5.5$. This is an
important consistency check on the currently more accurate
measurements of the cosmological parameters at lower redshift, based
on SnIa and on the CMB fluctuation spectrum at high redshift. 

Our main conclusions are:

\begin{itemize}

\item{We have identified 106 multiply lensed images, corresponding
to 30 multiply lensed galaxies behind A1689. These images
include tangential and radially stretched images and also tiny
de-magnified central images interior to the radial critical curve. Both
the radial and tangential critical curves are delineated in
unprecedented detail.  This is the first time that any radial
critical curve has been traced without a model, just by virtue of the
many radial arcs visible. Tiny point-like central images are also
found in considerable numbers, allowing for the first time the mass
distribution to be traced all the way in to the center of mass.}

\item{ The best fitting mass profile derived from the multiple images
has a continuously flattening projected slope towards the center of
mass with an average gradient of $d\log\Sigma(r)/d\log(r) =
-0.55\pm0.1$, in the observed range of $r<250kpc/h$ and is flatter than
the light profile. The profile derived from the positions of the
multiple images is checked independently by comparing the relative
fluxes of multiple images with the model prediction for the relative
magnifications and we also compare the photometric redshift of the
multiple images with the relative distances derived from the model. We
see that the relative magnifications are reproduced very well by our
best fitting model over the full range of radius. The
distance-redshift relation is found to follow accurately the form of
expected form of the cosmological relation for standard cosmology.
These consistencies add considerable confidence that our approach 
modeling has produced an accurate mass profile for the inner region of 
A1689.}

\item{ The shape of the mass distribution is approximately circular
in projection and much rounder than the clumpy spatial distribution of luminous
cluster galaxies. In particular the main sub-group is a relatively
small mass perturbation. The roundness of this mass distribution
argues against this cluster being a lengthly projected filament along
the line of sight, though does not rule out a favourable alignment of
a triaxial potential boosting the observed surface density.}

\item{ A softened isothermal profile can be made to match well the
modeled mass profile or to reproduce the locations of the critical
radii, but not both simultaneously. The profile is either too steep or
has too large a radial critical curve. Nevertheless this model comes
strikingly close to matching the data, for a core radius of $\simeq
20$\arcs. Given the potential physical importance of any core in the
mass distribution, this possibility should be examined more rigorously
to evaluate, for example, the influence of ellipticity on the location
and shape of the radial critical curve.}

\item{The mass to light ratio of A1689 is high,
$M/L(r<250kpc/h)\simeq 400h(M/L_B)_{\odot}$, 30\% larger than any other
well-studied cluster, continuing the general trend of higher M/L
with increasing mass. This large value means that the stellar mass is
a negligible contribution to the overall mass, even in the center
$r<50kpc/h$, where the light is more concentrated than the mass.  Only
for values exceeding $M/L>30h(M/L_B)_{\odot}$ do galaxies have any
noticeable effect on the slope of the inner mass profile. So unlike other
lensing work where the cD dominates central mass, we do
not suffer from the uncertain subtraction of the central galaxies.}

\item{ Our best fitting mass profile to an NFW profile is strikingly
good, accounting naturally for the shallow profile observed and its
continuous flattening towards the center of mass and the relative
radii of the tangential and radial critical curves. This is arguably
the most accurate measurement of the central mass profile of any
cluster in the region $r<r_s$, where the profile of a CDM dominated
halo is predicted to be relatively shallow. Interestingly, the
concentration parameter we derive for the inner region may be
surprisingly high for such a massive halo, and it will be important to
examine the form of the mass profile to larger radius with weak
lensing measurements to more directly constrain the degree to which
the profile as a whole is concentrated.}

\item{ The cosmological dependence of the bend-angle of light with
redshift scales purely geometrically with the lensing distance ratio,
$d_{ls}/d_s$. This trend is convincingly detected here, for the first
time. It is, however, not a strong function of redshift for the
interesting range of cosmological parameters but it is consistent with
the standard flat model, and excludes combinations with $\Omega_M
+\Omega_\lambda>1.2$. The agreement need not have been so clear - the
modeling does not include the redshift information, so therefore the
fact that the bend-angle-redshift trend follows the expected
geometric relation add considerable confidence to our approach to 
lens modeling.}

\end{itemize}

Clearly it is also important to explore the generality of our results
by observing other clusters. Our GTO survey will include a few more
massive lensing clusters, but the images will be shallower than for
A1689, with fewer than 20 orbits each. A more dedicated effort,
equivalent in depth to the deep field programs would be very well
rewarded, increasing our knowledge of the nature of dark matter, and
provide samples of the most distant galaxies and should also permit an
independent measurement of the cosmological curvature in a redshift
range not covered by SNIa or CMB observations.
 
\acknowledgements

TJB acknowledges grant ISF03310211 from the Israeli Science
Foundation, and useful conversations with Bernard Fort, Renan Barkana,
Eran Ofek, Nir Shaviv, Avishai Dekel, Keiichi Umetsu and Masahiro Takada. 
We thank the referee for useful suggestions.  ACS was developed under NASA
contract NAS 5-32865, and this research is supported by NASA grant
NAG5-7697.  We are grateful for an equipment grant from the Sun
Microsystems, Inc.

{}

\clearpage
\begin{deluxetable}{lllllll}
\tabletypesize{\scriptsize} 
\tablecaption{ACS WFC imaging of ABELL 1689} 
\tablewidth{0pt} 
\tablehead{
\colhead{ACS WFC filter} & 
\colhead{Exposure time}  & 
\colhead{N exposures\tablenotemark{a}}&
\colhead{Area (arcmin$^2$)} &
\colhead{Extinction\tablenotemark{b}} &
\colhead{$m_{lim}$\tablenotemark{c}} & 
\colhead{Observation date}
}
\startdata 
F475W    & $9500$  & $8$  & 11.73 & 0.107 & 27.24 & Jun 15th,16th 2002 \\ 
F625W    & $9500$  & $8$  & 11.70 & 0.076 & 27.00 & Jun 14th 2002      \\
F775W    & $11800$ & $10$ & 11.63 & 0.057 & 26.92 & Jun 13th,20th 2002 \\ 
F850LP  & $16600$ & $14$ & 11.79 & 0.042 & 26.50 & Jun 12th,16th 2002 \\
\enddata
\tablenotetext{a}{This includes pointings and CR splits}
\tablenotetext{b}{From Schlegel, Finkbeiner, \& Davis 1998}
\tablenotetext{c}{This is the 10$\sigma$ limiting magnitude within a 0.2$\arcsec$ aperture. 
Because of the high quality psf, the limiting magnitude for point sources within a 
$4\times$FWHM aperture is $\approx 0.2$ magnitudes fainter than this value.}
\end{deluxetable}

\newpage

\newpage

\begin{deluxetable}{lllllllll}
\tabletypesize{\scriptsize}
\tablecaption{Multiple arc systems in ABELL 1689} 
\tablewidth{0pt}
\tablehead{
\colhead{Arc ID} & 
\colhead{RA (J2000)}  &
\colhead{Dec (J2000)}  &
\colhead{X} \tablenotemark{a}&
\colhead{Y} \tablenotemark{a}&
\colhead{$i_{F775W}$\tablenotemark{b}} & 
\colhead{$z_{phot}$\tablenotemark{c}} &
\colhead{$z_{best}$\tablenotemark{d}} &
\colhead{$z_{spec}$\tablenotemark{e}} 
} 
\startdata
1.1\tablenotemark{*} & 13:11:26.4498 & -1:19:56.753 & 8.72  & -54.80 & 23.44 & $3.03^{0.53}_{0.53}$ & $2.99^{0.48}_{0.48}$ & 3.04\tablenotemark{f,g} \\
1.2 & 13:11:26.2812 & -1:20:00.261 & 4.48  & -55.60 & 23.69 & $3.04^{0.53}_{0.53}$ & $2.99^{0.48}_{0.48}$ & - \\
1.3 & 13:11:29.7771 & -1:21:07.475 & -34.12  & 20.40 & 24.52 & $3.27^{0.56}_{0.56}$ & $2.99^{0.48}_{0.48}$ & - \\
1.4 & 13:11:33.0629 & -1:20:27.403 & 23.08  & 48.00 & 24.16 & $2.94^{0.52}_{0.52}$ & $2.99^{0.48}_{0.48}$ & 3.05\tablenotemark{f} \\
1.5 & 13:11:31.9350 & -1:20:05.998 & 35.27  & 23.60 & 24.62 & $3.35^{0.57}_{0.57}$ & $2.99^{0.48}_{0.48}$ & - \\
1.6 & 13:11:29.8536 & -1:20:38.413 & -7.33  & 9.10 & 25.82 & $1.06^{1.91}_{0.27}$ & $2.99^{0.48}_{0.48}$ & - \\
\hline
2.1 & 13:11:26.5237 & -1:19:55.450 & 10.38  & -54.35 & 23.37 & $2.62^{0.47}_{0.48}$ & $2.54^{0.42}_{0.42}$ & - \\
2.2 & 13:11:32.9642 & -1:20:25.494 & 24.18  & 45.85 & 24.18 & $2.57^{0.47}_{0.47}$ & $2.54^{0.42}_{0.42}$ & - \\
2.3 & 13:11:31.9728 & -1:20:07.147 & 34.48  & 24.60 & 24.36 & $2.64^{0.48}_{0.48}$ & $2.54^{0.42}_{0.42}$ & - \\
2.4 & 13:11:29.8119 & -1:21:06.008 & -32.58  & 20.25 & 24.48 & $2.36^{0.44}_{0.44}$ & $2.54^{0.42}_{0.42}$ & - \\
2.5 & 13:11:29.8790 & -1:20:39.365 & -8.03  & 9.85 & 25.63 & $1.59^{0.86}_{0.34}$ & $2.54^{0.42}_{0.42}$ & - \\
\hline
3.1 & 13:11:32.0428 & -1:20:27.520 & 16.48  & 34.20 & 26.65 & $5.48^{0.85}_{0.85}$ & $5.47^{0.78}_{0.78}$ & - \\
3.2 & 13:11:32.1718 & -1:20:33.287 & 12.08  & 38.40 & 26.85 & $5.45^{0.85}_{0.85}$ & $5.47^{0.78}_{0.78}$ & - \\
3.3 & 13:11:31.6850 & -1:20:56.040 & -11.62 & 41.45 & - & - & $5.47^{0.78}_{0.78}$ & - \\
\hline
4.1 & 13:11:32.1705 & -1:20:57.355 & -9.73  & 48.60 & 24.62 & $1.06^{0.27}_{0.27}$ & $1.33^{0.28}_{0.28}$ & - \\
4.2 & 13:11:30.5186 & -1:21:12.026 & -33.52  & 32.40 & 23.91 & $1.32^{0.30}_{0.31}$ & $1.33^{0.28}_{0.28}$ & - \\
4.3 & 13:11:30.7576 & -1:20:08.322 & 25.68  & 8.60 & 25.12 & $1.47^{0.32}_{0.33}$ & $1.33^{0.28}_{0.28}$ & - \\
4.4 & 13:11:26.2866 & -1:20:35.422 & -27.33  & -40.60 & 24.67 & $1.33^{0.31}_{0.31}$ & $1.33^{0.28}_{0.28}$ & - \\
4.5 & 13:11:29.8458 & -1:20:29.357 & 0.83  & 5.15 & - & - & $1.33^{0.28}_{0.28}$ & - \\
\hline
5.1 & 13:11:29.0655 & -1:20:48.776 & -21.73  & 2.80 & 24.42 & $3.29^{0.56}_{0.56}$ & $3.27^{0.51}_{0.51}$ & - \\
5.2 & 13:11:29.2248 & -1:20:44.153 & -16.53  & 3.00 & 24.92 & $3.16^{0.55}_{0.55}$ & $3.27^{0.51}_{0.51}$ & - \\
5.3 & 13:11:34.1199 & -1:20:20.919 & 35.68  & 59.60 & 25.26 & $2.15^{0.67}_{0.41}$ & $3.27^{0.51}_{0.51}$ & - \\
\hline
6.1 & 13:11:30.7484 & -1:19:37.995 & 53.08  & -4.40 & 23.61 & $1.22^{0.29}_{0.29}$ & $1.17^{0.26}_{0.26}$ & - \\
6.2 & 13:11:33.3475 & -1:20:12.174 & 38.68  & 45.40 & 23.85 & $1.31^{0.30}_{0.30}$ & $1.17^{0.26}_{0.26}$ & - \\
6.3 & 13:11:32.7510 & -1:19:54.506 & 50.88  & 29.80 & 23.02 & $0.94^{0.26}_{0.25}$ & $1.17^{0.26}_{0.26}$ & - \\
6.4 & 13:11:32.4824 & -1:19:58.857 & 45.23  & 28.00 & 24.00 & $1.09^{0.27}_{0.27}$ & $1.17^{0.26}_{0.26}$ & - \\
\hline
7.1 & 13:11:25.4491 & -1:20:51.843 & -47.53  & -45.00 & 23.31 & $4.92^{0.78}_{0.78}$ & $4.92^{0.71}_{0.71}$ & 4.87\tablenotemark{g} \\
7.2 & 13:11:30.6715 & -1:20:13.902 & 20.08  & 9.80 & 24.19 & $5.20^{0.81}_{0.81}$ & $4.92^{0.71}_{0.71}$ & - \\
7.3 & 13:11:29.8204 & -1:20:24.870 & 4.73  & 2.90 & 28.23 & $0.77^{4.01}_{0.23}$ & $4.92^{0.71}_{0.71}$ & - \\
\hline
8.1 & 13:11:32.2984 & -1:20:50.909 & -3.08  & 47.60 & 24.54 & $2.63^{0.48}_{0.48}$ & $2.67^{0.48}_{0.74}$ & - \\
8.2 & 13:11:31.4027 & -1:21:05.541 & -22.03  & 41.65 & 24.31 & $2.77^{0.50}_{0.50}$ & $2.67^{0.48}_{0.74}$ & - \\
8.3 & 13:11:31.5056 & -1:20:14.078 & 25.23  & 21.20 & 25.71 & $2.75^{0.49}_{0.89}$ & $2.67^{0.48}_{0.74}$ & - \\
8.4 & 13:11:25.5301 & -1:20:20.162 & -18.32  & -57.35 & 23.97 & $0.70^{0.22}_{0.22}$ & $2.67^{0.48}_{0.74}$ & - \\
8.5 & 13:11:30.3295 & -1:20:30.494 & 2.88  & 12.20 & 27.42 & $0.77^{2.53}_{0.23}$ & $2.67^{0.48}_{0.74}$ & - \\
\hline
9.1 & 13:11:30.3084 & -1:19:48.652 & 40.62  & -5.85 & 25.92 & $4.97^{0.78}_{0.78}$ & $5.16^{0.74}_{0.74}$ & - \\
9.2 & 13:11:33.5206 & -1:20:50.335 & 5.23  & 63.95 & 27.53 & $1.06^{0.27}_{0.27}$ & $5.16^{0.74}_{0.74}$ & - \\
9.3 & 13:11:28.7470 & -1:21:15.805 & -48.23  & 9.95 & 25.72 & $5.16^{0.81}_{0.81}$ & $5.16^{0.74}_{0.74}$ & - \\
9.4 & 13:11:26.2723 & -1:20:26.927 & -19.73  & -44.40 & 27.15 & $5.17^{0.81}_{0.81}$ & $5.16^{0.74}_{0.74}$ & - \\
\hline
10.1 & 13:11:33.9794 & -1:20:50.855 & 7.68  & 70.40 & 23.66 & $1.75^{0.74}_{0.36}$ & $1.57^{0.31}_{0.31}$ & 1.37\tablenotemark{f} \\
10.2 & 13:11:28.0499 & -1:20:12.477 & 4.67  & -26.40 & 23.30 & $1.54^{0.33}_{0.33}$ & $1.57^{0.31}_{0.31}$ & - \\
10.3 & 13:11:29.3180 & -1:20:27.744 & -1.07  & -2.70 & 25.15 & $2.57^{0.47}_{0.63}$ & $1.57^{0.31}_{0.31}$ & - \\
\hline
11.1 & 13:11:33.3423 & -1:21:06.754 & -10.78  & 68.50 & 24.03 & $2.91^{0.51}_{0.51}$ & $2.90^{0.47}_{0.47}$ & - \\
11.2 & 13:11:29.0588 & -1:20:01.292 & 21.23  & -17.45 & 23.36 & $2.87^{0.51}_{0.51}$ & $2.90^{0.47}_{0.47}$ & - \\
11.3 & 13:11:29.4933 & -1:20:26.381 & 1.28  & -0.90 & 26.78 & $1.58^{0.73}_{0.52}$ & $2.90^{0.47}_{0.47}$ & - \\
\hline
12.1 & 13:11:30.3638 & -1:19:51.471 & 38.43  & -3.90 & 25.39 & $1.87^{0.38}_{0.38}$ & $1.87^{0.34}_{0.34}$ & 1.82\tablenotemark{f} \\
12.2 & 13:11:27.3591 & -1:20:54.946 & -38.18  & -17.75 & 24.30 & $1.99^{0.39}_{0.39}$ & $1.87^{0.34}_{0.34}$ & 1.82\tablenotemark{f} \\
12.3 & 13:11:27.2257 & -1:20:51.910 & -36.27  & -20.85 & 23.97 & $1.99^{0.39}_{0.39}$ & $1.87^{0.34}_{0.34}$ & - \\
12.4 & 13:11:28.9644 & -1:21:10.265 & -41.83  & 10.55 & 25.93 & $1.92^{0.38}_{0.38}$ & $1.87^{0.34}_{0.34}$ & - \\
12.5 & 13:11:33.4898 & -1:20:51.555 & 3.93  & 64.05 & 25.55 & $1.87^{0.38}_{0.38}$ & $1.87^{0.34}_{0.34}$ & - \\
\hline
13.1 & 13:11:32.8241 & -1:19:24.371 & 78.62  & 18.00 & 24.10 & $1.02^{0.28}_{0.27}$ & $1.02^{0.24}_{0.24}$ & - \\
13.2 & 13:11:32.9847 & -1:19:25.831 & 78.33  & 20.80 & 23.68 & $0.72^{0.23}_{0.23}$ & $1.02^{0.24}_{0.24}$ & - \\
13.3 & 13:11:33.3934 & -1:19:31.134 & 76.12  & 28.60 & 24.12 & $1.10^{0.28}_{0.28}$ & $1.02^{0.24}_{0.24}$ & - \\
\hline
14.1 & 13:11:29.0279 & -1:21:41.802 & -69.98  & 24.80 & 24.98 & $3.37^{0.57}_{0.82}$ & $3.41^{0.53}_{0.53}$ & - \\
14.2 & 13:11:29.4588 & -1:21:42.623 & -67.98  & 31.00 & 25.99 & $3.64^{0.61}_{0.61}$ & $3.41^{0.53}_{0.53}$ & - \\
\hline
15.1 & 13:11:28.0753 & -1:20:15.196 & 2.38  & -24.90 & 25.69 & $1.99^{0.39}_{0.39}$ & $1.99^{0.36}_{0.36}$ & - \\
15.2 & 13:11:34.0756 & -1:20:51.311 & 7.88  & 71.90 & 25.79 & $2.00^{0.39}_{0.39}$ & $1.99^{0.36}_{0.36}$ & - \\
15.3 & 13:11:29.2387 & -1:20:27.573 & -1.43  & -3.85 & 27.16 & $1.97^{0.39}_{0.43}$ & $1.99^{0.36}_{0.36}$ & - \\
\hline
16.1 & 13:11:27.9835 & -1:20:25.319 & -7.38  & -21.85 & 23.68 & $1.81^{0.37}_{0.37}$ & $2.01^{0.36}_{0.36}$ & - \\
16.2 & 13:11:28.9137 & -1:20:28.546 & -4.38  & -7.85 & 25.06 & $2.26^{0.43}_{0.43}$ & $2.01^{0.36}_{0.36}$ & - \\
16.3 & 13:11:34.3985 & -1:20:46.402 & 14.38  & 74.20 & 25.35 & $1.80^{0.71}_{0.37}$ & $2.01^{0.36}_{0.36}$ & - \\
\hline
17.1 & 13:11:30.6557 & -1:20:24.890 & 10.03  & 14.25 & 24.19 & $2.74^{0.49}_{0.49}$ & $2.22^{0.39}_{0.39}$ & - \\
17.2 & 13:11:30.3890 & -1:20:27.765 & 5.73  & 11.85 & 25.11 & $2.02^{0.40}_{0.40}$ & $2.22^{0.39}_{0.39}$ & - \\
17.3 & 13:11:24.9804 & -1:20:41.865 & -41.48  & -55.60 & 24.31 & $2.25^{0.43}_{0.43}$ & $2.22^{0.39}_{0.39}$ & - \\
\hline
18.1 & 13:11:28.2448 & -1:20:09.540 & 8.58  & -25.00 & 25.00 & $2.56^{0.47}_{0.47}$ & $2.54^{0.42}_{0.42}$ & - \\
18.2 & 13:11:33.8200 & -1:20:54.539 & 3.33  & 69.80 & - & - & $2.54^{0.42}_{0.42}$ & - \\
18.3 & 13:11:29.3622 & -1:20:27.392 & -0.48  & -2.25 & 26.78 & $1.58^{0.73}_{0.52}$ & $2.54^{0.42}_{0.42}$ & - \\
\hline
19.1 & 13:11:31.6333 & -1:20:22.597 & 18.32  & 26.55 & 24.83 & $1.72^{0.36}_{0.36}$ & $2.63^{0.44}_{0.44}$ & - \\
19.2 & 13:11:25.2404 & -1:20:20.003 & -20.03  & -61.35 & 25.23 & $2.74^{0.49}_{0.49}$ & $2.63^{0.44}_{0.44}$ & - \\
19.3 & 13:11:31.9546 & -1:20:59.315 & -12.88  & 46.50 & 25.88 & $1.57^{0.34}_{0.34}$ & $2.63^{0.44}_{0.44}$ & - \\
19.4 & 13:11:32.0523 & -1:20:57.131 & -10.28  & 46.90 & 24.76 & $2.58^{0.47}_{0.47}$ & $2.63^{0.44}_{0.44}$ & - \\
19.5 & 13:11:30.2100 & -1:20:33.961 & -1.03  & 12.05 & 27.94 & $4.54^{1.66}_{0.73}$ & $2.63^{0.44}_{0.44}$ & - \\
\hline

20.1 & 13:11:31.1430 & -1:21:40.935 & -55.73  & 53.15 & 22.80 & $1.63^{0.35}_{0.34}$ & $1.12^{0.25}_{0.25}$ & - \\
20.2 & 13:11:32.2212 & -1:21:29.299 & -38.33  & 62.85  & - & - & $1.12^{0.25}_{0.25}$ & - \\
\hline
21.1 & 13:11:31.0264 & -1:20:45.776 & -6.53  & 28.15 & 25.15 & $1.79^{0.37}_{0.37}$ & $1.78^{0.33}_{0.33}$ & - \\
21.2 & 13:11:30.8009 & -1:20:44.743 & -7.03  & 24.65 & 26.70 & $1.59^{0.34}_{0.34}$ & $1.78^{0.33}_{0.33}$ & - \\
21.3 & 13:11:25.2540 & -1:20:11.207 & -11.98  & -64.90 & 25.59 & $1.78^{0.36}_{0.36}$ & $1.78^{0.33}_{0.33}$ & - \\
\hline
22.1 & 13:11:29.6859 & -1:20:08.794 & 18.43  & -5.75 & 23.66 & $1.99^{0.39}_{0.39}$ & $1.98^{0.36}_{0.36}$ & - \\
22.2 & 13:11:29.6156 & -1:20:23.762 & 4.42  & -0.35 & 25.49 & $1.99^{0.59}_{0.39}$ & $1.98^{0.36}_{0.36}$ & - \\
22.3 & 13:11:32.4150 & -1:21:15.917 & -24.98  & 59.80 & 23.27 & $1.96^{0.39}_{0.39}$ & $1.98^{0.36}_{0.36}$ & - \\
\hline
23.1 & 13:11:29.5299 & -1:20:10.016 & 16.32  & -7.35 & 24.63 & $2.03^{0.40}_{0.40}$ & $2.00^{0.36}_{0.36}$ & - \\
23.2 & 13:11:29.5545 & -1:20:22.891 & 4.83  & -1.55 & 25.99 & $1.99^{0.62}_{0.39}$ & $2.00^{0.36}_{0.36}$ & - \\
23.3 & 13:11:32.6585 & -1:21:15.199 & -22.78  & 62.80 & 24.72 & $2.00^{0.39}_{0.39}$ & $2.00^{0.36}_{0.36}$ & - \\
\hline
24.1 & 13:11:29.1913 & -1:20:56.177 & -27.62  & 7.65 & 25.24 & $2.63^{0.48}_{0.48}$ & $2.52^{0.42}_{0.42}$ & - \\
24.2 & 13:11:32.0642 & -1:19:50.560 & 50.08  & 18.80 & 24.80 & $2.50^{0.46}_{0.46}$ & $2.52^{0.42}_{0.42}$ & - \\
24.3 & 13:11:30.2944 & -1:19:34.140 & 53.68  & -12.20 & 24.33 & $2.43^{0.45}_{0.45}$ & $2.52^{0.42}_{0.42}$ & - \\
24.4 & 13:11:33.7184 & -1:20:19.863 & 34.08  & 53.70 & 24.99 & $2.81^{0.50}_{0.69}$ & $2.52^{0.42}_{0.42}$ & - \\
24.5 & 13:11:29.6289 & -1:20:36.999 & -7.48  & 5.45 & 27.27 & $4.55^{0.80}_{0.73}$ & $2.52^{0.42}_{0.42}$ & - \\
\hline
25.1 & 13:11:28.4949 & -1:20:34.990 & -12.88  & -10.80 & 25.44 & $4.59^{0.73}_{0.73}$ & $4.53^{0.66}_{0.66}$ & - \\
25.2 & 13:11:34.6484 & -1:20:33.581 & 27.58  & 72.15 & 26.62 & $4.42^{0.71}_{0.71}$ & $4.53^{0.66}_{0.66}$ & - \\
\hline
26.1 & 13:11:25.1545 & -1:20:32.763 & -32.12  & -57.10 & 24.67 & $1.08^{0.39}_{0.27}$ & $1.07^{0.25}_{0.25}$ & - \\
26.2 & 13:11:31.3260 & -1:20:25.241 & 13.98  & 23.50 & 25.39 & $1.04^{0.27}_{0.27}$ & $1.07^{0.25}_{0.25}$ & - \\
26.3 & 13:11:30.2420 & -1:20:32.585 & 0.43  & 11.90 & 27.42 & $0.77^{2.53}_{0.23}$ & $1.07^{0.25}_{0.25}$ & - \\
\hline
27.1 & 13:11:25.1730 & -1:20:33.113 & -32.33  & -56.70 & 25.22 & $1.81^{0.37}_{0.37}$ & $1.74^{0.33}_{0.33}$ & - \\
27.2 & 13:11:31.3668 & -1:20:24.644 & 14.78  & 23.80 & 26.19 & $1.58^{0.48}_{0.34}$ & $1.74^{0.33}_{0.33}$ & - \\
27.3 & 13:11:30.1918 & -1:20:32.894 & -0.18  & 11.35 & 29.82 & $4.55^{1.63}_{0.73}$ & $1.74^{0.33}_{0.33}$ & - \\
\hline
28.1 & 13:11:28.2978 & -1:20:10.907 & 7.68  & -23.70 & 27.20 & $1.17^{4.29}_{0.29}$ & $5.45^{0.77}_{0.77}$ & - \\
28.2 & 13:11:34.2602 & -1:21:00.009 & 1.18  & 78.10 & 26.47 & $2.00^{0.43}_{1.23}$ & $5.45^{0.77}_{0.77}$ & - \\
\hline
29.1 & 13:11:29.2256 & -1:20:57.909 & -28.98  & 8.85 & 25.97 & $2.47^{0.57}_{0.46}$ & $2.49^{0.42}_{0.42}$ & - \\
29.2 & 13:11:30.0380 & -1:19:34.215 & 51.98  & -15.65 & 25.00 & $3.40^{0.58}_{0.58}$ & $2.49^{0.42}_{0.42}$ & - \\
29.3 & 13:11:32.1451 & -1:19:52.565 & 48.78  & 20.75 & 24.80 & $2.50^{0.46}_{0.46}$ & $2.49^{0.42}_{0.42}$ & - \\
29.4 & 13:11:33.6261 & -1:20:20.815 & 32.62  & 52.85 & 25.84 & $3.35^{0.57}_{0.57}$ & $2.49^{0.42}_{0.42}$ & - \\
29.5 & 13:11:29.7297 & -1:20:36.603 & -6.48  & 6.65 & 27.97 & $4.59^{1.66}_{0.73}$ & $2.49^{0.42}_{0.42}$ & - \\
\hline
30.1 & 13:11:32.4212 & -1:19:19.826 & 80.18  & 10.60 & 25.91 & $4.49^{0.72}_{0.72}$ & $3.28^{0.51}_{0.51}$ & - \\
30.2 & 13:11:33.1835 & -1:19:26.069 & 79.38  & 23.60 & 25.80 & $3.23^{0.56}_{0.76}$ & $3.28^{0.51}_{0.51}$ & - \\
30.3 & 13:11:33.6540 & -1:19:32.691 & 76.38  & 32.80 & 25.73 & $3.30^{0.56}_{0.56}$ & $3.28^{0.51}_{0.51}$ & -
\enddata
\tablenotetext{a}{Coordinates centered on the cluster cD, in arcsec}
\tablenotetext{b}{Observed  i-band magnitude}
\tablenotetext{c}{Bayesian photometric redshift}
\tablenotetext{d}{``Best'' combined photometric redshift for the system of arcs}
\tablenotetext{e}{Spectroscopic redshifts}
\tablenotetext{f}{Fort et al.}
\tablenotetext{g}{Frye et al.}
\tablenotetext{h}{Magellan}
\tablenotetext{i}{Duc et al.}
\tablenotetext{*}{double image}
\end{deluxetable}

\clearpage
\begin{figure}[ht]\label{stamps1}
\epsscale{0.65}
\plotone{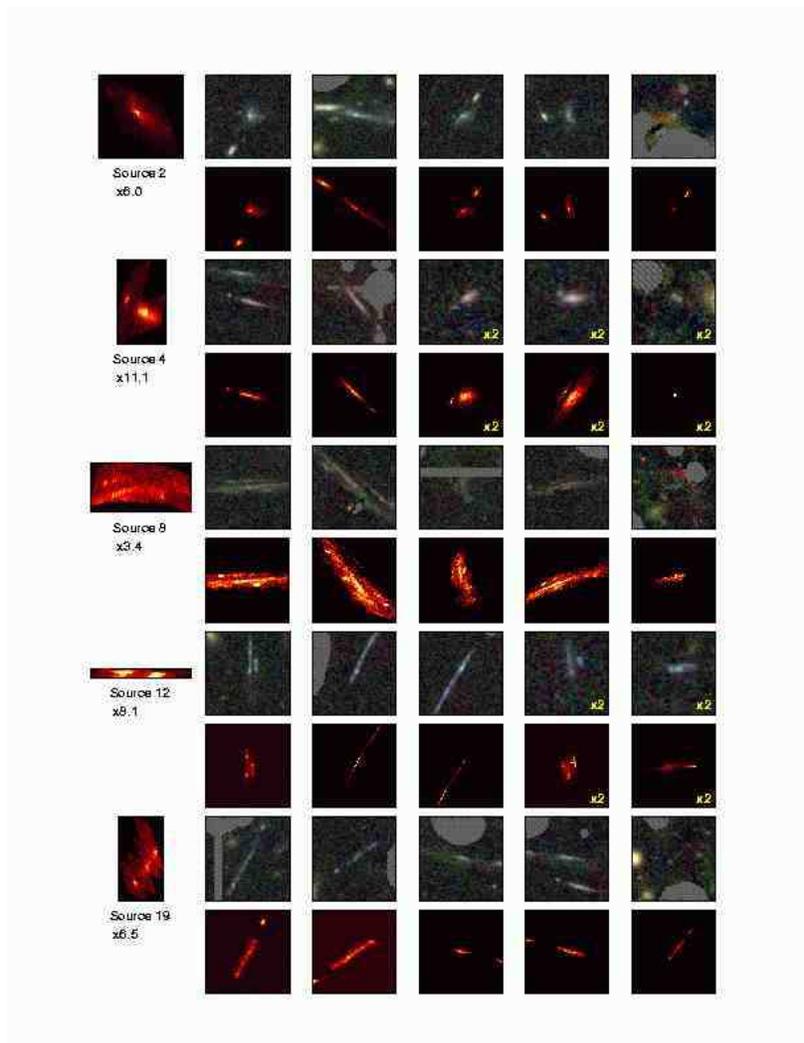}
\caption{ \footnotesize{All sources lensed into five images are shown
here in GRZ colour. Each set of multiple images is displayed along a
single row. On the left of this row is the reconstructed source
derived from delensing the first image in each row, usually the most
magnified image. The source image is usally much smaller than its
lensed images, so we show a magnified version relative to the lensed
images (the magnification factor is printed on the stamp of the
source).  For comparison we generate the corresponding model images in
a row below each set of observed images by applying the model to
de-lens one of the observed images. We choose the R-band data of the
most magnified multiple image - usually the first one in each row - to
generate all the other images using the model. These model images
generally match well their corresponding counter images, with little
loss of resolution, demonstrating the accuracy of the model. This
comparison is important to show when claiming a convincing model.
Note that the model stamps are centered on the reconstructed image
position for easy comparison with the data, and typically there is a
small offset of the model image of $1-3$\arcs~ with respect to the
observed image position, when using one image for generating the
others (see section 8.4)} }
\end{figure}

\clearpage
\begin{figure}[ht]\label{stamps2}
\epsscale{0.75}
\plotone{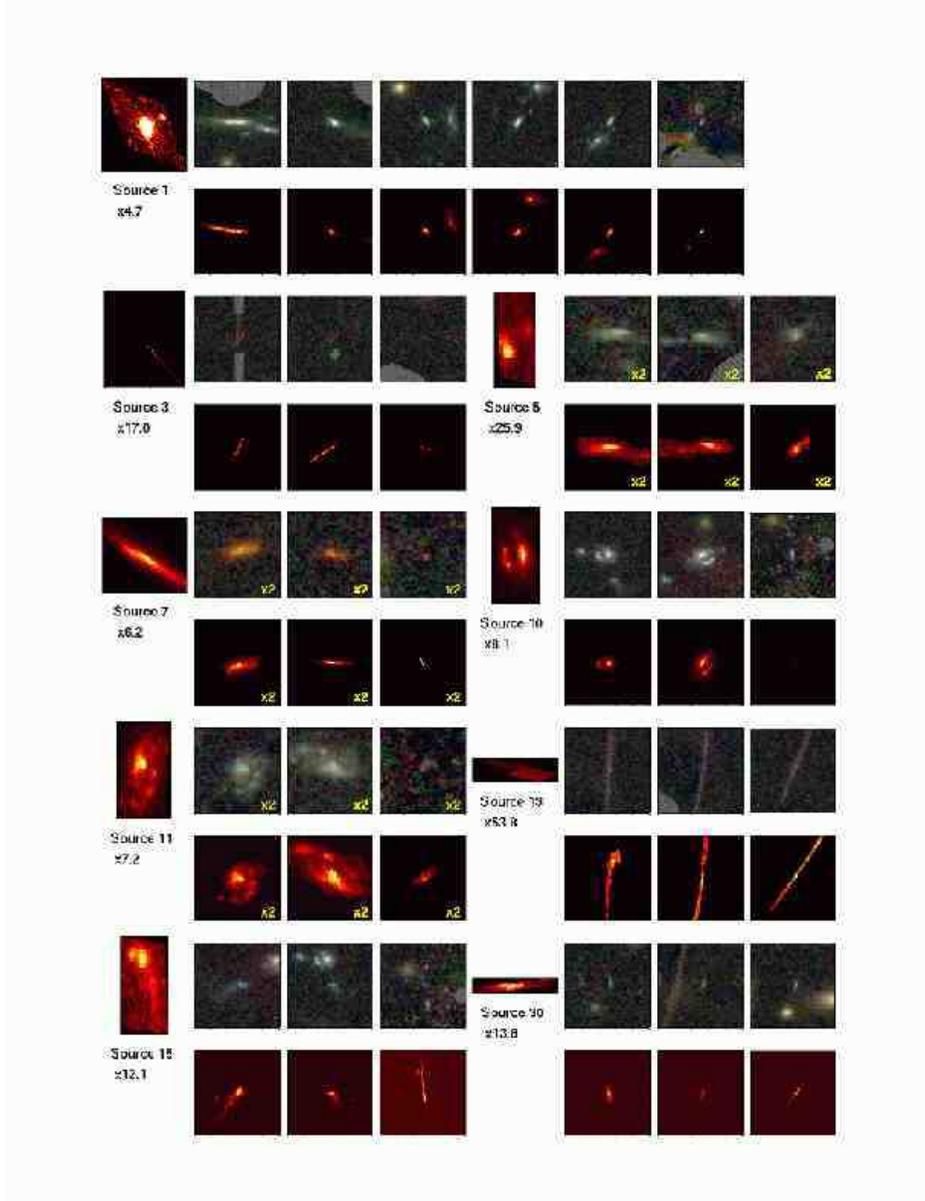}
\caption{ Similar to Figure 1. The first row here includes the source
with the largest number of lensed images, 7 in total. Note the first
postage stamp image of source 1 (labelled 1.1 in figure 5 and Table 2)
is really two very closely spaced images which are split locally by a nearby
cluster sequence galaxy. Interestingly, source 11 has a photometric
redshift of z=2.9, and is clearly a well resolved spiral galaxy,
magnified by a factor of $\sim 7$, and if the redshift is confirmed
would be the highest redshift example of a galaxy displaying spiral
structure. }
 \end{figure}

\clearpage
\begin{figure}[ht]\label{stamps3}
\epsscale{0.85}
\plotone{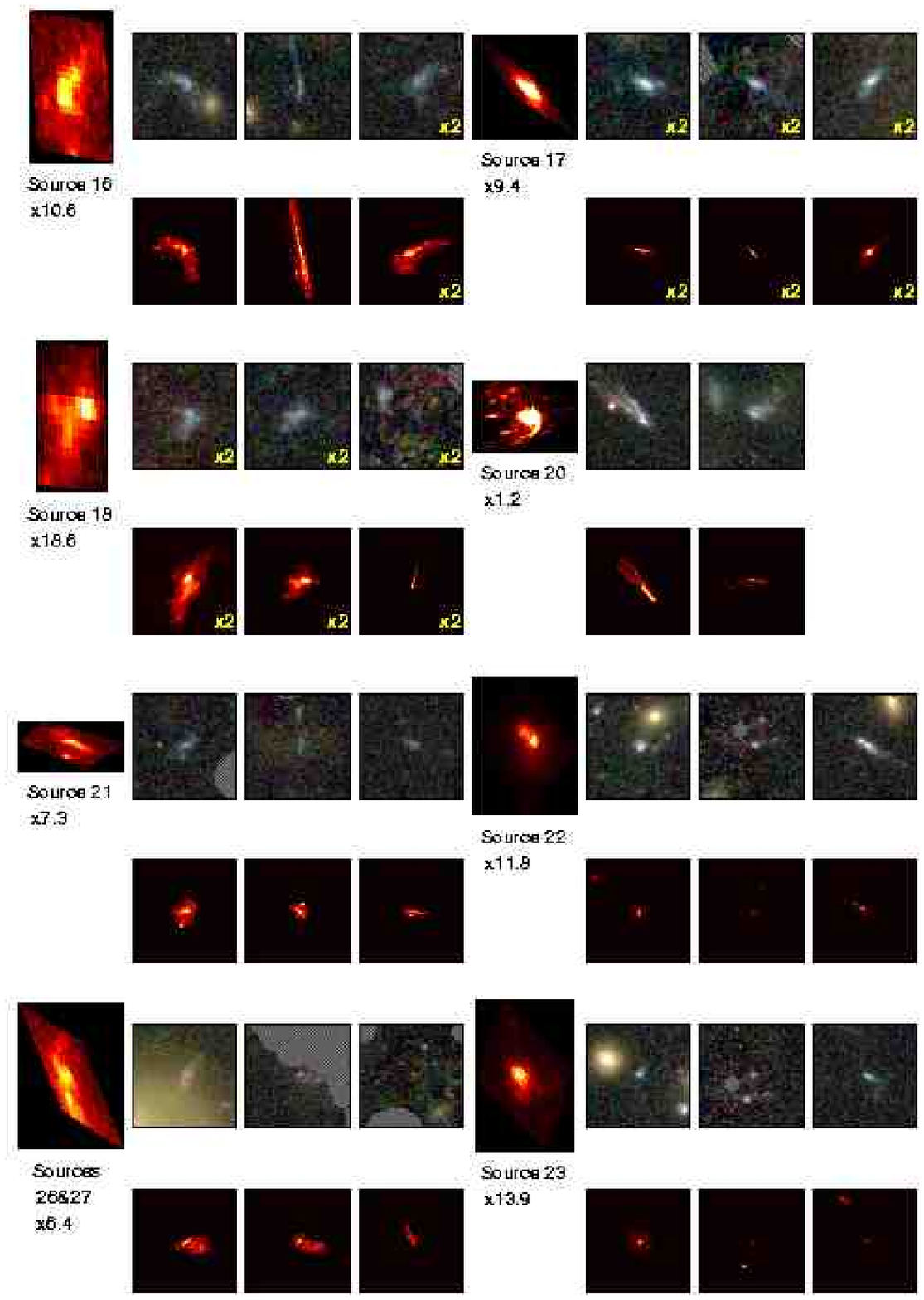}
\caption{Similar to Figure 1. Comparison of observed and modeled multiple
images found by our model. Observed images are shown above the 
corresponding model images and the reconstructed source is on the left.}
 \end{figure}

\clearpage
\begin{figure}[ht]\label{stamps4}
\epsscale{0.85}
\plotone{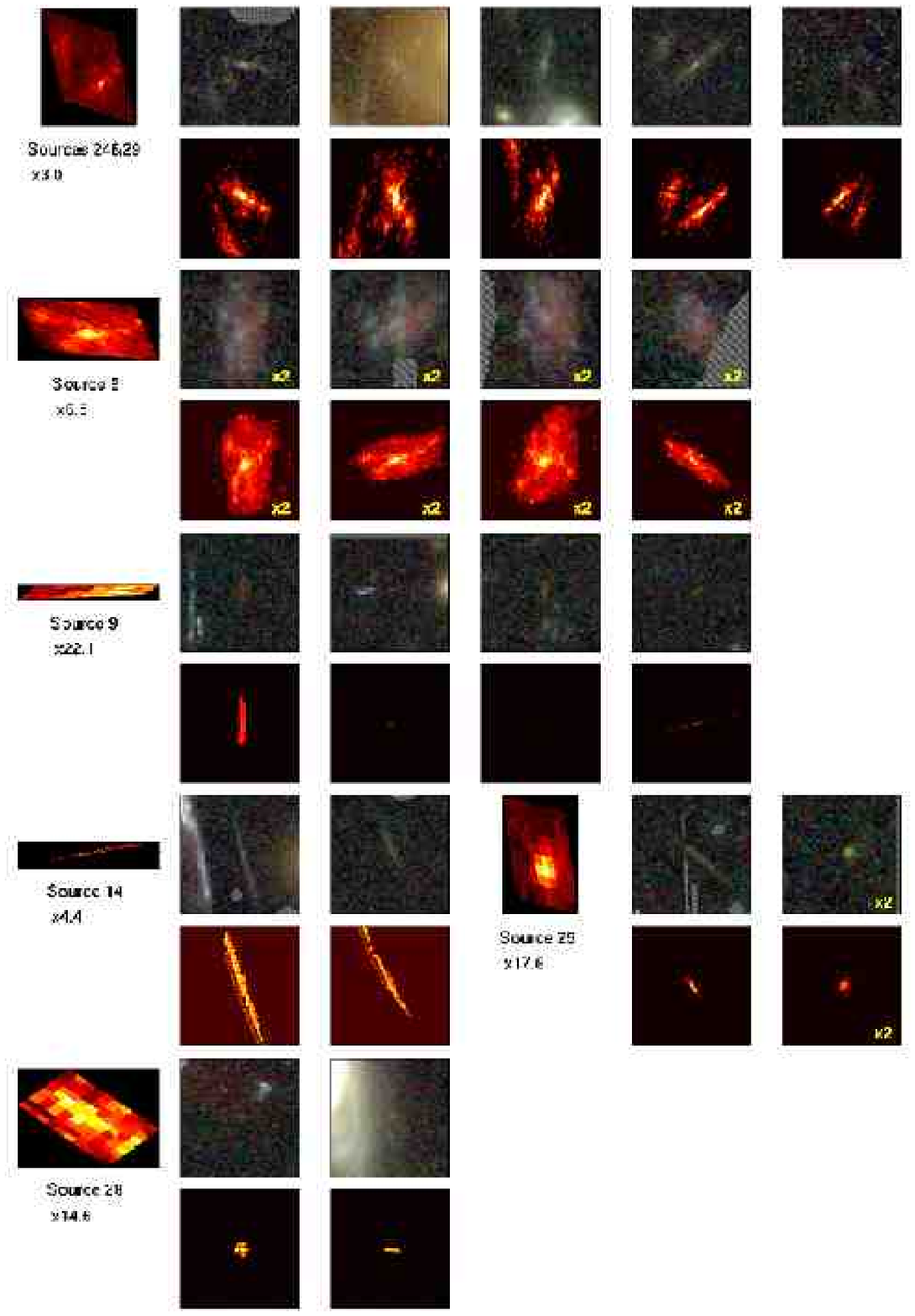}
\caption{ Similar to Figure 1. Comparison of observed and modeled
multiple images found by our model. Observed images are shown above the 
corresponding model images and the reconstructed source is on the left.}
 \end{figure}

\clearpage
\begin{figure}[ht]\label{catalog_fig}
\epsscale{0.95}
\plotone{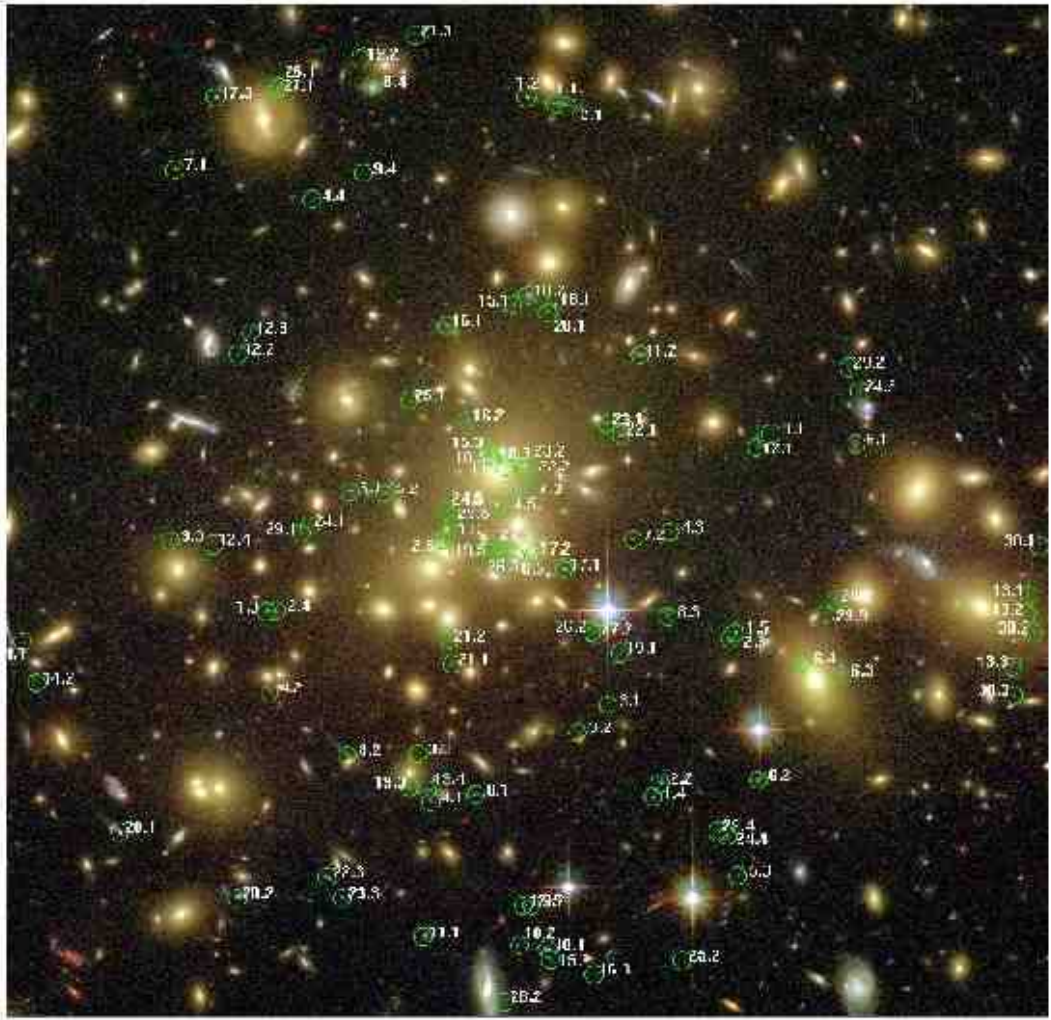}
\caption{ All 106 images identified to date with the help of the model
are marked on the field of the data. Labels correspond to Table 2,
where details of the photometry, coordinates, and redshift estimates
are listed. The multiple images cover the cluster fairly evenly
including the central region, interior to the radial critical curve. A closer
view of the central region is provided below with the cluster galaxies subtracted. }
\end{figure}

\clearpage
\begin{figure}[ht]\label{zoom_con}
\epsscale{0.95}
\plotone{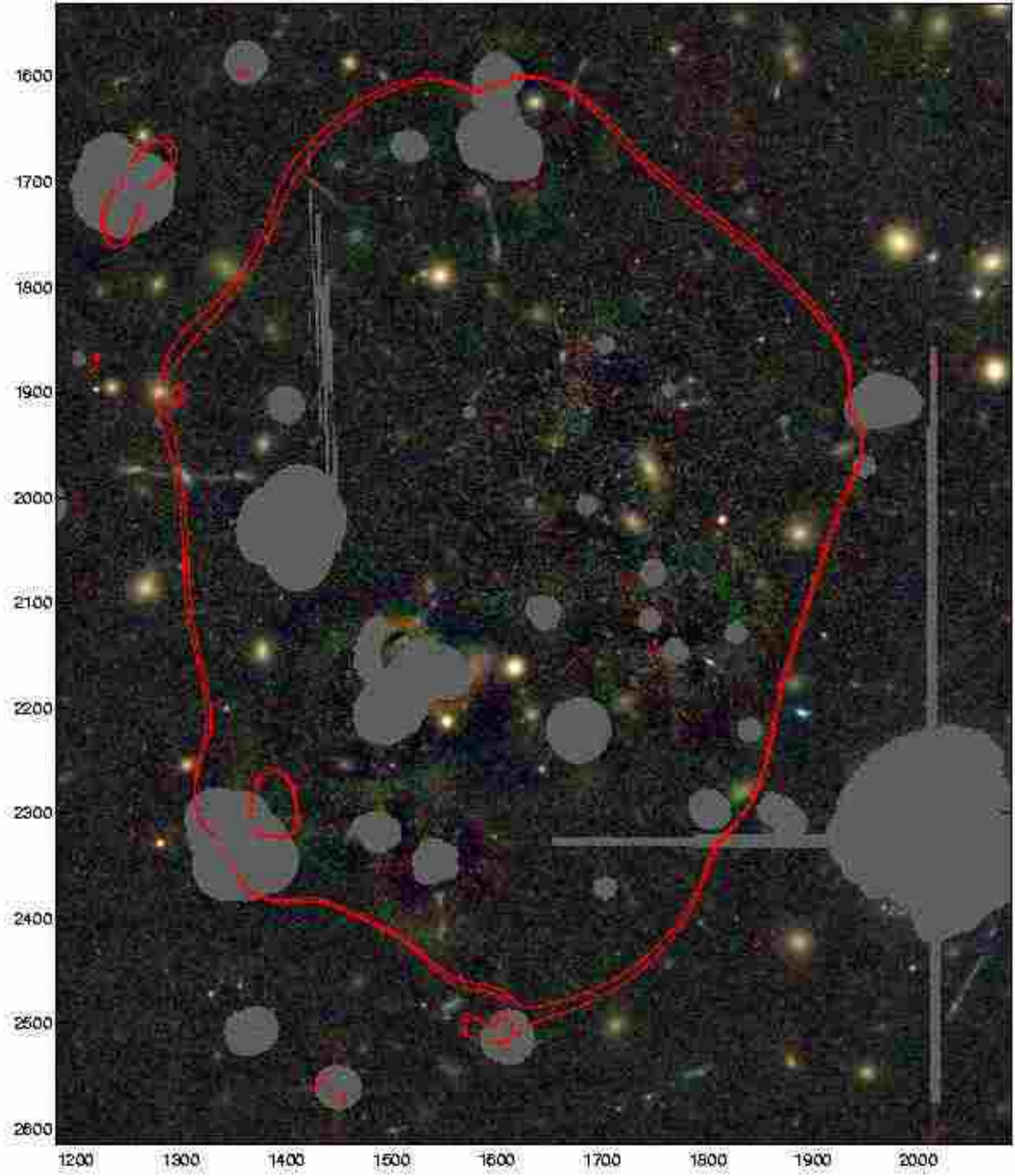}
\caption{ A blow-up of the central region with the bright elliptical
galaxies subtracted. Significant residuals are masked out, as are
stars and image artifacts.  Overlaid is the model radial critical
curve for $z_s=3$ ($f_k \equiv 1$). Several highly elongated
spoke-like images can be seen close to this curve, in some cases
bisecting the critical curve. Other images close to the curve
generally point in a radial direction towards the center of mass (in
the middle of the radial critical curve), and others are split on
either side.}

\end{figure}

\clearpage
\begin{figure}[ht]\label{catalog_zoom_con}
\epsscale{0.95}
\plotone{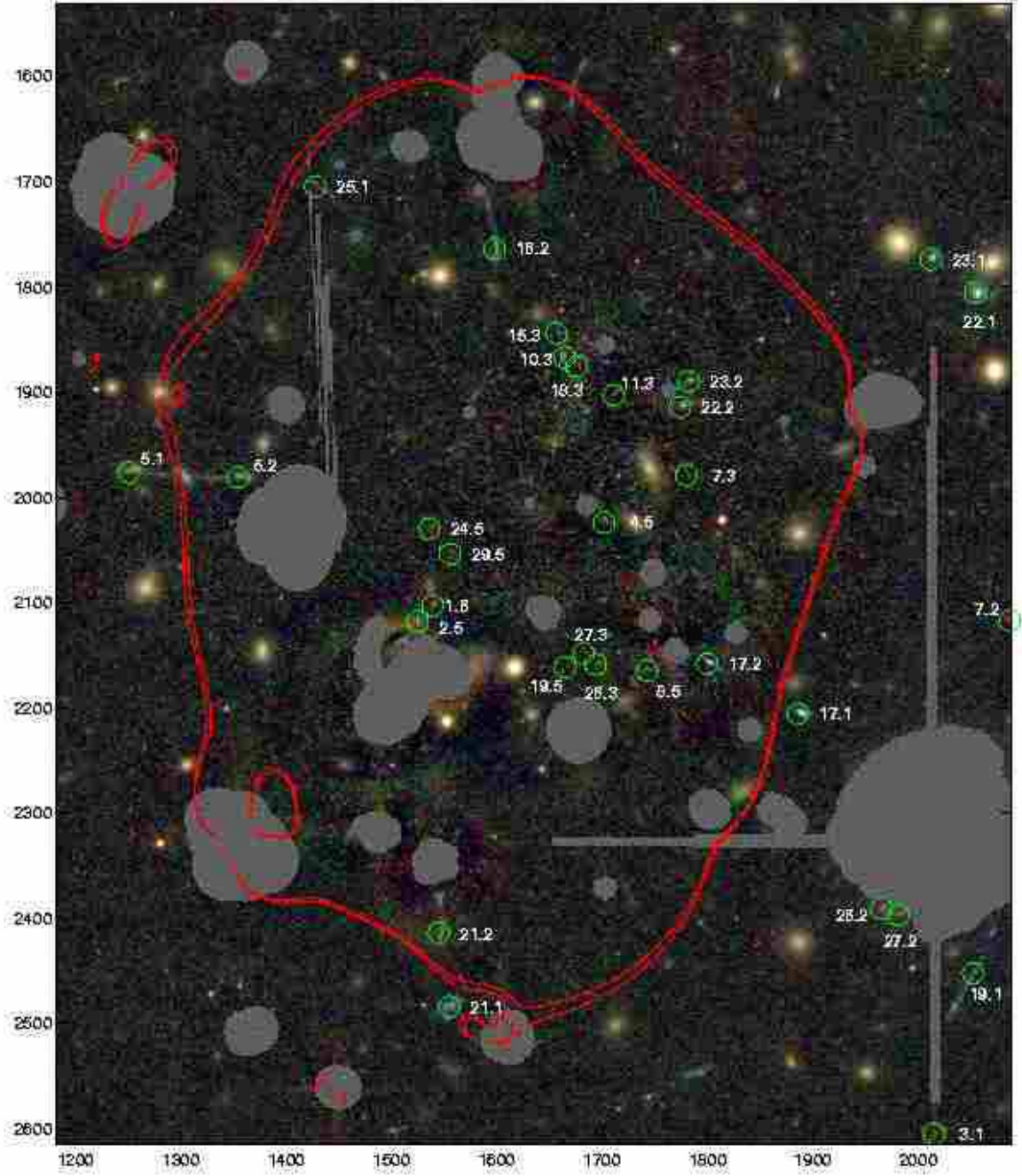}
\caption{ Blow-up of the central region with the bright elliptical
galaxies subtracted with the counter-images marked with their ID 
corresponding to Table 2. These images are either long and point 
radially or are small images interior to the radial critical curve, some of
which are demagnified and seen here for the first time in appreciable
numbers. Such central images are very helpful in constraining the
inner mass profile. }
 \end{figure}

\clearpage
\begin{figure}[ht]\label{lightmap}
\epsscale{0.95}
\plotone{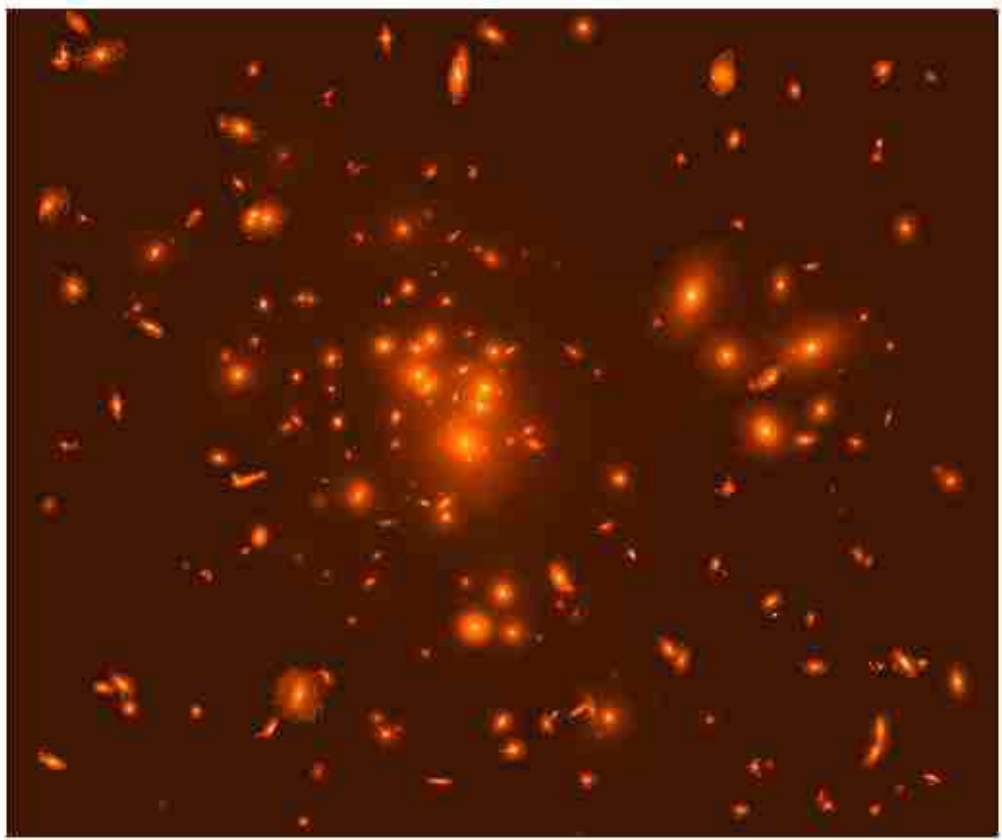}
\caption{ A 2-D map of the light generated by replacing the luminous
early-type cluster galaxies with model profiles to trace their light
to large radius free of sky noise.  Also added to this map are
luminous later type galaxies. These are directly cut out from the
observed image rather than replaced with model profiles, because
of their more complex structure. This 2D image is sued in calculating
M/L profiles.}

 \end{figure}

\clearpage
\begin{figure}[ht]\label{inital_mass}
\epsscale{0.95}
\plotone{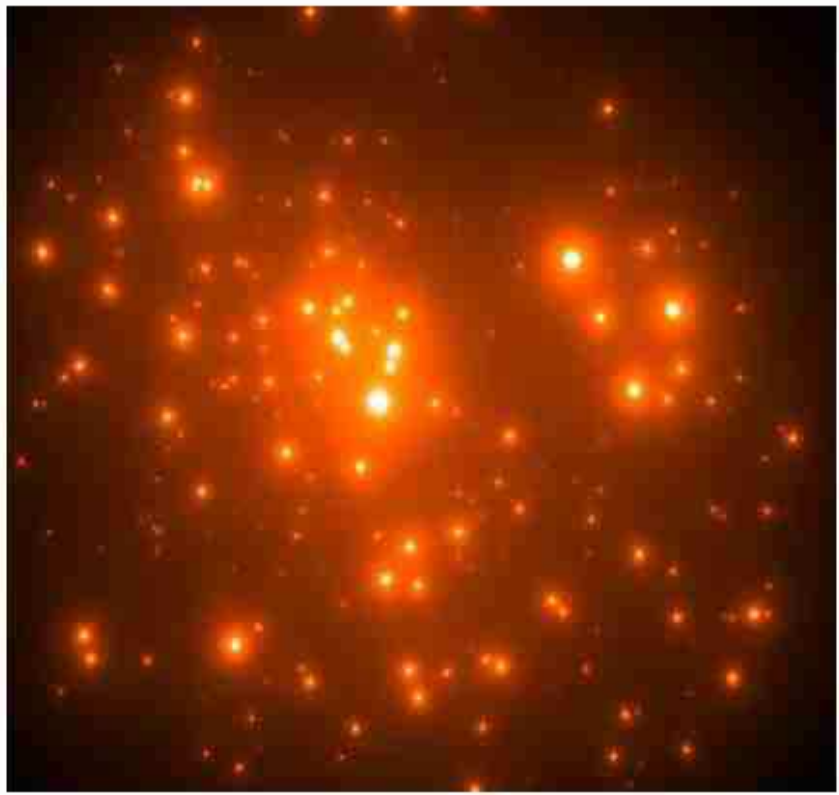}
\caption{ This is the starting point for our lens model. A plausible
projected surface density map is constructed by summing up
profiles assigned to all cluster sequence galaxies. Each galaxy is
modeled simply as an extended power-law density profile (in this
figure $q=1$) with a normalization proportional to luminosity. The
result is a surface density map which scales linearly with $M/L$, and
has an overall smooth appearance, punctuated by galaxy peaks.}

\end{figure}

\clearpage
\begin{figure}[ht]\label{smooth_mass}
\epsscale{0.95}
\plotone{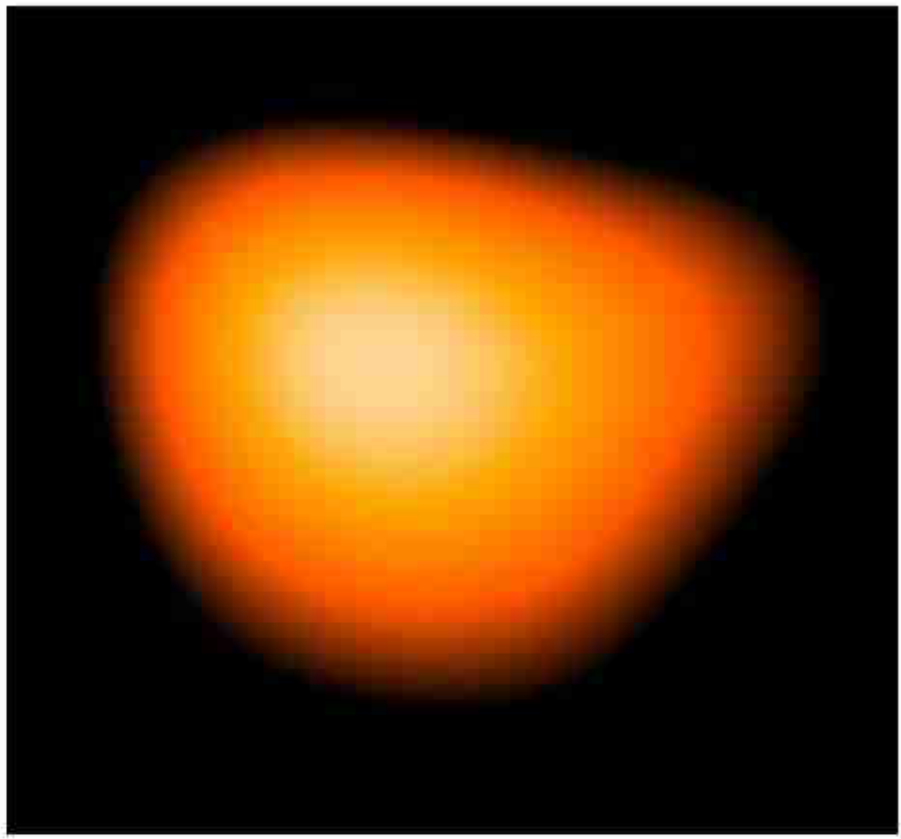}
\caption{ A low order cubic-spline fit to the surface mass density in
the previous figure produces a smooth component. After
calculating its corresponding deflection field, we will add low order
perturbations in order to produce a smooth deflection field
for fitting the multiple-image locations. The main departure from circularity
visible on the left follows a the most luminous sub-group of galaxies.}
\end{figure}

\clearpage
\begin{figure}[ht]\label{residual_mass}
\epsscale{0.95}
\plotone{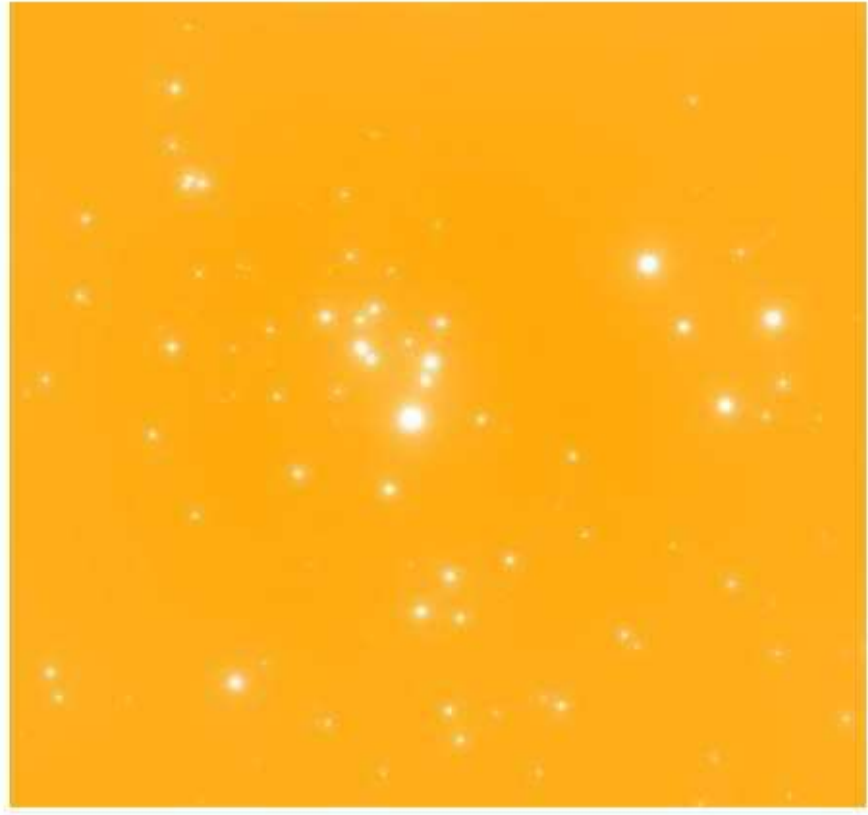}
\caption{ A ``lumpy'' residual map is formed by subtracting the previous smooth
component from the initial surface mass distribution. This map
highlights the galaxies, and we use it to crudely represent them as
concentrations of mass with the correct locations.  More accurate
descriptions of galaxies are not justified in the context of cluster
lensing as the galaxy contribution is very small and only important for
cases where a multiply lensed image falls near a massive galaxy. Our
modeling allows only the amplitude of deflection field corresponding
to this map to vary with respect to the smooth component; this
approximates varying M/L of the galaxy component relative to the
cluster component.}
\end{figure}



\clearpage
\begin{figure}[ht]\label{fk_compare}
\epsscale{0.95}
\plotone{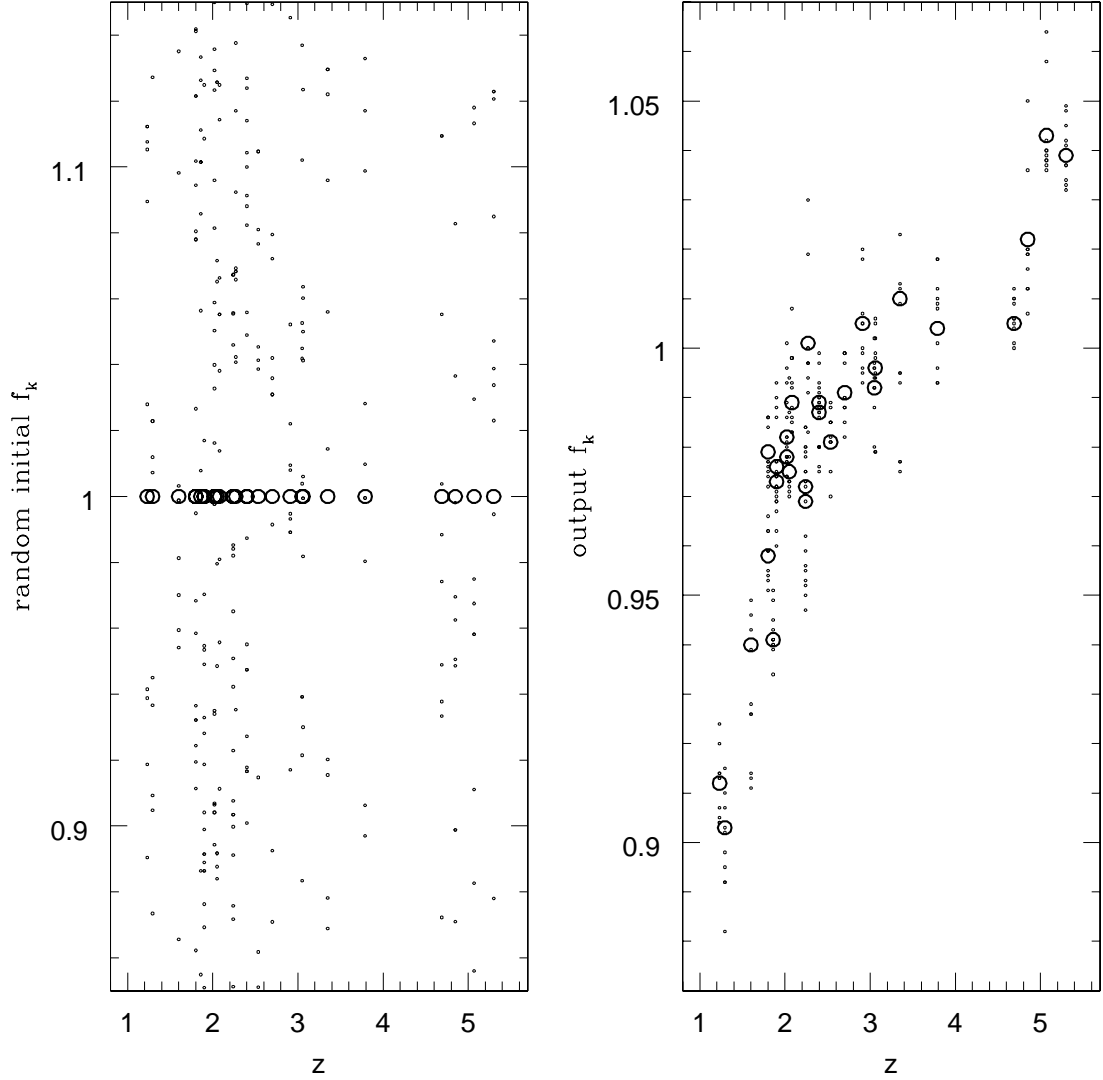}
\caption{ The left-hand panel shows the random values of the relative
distance ratio,$f_k$, which we input to test the modeling procedure.
$f_k$ is plotted against redshift, showing no inherent trend. On the
right are the output values of $f_k$ from the model, which are found
to follow a tight relation with redshift. A constant input value of
$f_k$ (open circles) is used in our modeling to avoid specifying many
additional free parameters, and can be seen to behave in the same way
with redshift as the random input. The relative deflection angles
scale in a way which is redshift-dependent with a small scatter in
$f_k$ that varies between sources slightly, depending on the
the location of the multiple images of each source. This redshift dependence
is compared to the expected behaviour of cosmological models (Section
13).  }
\end{figure}

\clearpage
\begin{figure}[ht]\label{omlam_compare}
\epsscale{0.75}
\plotone{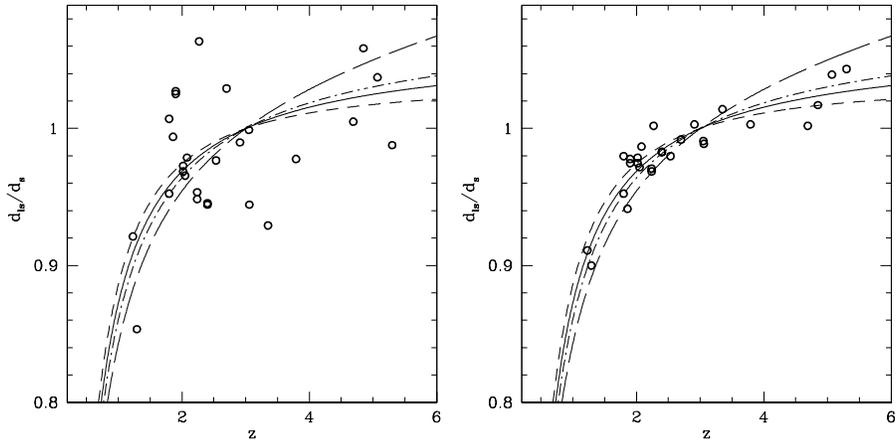}
\caption{ The relative distance ratios, $f_k$ for 28 multiply-lensed
 galaxies with redshift measurements or reliable photo-z's. The
 left-hand plot shows the calculated value for each set of images
 before the model is iterated using the input mass distribution, and
 the right-hand shows the marked improvement achieved when the
 iteration procedure is applied to minimise the overall error on the
 model image positions, described in section 8.3. The
 bend-angle is scaled to $z=3$ ($f_k \equiv 1$). The absolute
 bend-angles are not known independently as they are products of the
 relative distances and the normalization of the mass map. The three
 close curves cover the range of currently interesting cosmologies,
 which in order of increasing gradient are ($\Omega_m,
 \Omega_\lambda$) = (0.1, 0), (0.3, 0.7), (1, 0), and (0.1, 1.2), this
 final (upper) curve being discrepant.  This plot demonstrates that
 the bend-angles predicted by the model are fully consistent with the
 purely geometric effect expected for any reasonable cosmology.
 Higher redshift clusters must be examined in order to constrain the
 cosmological parameters with more accuracy. Note that two very close
 pairs of images (sources 13 \& 14) are excluded from this plot as
 their predicted locations are too sensitive to the location of the
 critical curve which in both cases passes between producing an
 unreliable distance estimate. The remarkable point here is that the
 redshift information is {\it not} used in generating the best fit
 model - only the angular information on the image locations, and
 hence the sensible trend found here of bend-angle vs. redshift adds
 considerable confidence regarding the model.  }
\end{figure}

\clearpage
\newpage
\begin{figure}[ht]\label{grid_best}
\epsscale{0.95}
\plotone{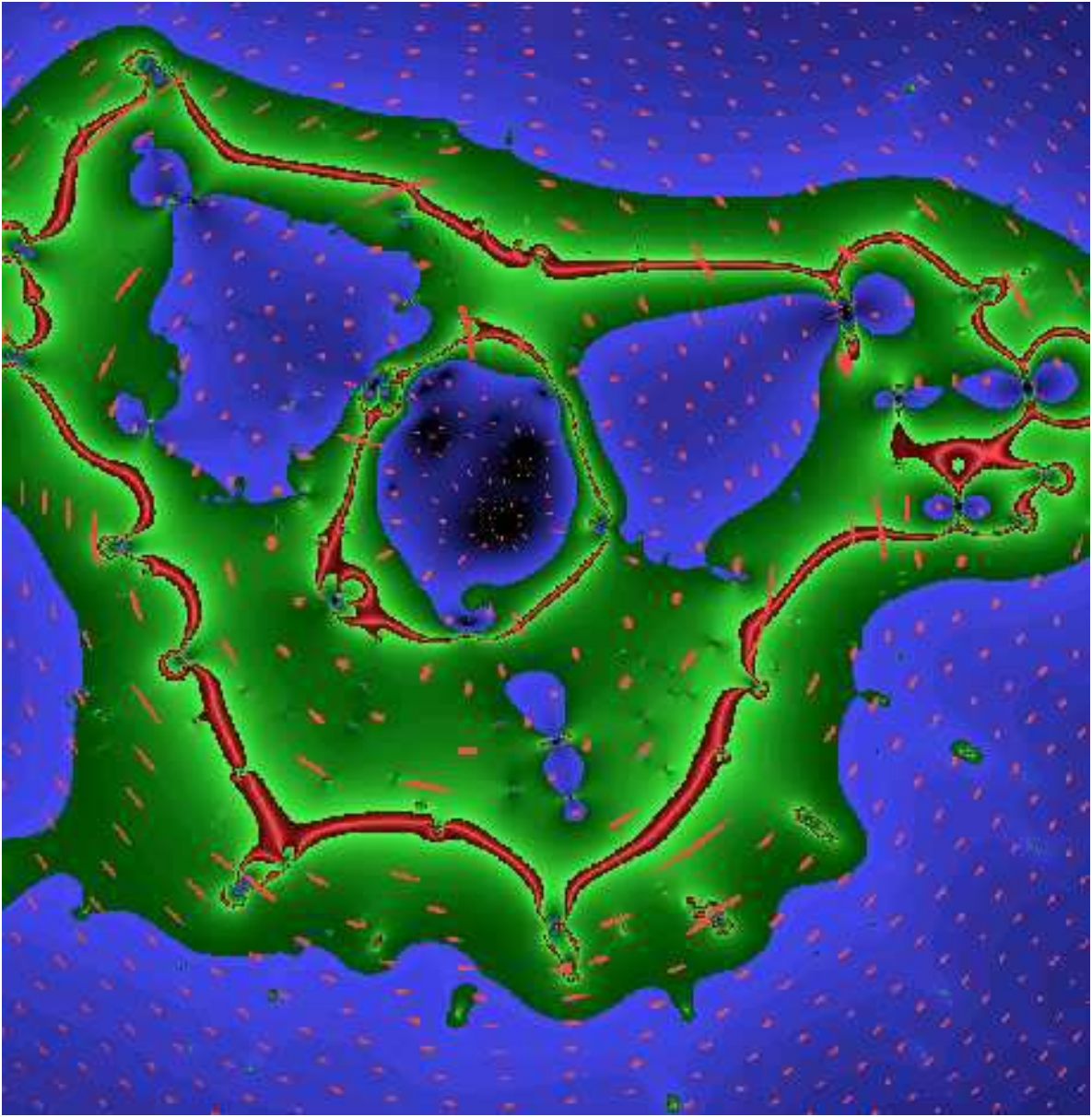}
\caption{ This figure illustrates the way images will appear to be
distorted and magnified by our best fitting lens model. We generate
the images from a grid of small circular sources using our best fit
lens to re-map them onto the image plane. Overlayed on this plot is
the magnification map, which is colour-coded to indicate the location
of the critical curves and the relative strength of the magnification
(for comparison with the next two similar figures). Note how a well
defined radial critical curve appears here, highlighted by spoke-like
images which bisect it, with much smaller less magnified images
inside the radial critical curve.}
 \end{figure}

\clearpage
\begin{figure}[ht]\label{grid_steep}
\epsscale{0.95}
\plotone{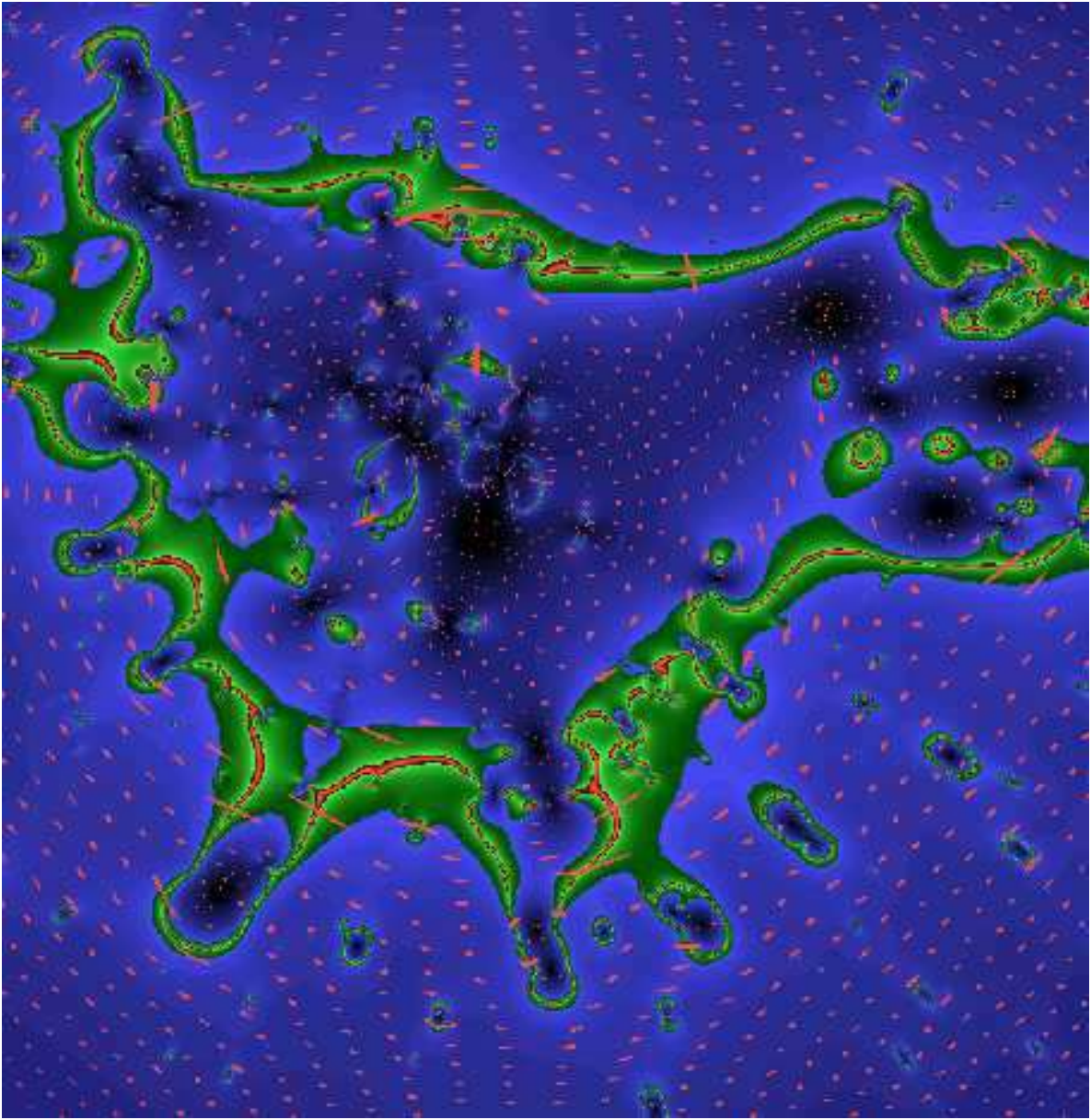}
\caption{ Same as the previous figure, but for a steeper mass profile,
corresponding to the right-hand column of Figure 31. Here the
magnifications of images interior to the tangential critical curve are
relatively small and no distinct radial critical curve is generated,
unlike the observations. The generally higher surface density of image
reflects the generally small magnification by a steeper mass profile
(for a fixed Einstein radius.)}
 \end{figure}

 \clearpage
\begin{figure}[ht]\label{grid_shallow}
\epsscale{0.95}
\plotone{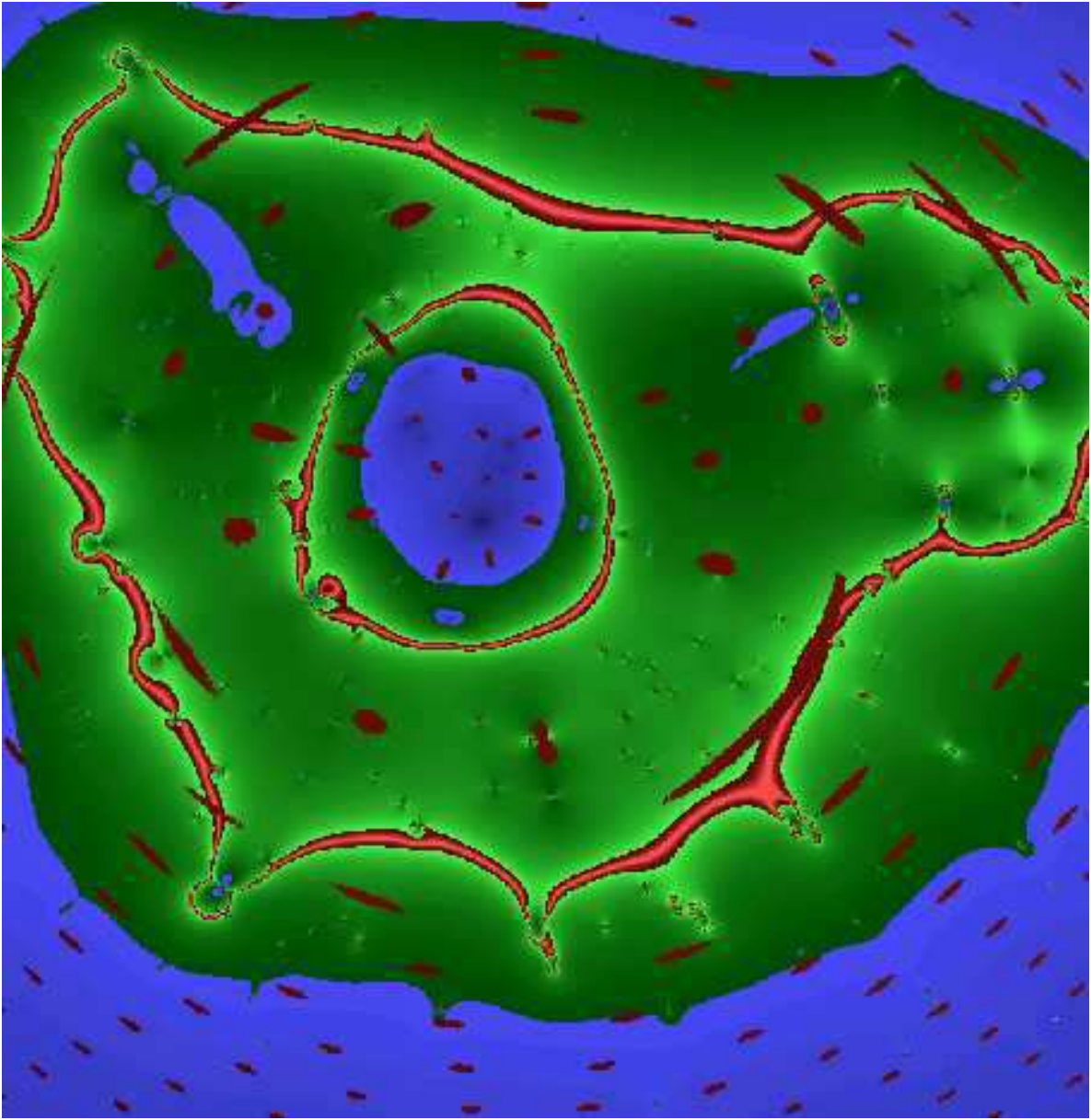}
\caption{ The profile here is relatively shallow, corresponding to the
profile in the left-hand column of Figure 31.  A
larger radial critical curve forms and the relative magnification of
images interior to the radial critical curve are larger than for
steeper profiles, and the arcs are relatively longer.}
 \end{figure}

 \clearpage
\begin{figure}[ht]\label{coloured_lens}
\epsscale{0.95}
\plotone{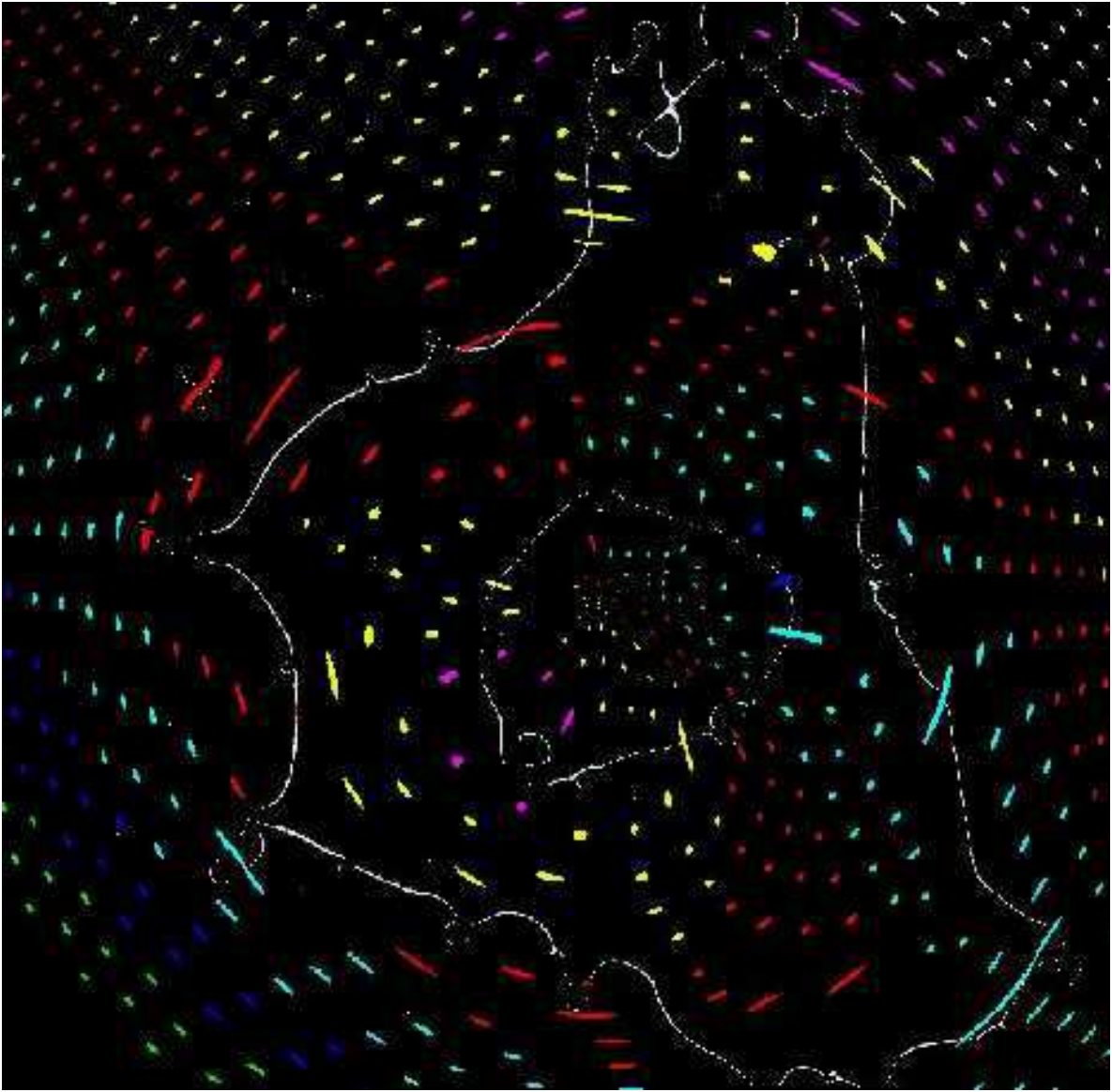}
\caption{ This figure shows the images generated by the best fit lens
model of Figure \ref{omlam_compare}, where the sources are colour
coded in diagonal stripes across the image, to demonstrate where
counter images of the same source may be found. For example, the blue
and purple bands of sources form counter images on the opposite side
of the radial critical curve, and red coded images form in a ring
around the lens including the demagnified region in the center - these
correspond to the many 5-image systems identified in Figure 1. Counter
images are not expected to form for the outermost images in the field of view.}
\end{figure}

\clearpage
\begin{figure}[ht]\label{dan_before}
\epsscale{0.95}
\plotone{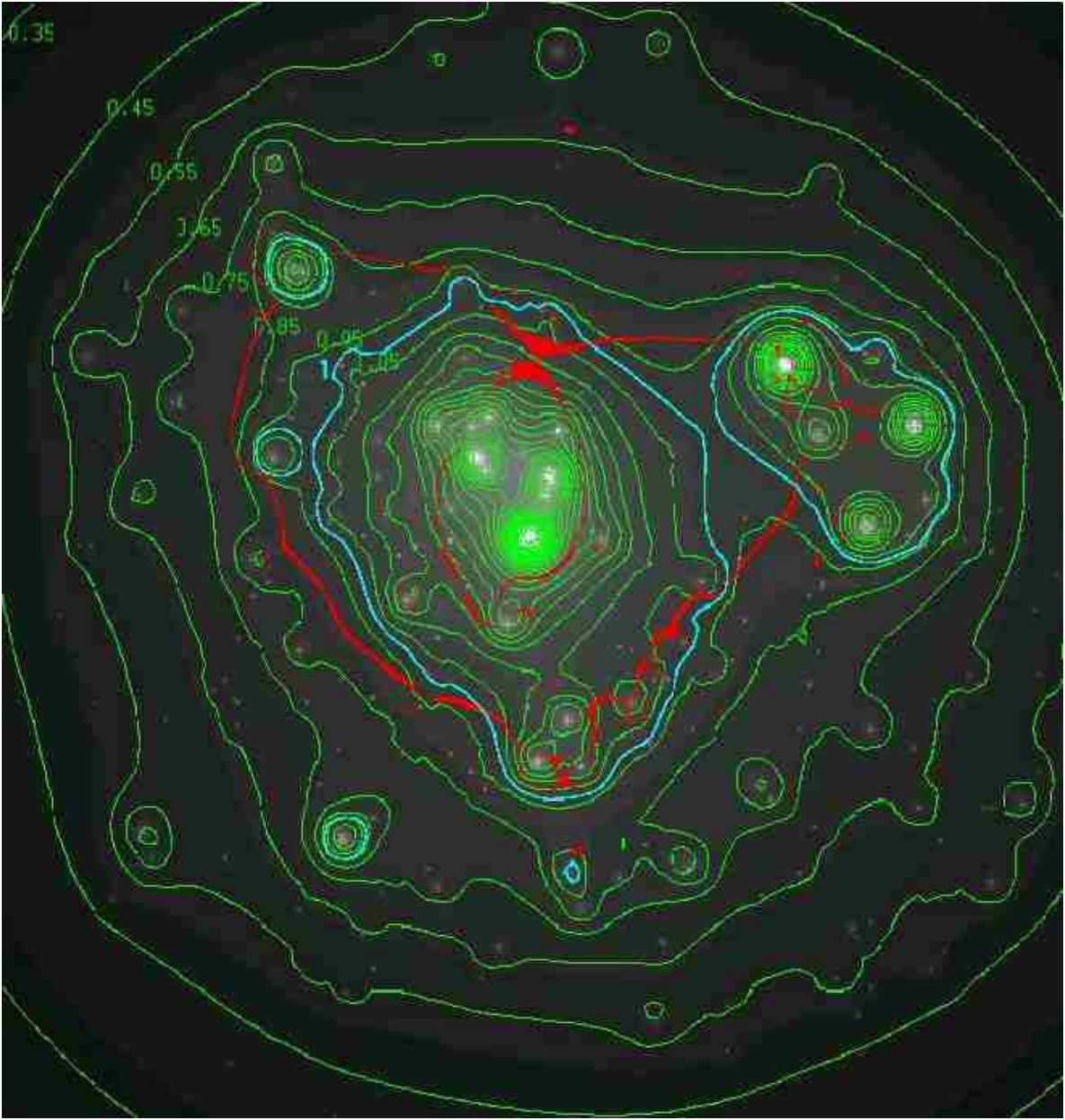}
\caption{ This is the mass distribution before iteration. The double
peaked structure reflects the distribution of luminous cluster
galaxies used as the starting point for the mass model. The blue
contour corresponding to the critical surface density, required to
generate multiple images.  Other mass contours (green) are labeled in
units of this critical density, and the critical curve is overlayed in
red.}

 \end{figure}

\clearpage
\begin{figure}[ht]\label{dan_after}
\epsscale{0.95}
\plotone{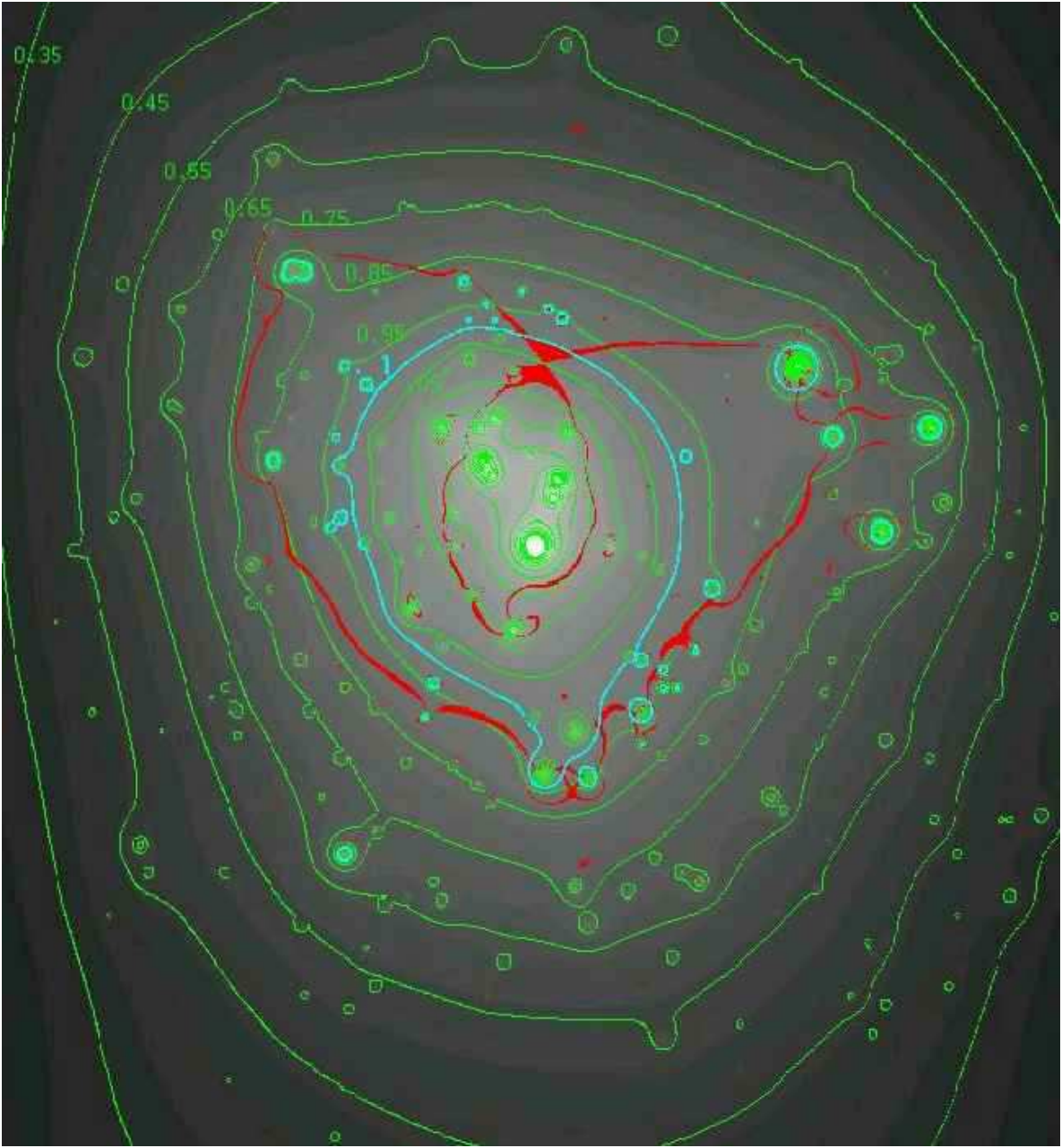}
\caption{ This is the best fitting mass distribution after iteration.
Our model is dominated by the main group of central luminous
ellipticals, yielding a fairly round mass distribution (rounder than
the previous figure) with one main peak centered near (6\arcs ~from)
the most luminous cD galaxy.  The main subgroup is found to have much
less mass in our model.  However, the tangential critical curve (red)
is greatly affected by the main subgroup, as image distortions are not
linearly related to mass in the strong lensing region but are rather
sensitive to perturbations near the Einstein ring. Note that the
critical surface mass density contour (blue) is now centered on the
main group of galaxies and does not form two ``islands'' as is the
case in the previous figure, for the starting position.  }

\end{figure}

\clearpage
\begin{figure}[ht]\label{dan_grzm4}
\epsscale{0.95}
\plotone{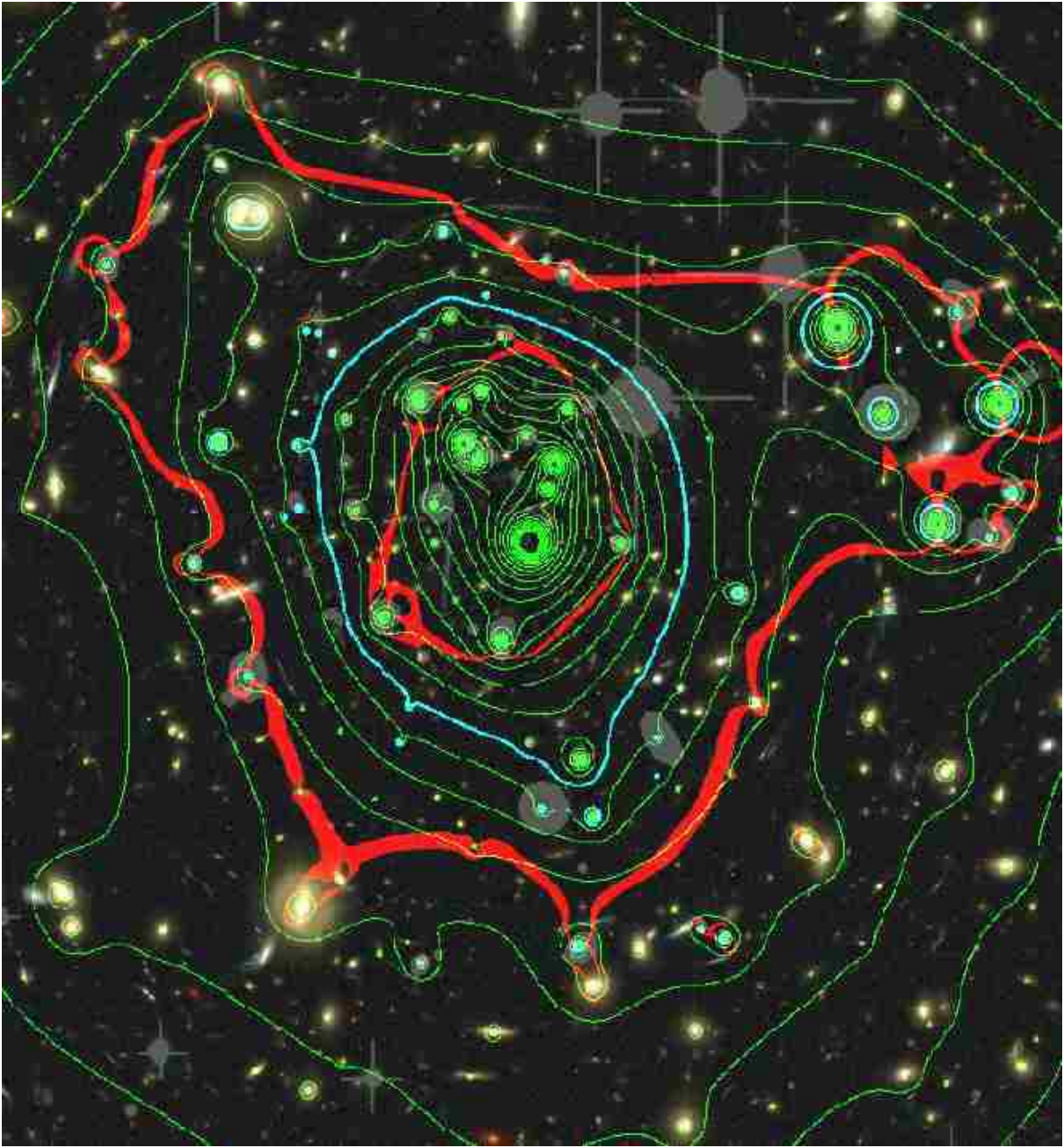}
\caption{ The resulting mass distribution and critical curves (scaled
to $z=3$ ($f_k\equiv 1$)) are plotted on the colour image of the arcs -
with the cluster galaxies removed, showing the relation between the
shape of the lensed images and the model. Note how the giant arc is
exactly bisected by the critical curve forming two large mirror images
(images 8.1 \& 8.2).}
 \end{figure}

\clearpage
\begin{figure}[ht]\label{m1}
\epsscale{0.95}
\plotone{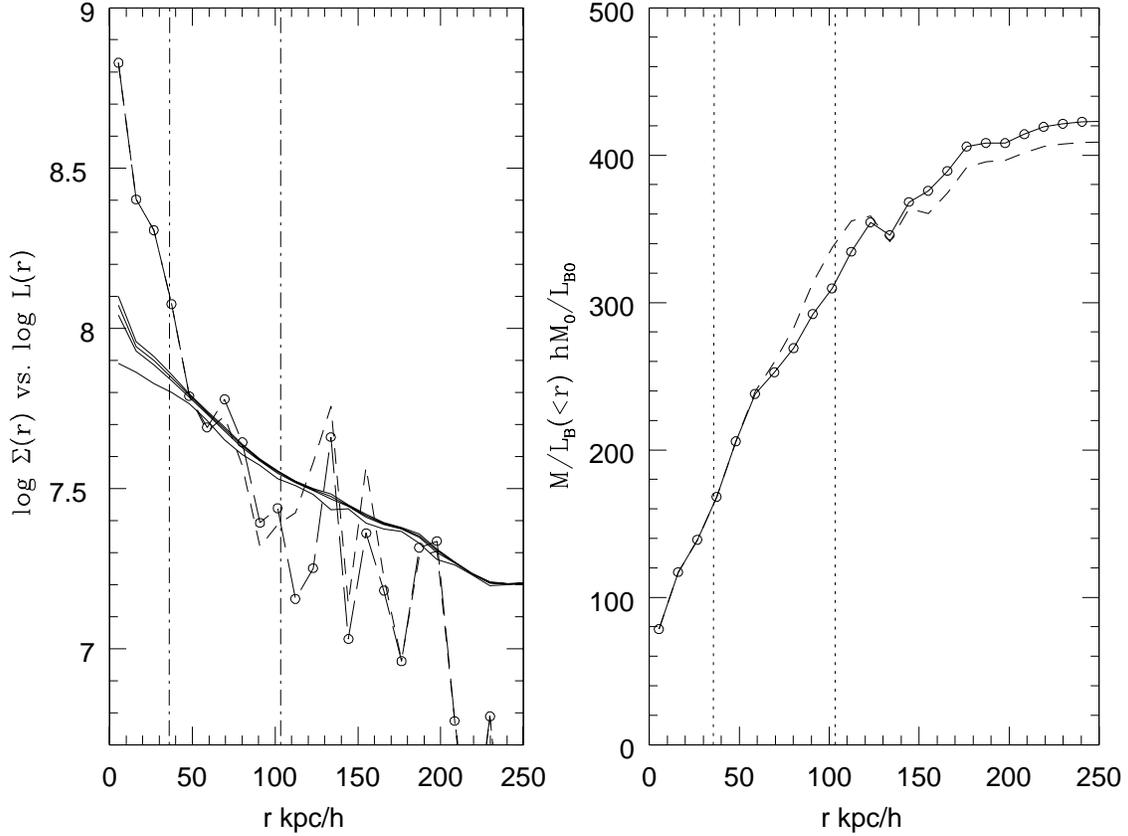}
\caption{ The mass and light profiles are compared in the left-hand
panel for our best fit model. The light profile (dashed curve with
circular points attached) is more concentrated than the mass (solid
lines). Four curves (solid) are shown for the mass distribution, which
differ in the amount of mass assigned to the galaxies and subtracted
from the total. This subtraction is performed in 2D using the light
map described in the text. The upper mass curve is the total mass, and
the three slightly lower curves have mass subtracted by scaling the
radial light curve (dished line) by the ratio $M/L= 5,10 \&
30h(M/L_B)_{\odot}$. This has negligible effect, even in the center
where the light peaks, due to the very high M/L of the cluster as a
whole. Excluding the light of the subgroup from the overall light
profile is shown too (small dashes) and this has a small effect on the
light profile, in particular it raises the overall $M/L$ (right hand
panel - dashed curve).}

 \end{figure}

\clearpage
\begin{figure}[ht]\label{m1_compare}
\epsscale{0.95}
\plotone{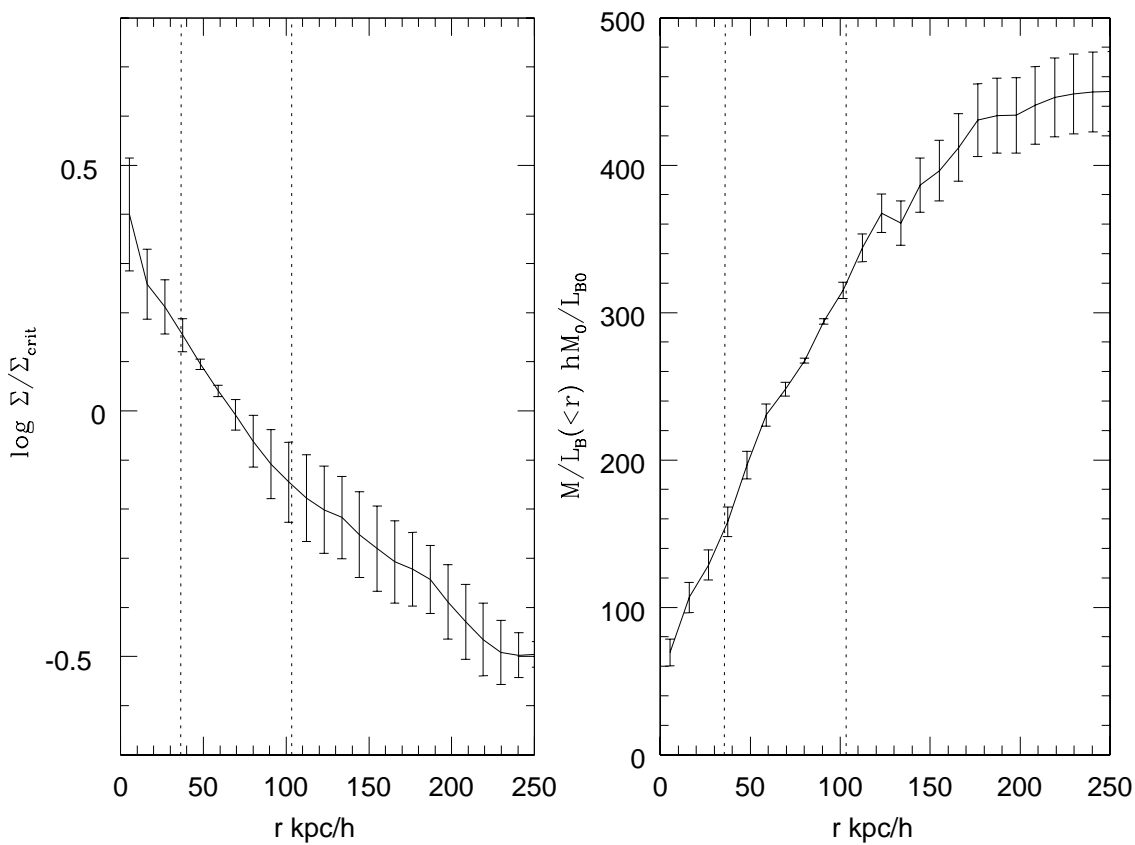}
\caption{ This shows the range of acceptable profiles corresponding to
the 1$\sigma$ limits on fitting the image locations. The corresponding
range of $M/L$ profiles is plotted in the right-hand panels - the
steeper slope has the lower overall M/L at large radius. Note that
profiles of differing slopes must cross within the Einstein radius, as
can be seen in the left-hand panel, as the total mass is fixed by
definition for a given observed Einstein radius, and is independent
of the mass profile.}

\end{figure}

\begin{figure}[ht]\label{one_over_mu}
\epsscale{0.95}
\plotone{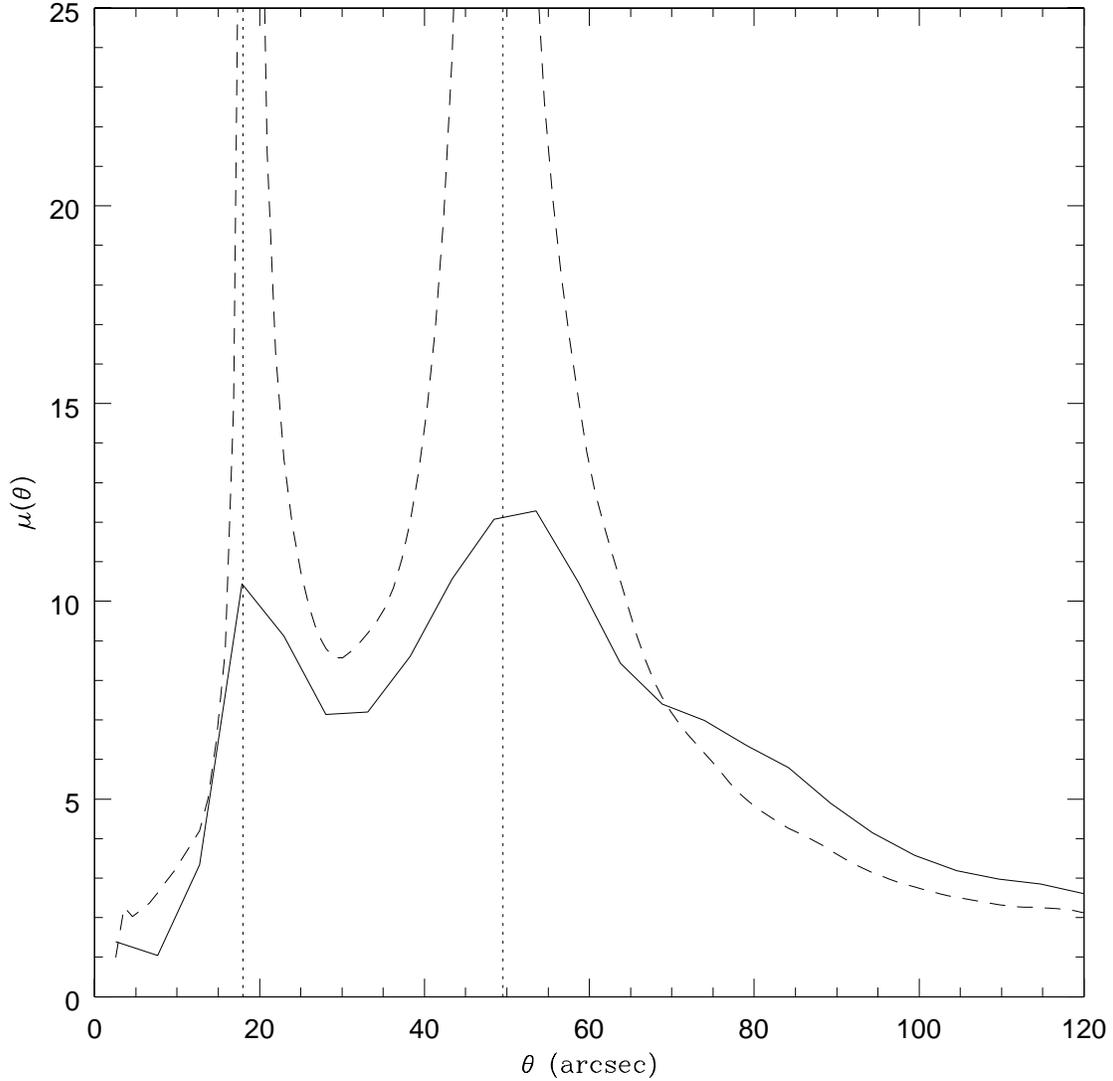}
\caption{ The radial magnification profile for the data determined
from the 2D map diverges over a wide range of radius because the
tangential critical curve is not circular but has excursions around
sub-groups. So to help better define critical curves we average
$1/\mu(r)$, and plot the inverse of this - solid curve and now the
the radial and tangential critical radii are well defined. If instead we calculate the
magnification profile form the radially average mass distribution
(using equation 16) then the profile is is better behaved, diverging
at only two points - dashed curve, and the radii of these divergent points
corresponds well the the maxima of the solid curve calculated from the
2D map. In the next series of figures we make radial comparisons of
theoretical predictions with the latter curve, derived by using the
radially binned mass maps.  }
 \end{figure}

\clearpage
\begin{figure}[ht]\label{mu_best}
\epsscale{0.75}
\plotone{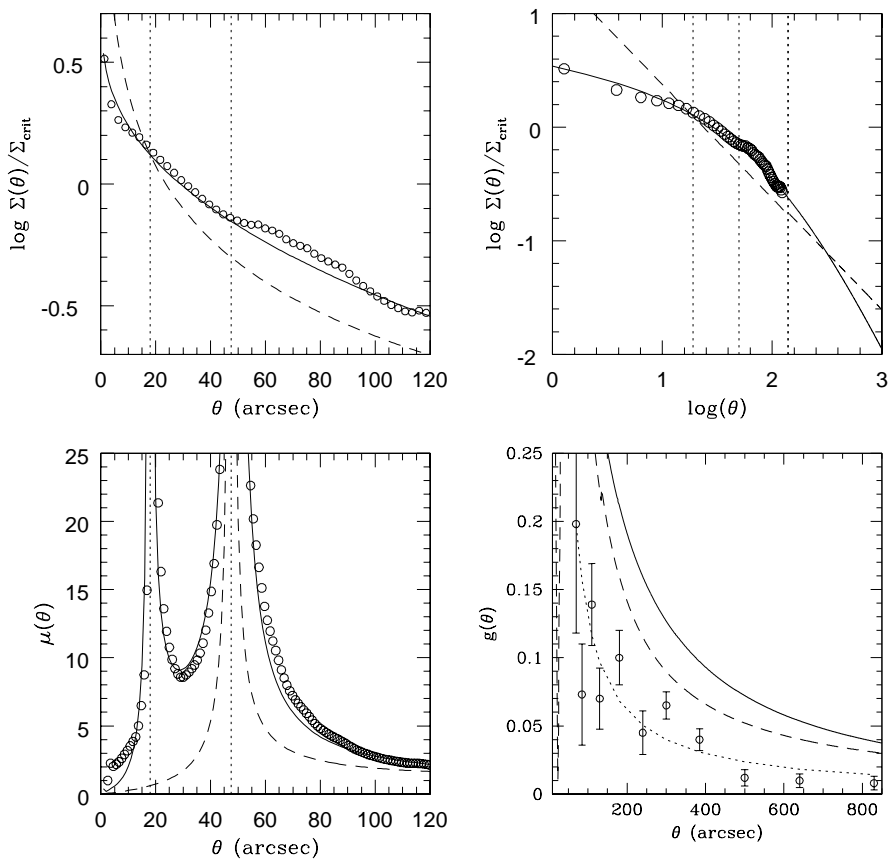}
\caption{ Comparison of the best fit model profile (upper left panel)
with the NFW prediction of CDM - shown is $r_s=310kpc/h$, $C=8.2$. The
dashed line is the singular isothermal case normalised to the observed
Einstein radius for comparison. The scale is linear in angle, to help
show the accuracy of the fit. The full logarithmic form is shown top
right compared with the full NFW profile demonstrating that our
lensing data covers only the shallow region of the profile interior to
$r<r_s$. The bottom left panel is the magnification derived by our
best-fitting model, compared with the NFW prediction, demonstrating
how well the locations of the critical curves are reproduced at radii
of $\sim17$\arcs ~and $\sim50$\arcs ~by the NFW model, and also that
the central magnification is accurately followed. There is a slight
excess outside the tangential critical radius which results from the
elongated shape of the tangential critical curve, enhancing the
azimuthally averaged magnification in circular bins. A much smaller
excess can be seen in the radial mass profile, which is less
pronounced because the mass distribution is inherently rounder than
the path of the tangential critical curve (see Figure 14
for a 2D comparison of mass contour with the path of the tangential
critical curve).  The weak lensing measurements of Clowe \&
Schneider are shown (bottom right panel) versus the ellipticity
expected for the best fitting NFW profile. The weak lensing data imply
a much smaller tangential critical radius than observed (20\arcs ~vs
50\arcs) indicating some problem. }
 \end{figure}

\begin{figure}[ht]\label{mu_toosmall}
\epsscale{0.95}
\plotone{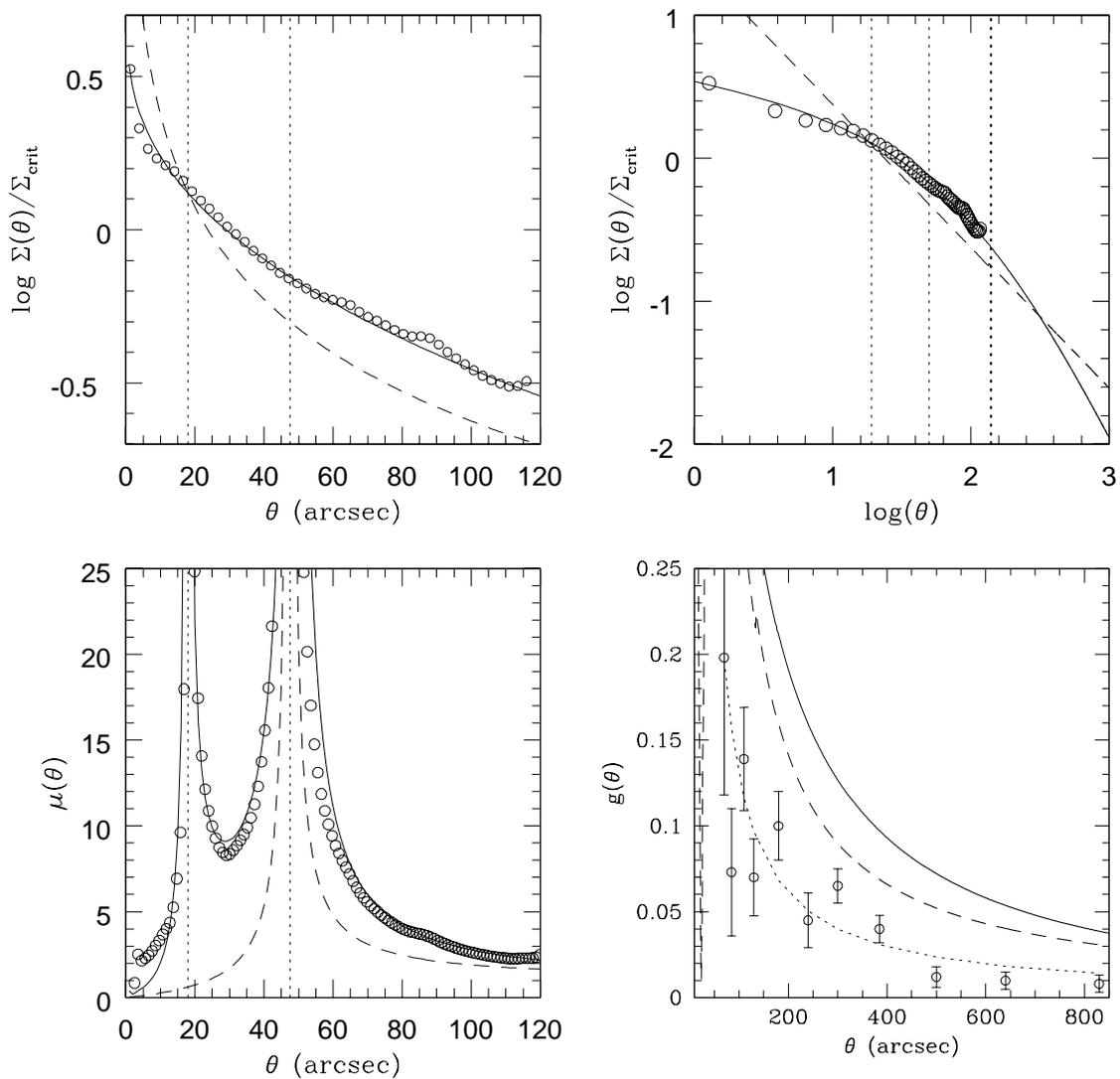}
\caption{This is the same as the previous figure but with
the main subgroup excluded (see text) when making all azimuthally 
averaged profiles. The fit to the NFW profile is very good, though the
concentration parameter, $C_{vir}=8.2$, is somewhat larger than 
predicted for cluster mass haloes.}
 \end{figure}

 \clearpage
\begin{figure}[ht]\label{mu_p6}
\epsscale{0.95}
\plotone{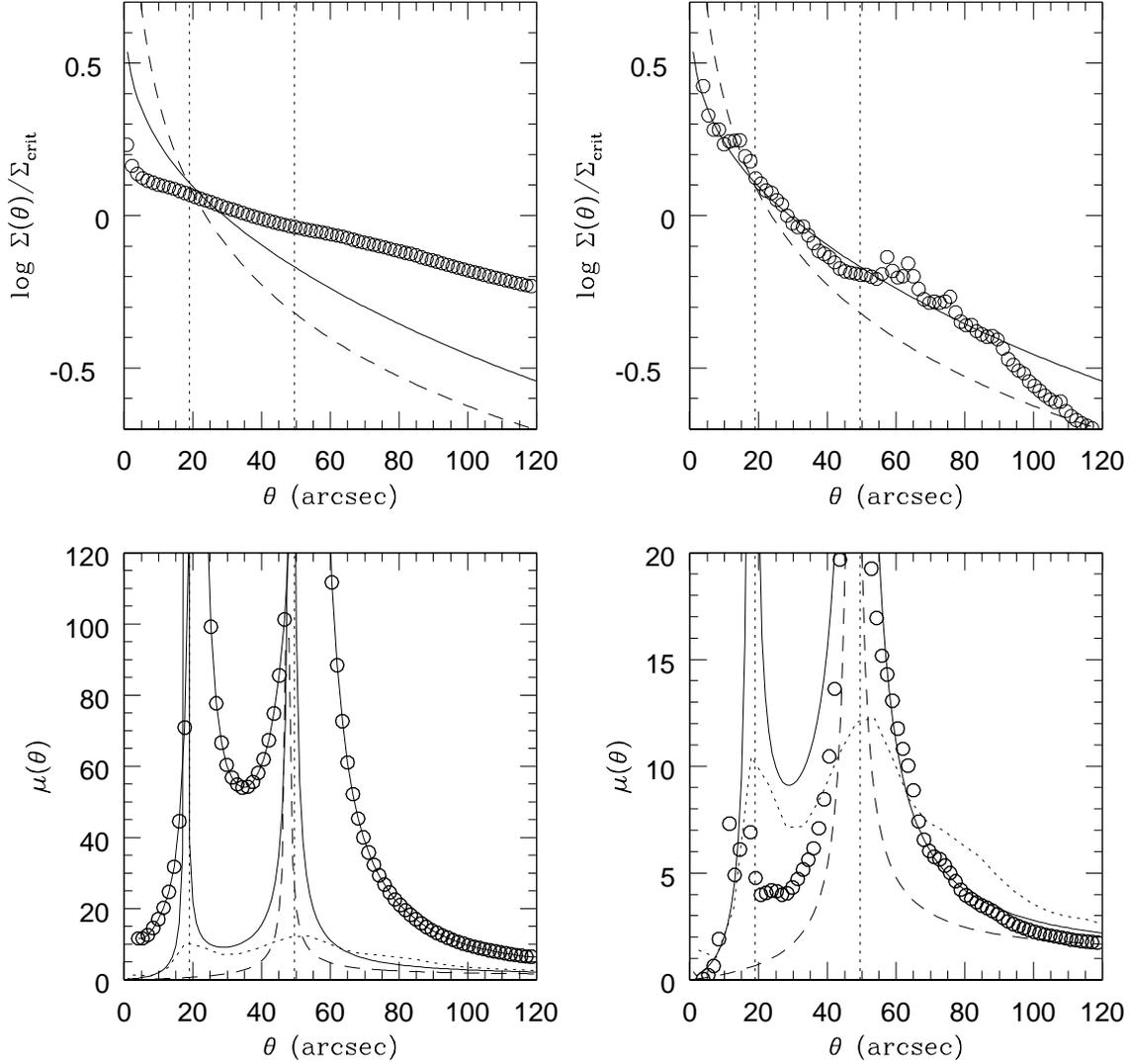}
\caption{ The left-hand panels show a shallow surface density profile
and the magnification for a shallow profile case corresponding to 2D
map of figure 16, and which poorly fits the multiple image
positions. The radial critical curve is too large and the
magnifications are enormous. The right-hand panels show a profile which
is steep to match well the multiple image locations, and can be seen
to have only a very small radial critical curve. These profiles
correspond to the 2D map shown in figure 15.  }
\end{figure}

 \clearpage
\begin{figure}[ht]\label{notready.eps}
\epsscale{0.95}
\plotone{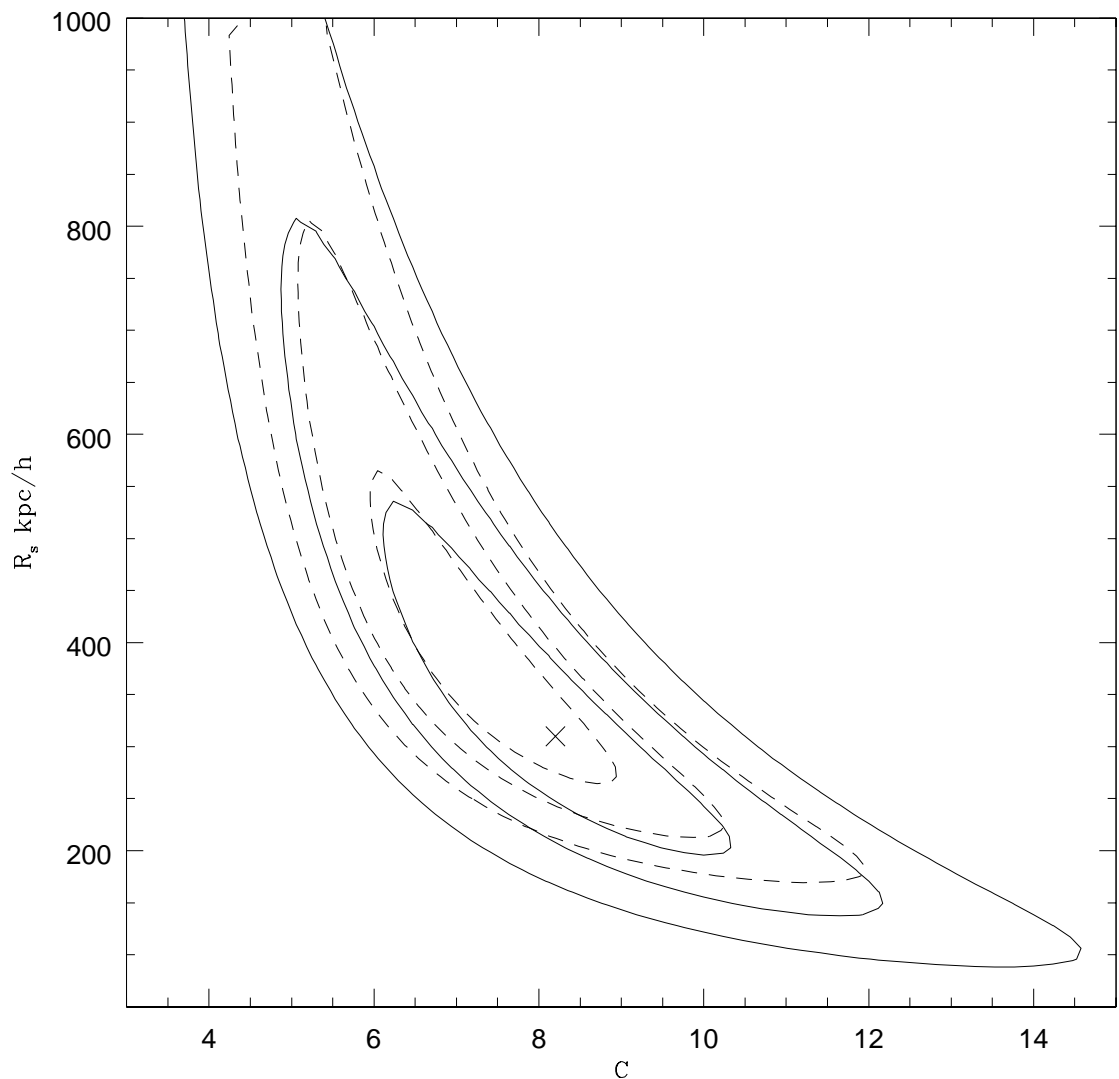}
\caption{ Chi-squared contours of the NFW profile fit for the
characteristic radius, $r_s$ and the concentration parameter,
$C=r_{vir}/r_s$. The best fit value
is marked at $C=8.2$ and $r_s=310kpc/h$. The contours correspond to 1, 2
\&3 $\sigma$ and are calculated using the dispersion about the best fit
model, shown in Figure 22. The dashed set of contours show the fit
when the relative magnifications of the lensed images are
included.}

\end{figure}

 \clearpage
\begin{figure}[ht]\label{mu_core_toosteep}
\epsscale{0.95}
\plotone{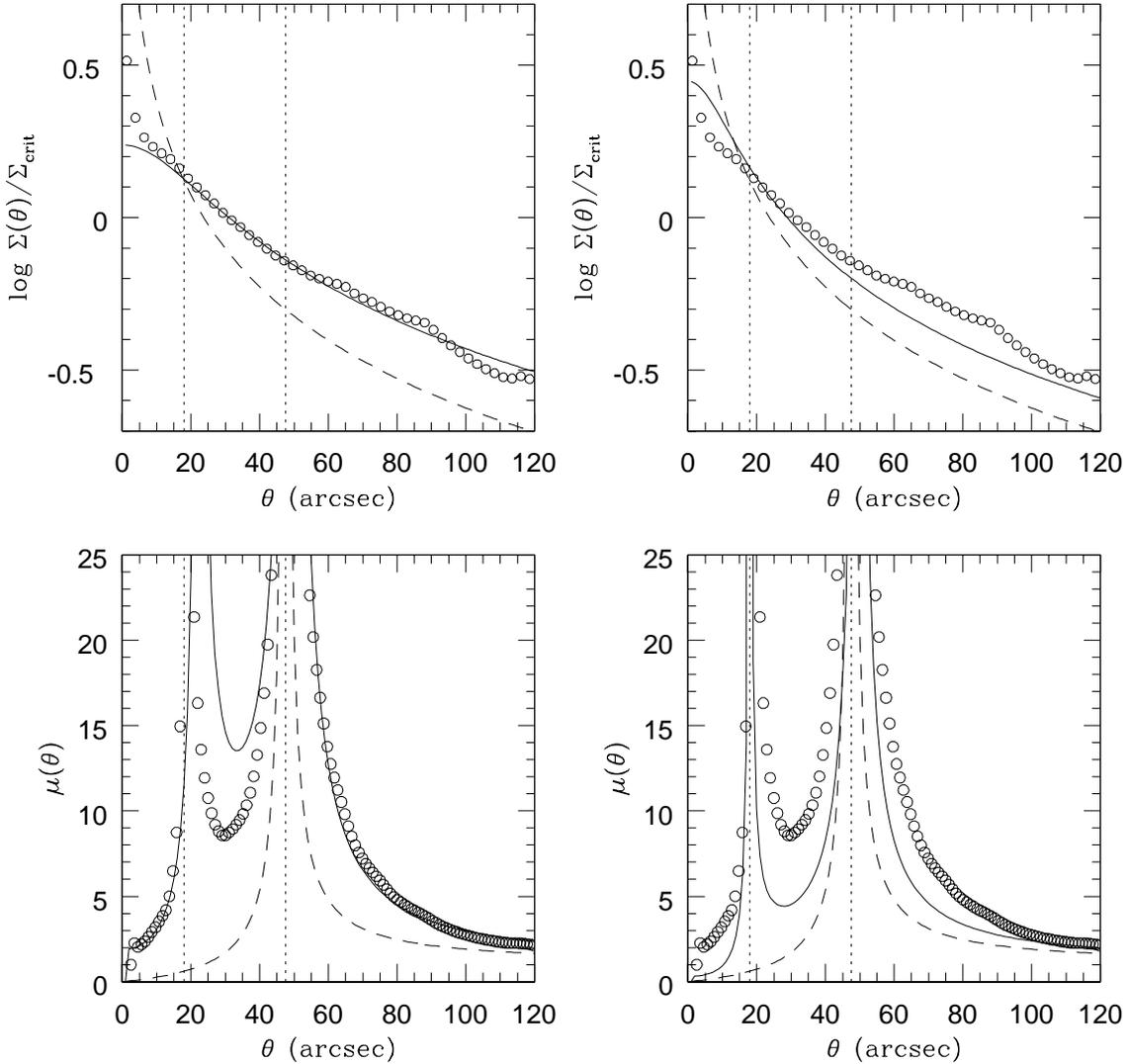}
\caption{ The solid curves are for a softened isothermal profile,
compared with the best fit solution for the mass profile and
magnification profile. The upper left-hand box shows that the mass
profile is well fitted by the model profile with a core radius of
22\arcs. However, the corresponding radius of the radial critical
curve significantly exceeds the observed radius - lower left box. If
instead the core radius of the softened isothermal model is reduced to
11\arcs ~to reproduce the observed radius of the radial critical curve - lower
right box - then the slope of the projected profile is too steep to
match the lens model - upper right box. }
 \end{figure}

\clearpage
\begin{figure}[ht]\label{comapare1}
\epsscale{0.95}
\plotone{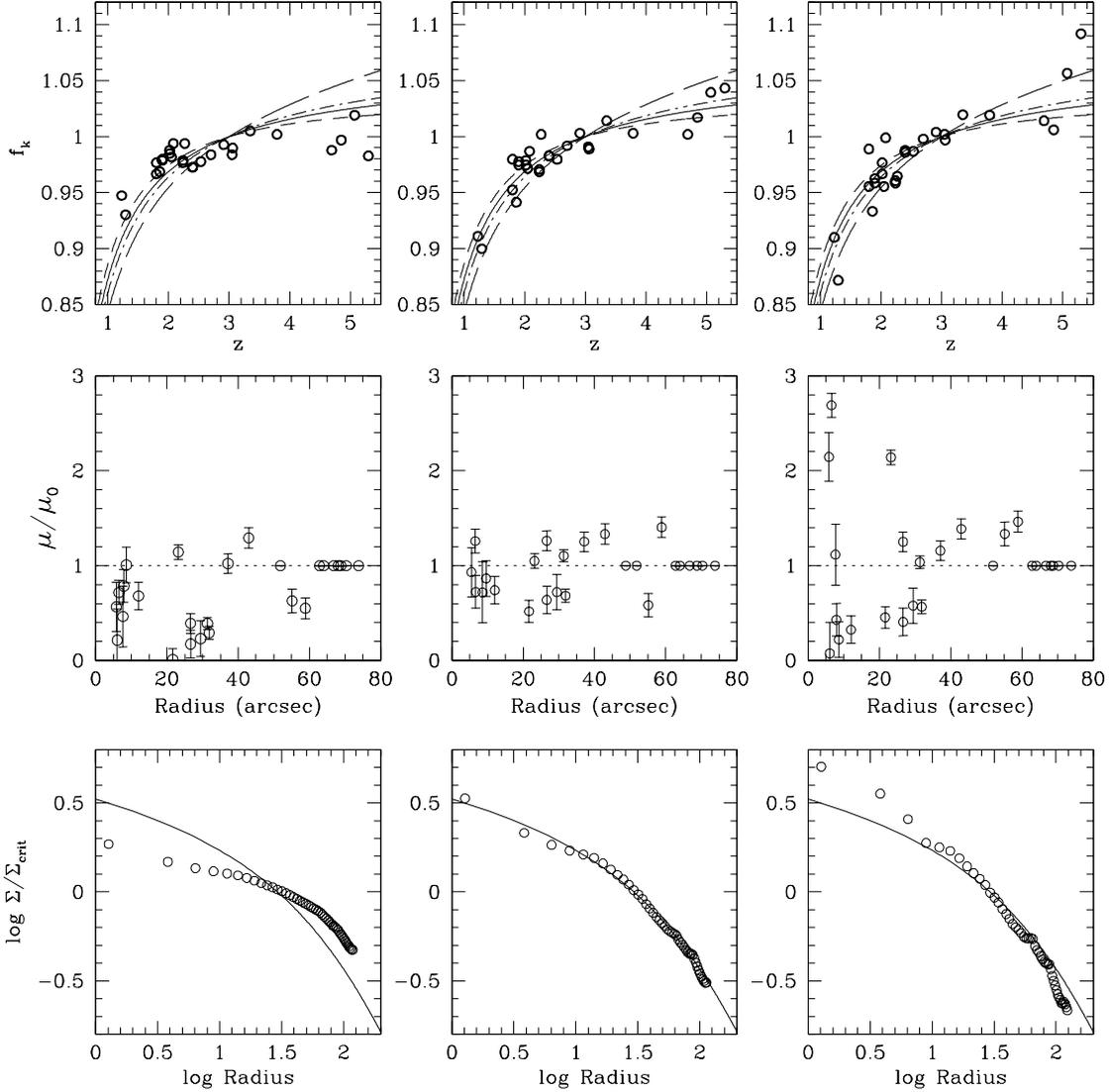}
\caption{ Three model outputs are shown for differing mass profiles -
from shallow to steep (bottom left to right panels) corresponding to
figures 14,15\&16. The upper panel (left to right) shows the
corresponding values of the relative distance ratio's ,$f_k$, for
these three choices of profile. The best fit to the cosmological
curvature is for the central case, $p\simeq -0.55$, where the relative
bend angles follow smoothly the expected cosmological behaviour
and are consistent with a flat universe (see also figure 13 right-hand
panel). The flatter profile generates a shallower trend with redshift
and the steeper profile does the opposite - the derived relation is
steeper. The central panel shows the relative image magnifications,
which again favour the model which best-fits the multiple image positions,
(center) with minimal dispersion and a mean value of unity.
}
\end{figure}

\clearpage
\begin{figure}[ht]\label{comapare2}
\epsscale{0.85}
\plotone{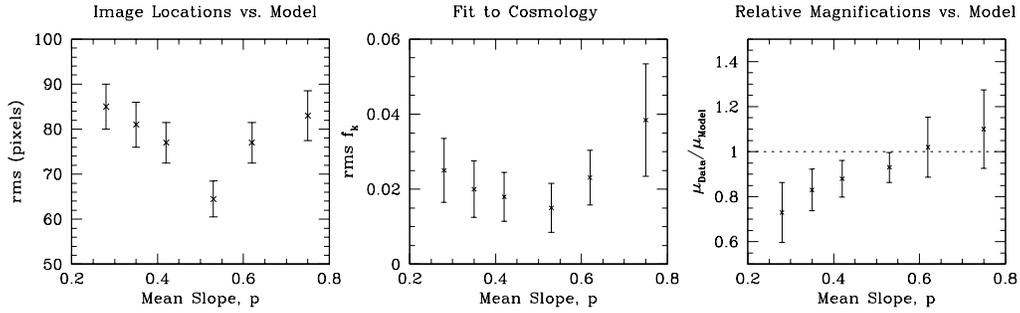}
\caption{Here we compare errors for differing output model profiles,
which we charactersise for simplicity by the mean output slope,
$p=d\log\Sigma/d\log r$ average over the observed range of
radius. The left-hand plot shows the difference between the model image
positions vs. the observed image positions, as a function of the
profile slope. There is clearly a pronounced minimum for $p \sim 0.55$
with an average positional error of 3\arcs (this is the model we have
referred to throughout as our best fit). The central panel shows the
error on fit to the standard $\Omega_m=0.3+\Omega_\lambda=0.7$
cosmology with a similar minimum using the predicted photo-z's and
relative distance ratios, $f_k$ (see upper panel of previous
figure). The right-hand panel is the dispersion of the relative
magnifications about the model (see middle panel of previous
figure). The data is most similar to the model for $p \sim 0.5-0.6$,
where it has a value consistent with unity and minimal
dispersion. These three estimates are largely independent and are in
good agreement. Since only the image locations were used in deriving
the model fits - the photo-z's used in the central plot and the fluxes
used in the right hand plot provide independent information for
constraining the model. The excellent agreement adds considerable
weight to our approach to modeling.
}
 \end{figure}

\clearpage
\begin{figure}[ht]\label{teague}
\epsscale{0.95} 
\plotone{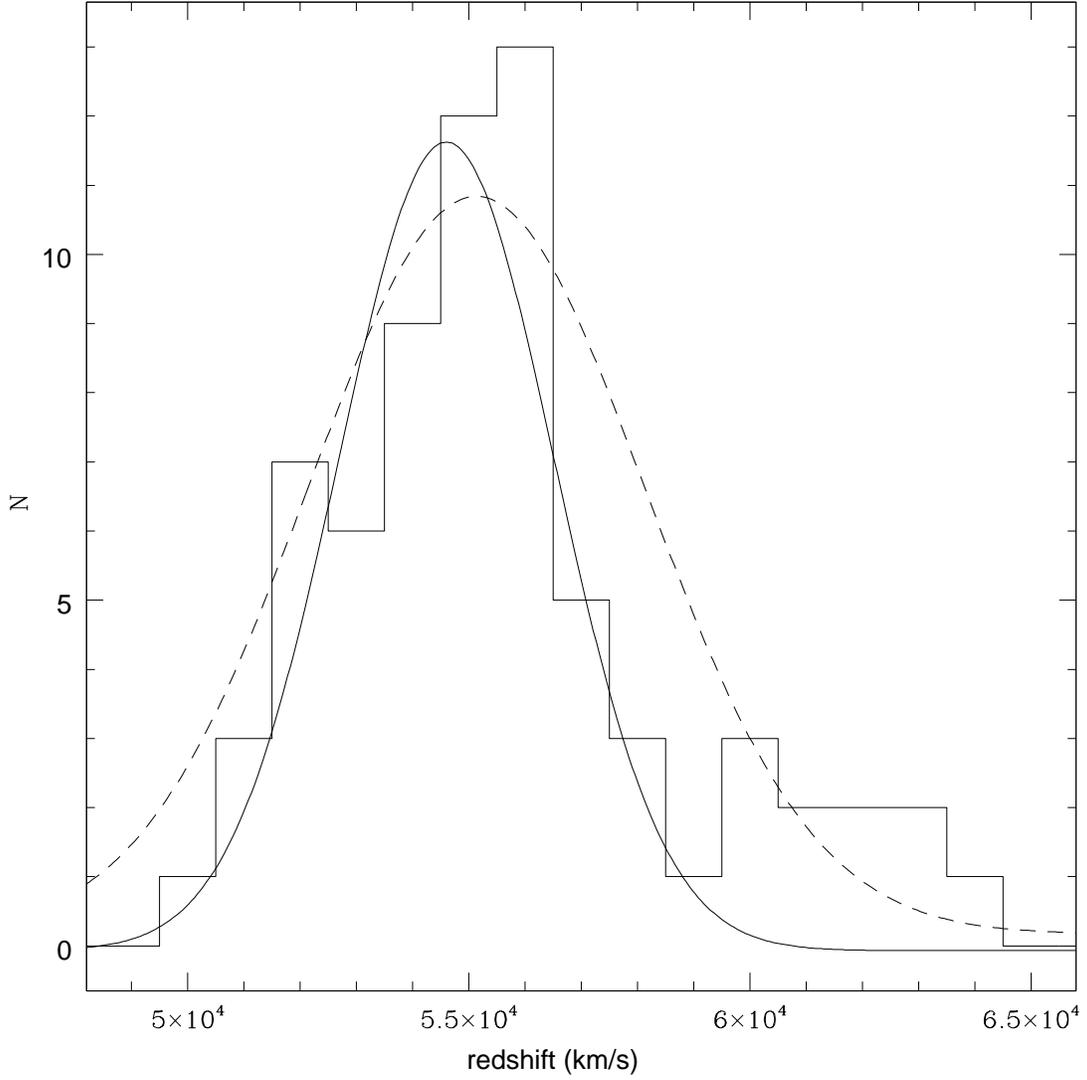}
\caption{ Redshift histogram for galaxies in field of A1689. The solid
curve is a Gaussian of width $1700km/s$ matched to the best fitting
NFW model (see text). The dashed curve is the original fit by Teague
et al., which includes the main subgroup of galaxies in the high
velocity tail of the histogram. This dispersion is too large to be
consistent with lensing if considered as a single relaxed object. The
preferred dispersion of $\sim 1700km/s$ for the best-fit NFW profile
to our data is over plotted (and corresponds to $\sim17Kev$), providing
a good fit to the bulk of the measured velocities and implies the
higher redshift group centered on $600km/s$ and corresponding to the
main subgroup of the cluster (seen displaced by 1.5' from the center of
mass in Figure 5) is not part of the relaxed body of the
cluster, but lies close in projection.}
\end{figure}

\end{document}